\documentclass[a4paper,10pt]{article}
\usepackage[utf8x]{inputenc}
\usepackage{amssymb}
\usepackage{amsmath}
\usepackage{amsthm}
\usepackage{mathrsfs}
\usepackage{amsfonts}
\usepackage{amsmath}
\usepackage{graphicx}
\usepackage{color}
\usepackage{subcaption}
\usepackage{soul}
\usepackage{cite}
 
\title{Photonic systems with two-dimensional landscapes of complex refractive index via time-dependent supersymmetry}
\author{A. Contreras-Astorga\\
C\'atedras CONACYT -- Departamento de F\'isica, Cinvestav, A.P. 14-740, 07000 \\ Ciudad de M\'exico, Mexico  \\ Department of Physics, Indiana
University Northwest, 3400 Broadway, Gary IN 46408, USA \\ alonso.contreras@conacyt.mx ; alonso.contreras.astorga@gmail.com
\and V. Jakubsk\'y\\
The Czech Academy of Science, Nuclear Physics Institute, Rez/Prague, Czech Republic\\
jakub@ujf.cas.cz ; v.jakubsky@gmail.com }

\begin{document}
\maketitle
\begin{abstract}
We present a framework for the construction of solvable models of optical settings with genuinely two-dimensional landscapes of refractive index. Solutions of the associated non-separable Maxwell equations in paraxial approximation are found using the time-dependent supersymmetry.
We discuss peculiar theoretical aspects of the construction.  In particular, we focus on the existence of localized solutions specific for the new systems. Sufficient conditions for their existence are discussed. Localized solutions vanishing for large $|\vec{x}|$, which we call light dots, as well as the guided modes that vanish exponentially outside the wave guides, are constructed. We consider different definitions of the parity operator and analyze general properties of the $\mathcal{PT}$-symmetric systems, e.g.  presence of localized states or existence of symmetry operators. Despite the models with parity-time symmetry are of the main concern, the proposed framework can serve for construction of non-$\mathcal{PT}$-symmetric systems as well. We explicitly illustrate the general results on 
a number of physically interesting examples, e.g. wave guides with periodic fluctuation of refractive index or with a localized defect, curved wave guides, two coupled wave guides or a uniform refractive index system with a localized defect. 
\end{abstract}

\section{Introduction}
In specific situations, propagation of light is governed by the same equations as matter waves in quantum mechanics. The coincidence of Maxwell equations  in paraxial approximation with the Schr\"odinger equation makes it possible to use methods of quantum mechanics in the analysis of the optical settings. 

This link proved to be particularly fruitful for investigation of optical systems where a complex refractive index representing balanced gain and loss prevents uncontrolled dimming or brightening of light \cite{Ruschhaupt,coupled,Makris,Christo1,ruter}. The Hamiltonian of the associated Schr\"odinger equation ceases to be Hermitian but possesses an antilinear symmetry. It was demonstrated two decades ago that such operators, having typically a $\mathcal{PT}$-symmetry with $\mathcal{P}$ and $\mathcal{T}$ being parity and time-reversal, can have purely real spectra \cite{BenderBoetcher}. It was showed later on that such models can provide consistent quantum mechanical predictions despite the non-Hermiticity of the Hamiltonian as long as the scalar product of the associated Hilbert space is redefined \cite{scalarproduct1,scalarproduct2,scalarproduct3,benderRev,AliMostRev}. As much as this task proved to be difficult to accomplish in explicit quantum systems, see e.g. \cite{Bila,Krejcirik}, it 
is non-existent in the realm of classical optics which, therefore, becomes an exciting field for the investigation of the systems described by $\mathcal{PT}$-symmetric (pseudo-Hermitian) Hamiltonians.

Supersymmetric quantum mechanics represents a highly efficient framework for construction of new exactly solvable models \cite{Cooper,Spectraldesign,Fernandez}. It is based on the Crum-Darboux transformation which is known in the analysis of Sturm-Liouville equations for a long time, see \cite{MatveevSale} and references therein. It allows to modify the potential term of the equation while preserving its solvability; the solutions of the new equation can be found by direct application of the Darboux transformation on the solutions of the original one. As by-product of the Crum-Darboux transformation, there could appear additional discrete energy levels in the spectrum of the new system. They are associated with the localized wave functions that were missing in the original system. i.e. these ``missing" states do not have preimage in the form of localized solutions. In this manner, the Crum-Darboux transformation can be employed in ``spectral design" of quantum systems \cite{Spectraldesign}. %In the framework of the supersymmetric quantum mechanics, the transformation forms the supercharge of the supersymmetric Hamiltonian.

Supersymmetry was utilized in the analysis of $\mathcal{PT}$-symmetric quantum models \cite{ZnojilSUSY,LevaiSUSY,AndrianovPTSUSY,CorreaLechtenfeld}.  It has been used in the construction of $\mathcal{PT}$-symmetric optical systems of required properties \cite{MiriPRL,Miri,LonghiEuroPhysLett,Mathias1,Mathias2,Tkachuk,MidyaInv,LonghiInv,Plyushchay,Correa15}.
For example, systems with invisible defects in crystal \cite{LonghiInv,Correa15}, with transparent interfaces \cite{LonghiEuroPhysLett} or unidirectional invisibility \cite{MidyaInv}  were constructed. Supersymmetry was also employed in the construction of random potentials with the energy bands in the spectrum and in the construction of $2D$ systems with potentials that are separable in Cartesian coordinates \cite{susyrandombands}, \cite{longhicrossroad}. It was also utilized in the context of coupled mode systems \cite{LonghiBragg,VJFCPTDirac}, in the formation of optical arrays \cite{mydia} or digital switching of spatially random waves \cite{park2}. Moreover, experiments with photonic lattices were performed  \cite{Mathias1,Mathias2}.

It is worth mentioning the increasing interest in the non-$\mathcal{PT}$-symmetric systems with complex refractive index where, however, the gain and loss can still support guided and non-decaying modes associated with real spectra of the associated Hamiltonians. Explicitly solvable models of these systems were constructed via supersymmetry in \cite{Miri}. They were also studied numerically \cite{Nixon,Turitsyna,Yang} and experimental setups were proposed in \cite{Hang}.

Vast majority of the settings considered in the literature are described by effectively one-dimensional Hamiltonians. In most cases, they possess translational symmetry, typically along the axis of propagation of the light beam. The two-dimensional, exactly (analytically) solvable models possessing separability in radial coordinates were considered in \cite{Agarwal,Macho}, the models separable in Cartesian coordinates were presented e.g. in \cite{susyrandombands,longhicrossroad}. $\mathcal{PT}$-symmetry breaking in two and three dimensions were considered in \cite{Stone}, scattering properties were studied in \cite{Ge}. Two-dimensional periodic arrays of localized gain and loss regions, called photonic crystals, were analyzed numerically in \cite{Mock}. %Solitons in two-dimensional nonlinear photonic systems were discussed in \cite{Xie,Wang}, see also \cite{Konotop}. 

In this work, we focus on the construction of  exactly solvable models of optical settings with non-separable complex refractive index with the use of the time-dependent Darboux transformation. Systems where the complex refractive index forms $\mathcal{PT}$-symmetric wave guides or that possess localized defects will be studied  with focus on existence of the missing states.  As the current experimental techniques \cite{Chen} seem to be ready for realization of such settings, analysis of exactly solvable models with genuinely two-dimensional complex inhomogeneities of the refractive index is desirable.

The work is organized as follows. In the next section, we present the framework of the time-dependent supersymmetry \cite{Samsonov1,Samsonov2} and discuss its peculiar properties. We find the missing state  for a broad family of systems, the localized solution of the new Schr\"odinger equation whose preimage\footnote{If the Crum-Darboux transformation $\mathcal{L}$ maps the state $f$ from the domain of the original Schr\"odinger operator $S_0$ into the state $g$ from the domain of the new Schr\"odinger operator $S_1$, i.e. $\mathcal{L}f=g$, we say that $f$ is preimage of $g$.} in the original system ceases to be localized. We show that it can be used in construction of symmetry operators of both the original and the new setting. Two definitions of the parity operator are introduced. They are distinctive for the models presented in the following sections. In the section \ref{tres}, we construct $\mathcal{PT}$-symmetric, exactly solvable models with a defect in the form of a localized gain and loss. We construct the missing state that represents a ``light dot'', the localized solution of the Maxwell equations in paraxial approximation. In the section \ref{cuatro}, we construct $\mathcal{PT}$-symmetric wave guides where fluctuations of refractive index are vanishing for $|x| \rightarrow\infty$ whereas they can be periodic along $z$-axis. We provide an alternative construction of the missing states that represent guided modes in the new system. We illustrate the general results on  explicit examples of periodically modulated $\mathcal{PT}$-symmetric wave guides supporting guided modes. In the subsection \ref{NonPT} we construct a model of a non-$\mathcal{PT}$-symmetric wave guide that possesses a guided mode despite the lack of $\mathcal{PT}$-symmetry.

\section{Mathematical framework}
In this section, we review briefly the main mathematical tools that will be used extensively in the forthcoming text. In particular, we present construction of the time-dependent Darboux transformation for the Schr\"odinger equation and discuss peculiarities of the construction for the $\mathcal{PT}$-symmetric systems. For the sake of completeness, let us start with a short review of the relation between the Schr\"odinger and the Maxwell equations.% with the time dependent Schr\"odinger equation and then the time-dependent version of the Darboux transformation applied to $\mathcal{PT}$-symmetric systems. 

\subsection{Paraxial approximation}
Consider a monochromatic light beam with wavelength in vacuum $\lambda$. Let $X,~Y$ and $Z$ be spatial coordinates. The Maxwell equations for electric and magnetic fields $\vec{E}=\vec{E}(X,Y,Z)$ and $\vec{H}=\vec{H}(X,Y,Z)$ of this monochromatic wave varying in time as $\exp(-i \omega t)$ are:
\begin{equation}
\nabla \times \vec{E} = i \omega \mu \vec{H}, \quad \nabla \times \vec{H}= - i  \omega \epsilon \vec{E}. \label{Maxwell}
\end{equation}
In this article we will focus on waves propagating mainly in the $Z$ direction in a medium with refractive index $n(X,Y,Z) = \sqrt{\mu \epsilon/\mu_0 \epsilon_0}=c \sqrt{\mu \epsilon}$. Under circumstances that will be discussed in this subsection, equations \eqref{Maxwell} can be written as a Schr\"odinger equation. This process is known as paraxial approximation \cite{Lax75,Permitin96,Miri,Cruz15,Cruz15b,Cruz17}. We revisit some important aspects of this approximation in this subsection as presented in \cite{Lax75}, with minor changes in notation.

Let us write the electric field $\vec{E}$ as 
\begin{eqnarray}
\vec{E}=\exp(i k n_0 Z)\left(\vec{\psi}_T + \hat{a}_Z \psi_Z \right), \label{ansatz}
\end{eqnarray}
where $k=2 \pi/ \lambda$ is the corresponding wave number in vacuum, $n_0$ is a reference value of the refractive index, $\hat{a}_Z$ is a unit vector in the $Z$ direction, $T$ stands for the transverse part of the field, $\vec{\psi}_T= \vec{\psi}_T(X,Y,Z)$ and $\psi_Z=\psi_Z(X,Y,Z)$. 
By taking the curl of the first equation in \eqref{Maxwell}, the equation that the electric field must satisfy is 
\begin{eqnarray}
\nabla (\nabla \cdot \vec{E}) - \nabla^2 \vec{E}= k^2 n^2 \vec{E}. \label{Eequation}
\end{eqnarray}
Then, substitution of ansatz \eqref{ansatz} in \eqref{Eequation} and the use of the notation $\nabla_T= \hat{a}_X \partial_X + \hat{a}_Y \partial_Y$ leads to the transverse equation 
\begin{eqnarray}
\nabla_T \left( \nabla_T \cdot \vec{\psi}_T+ i k n_0 \psi_Z + \partial_Z \psi_Z \right) - \nabla^2_T \vec{\psi}_T - \partial_Z^2 \vec{\psi}_T + k^2 n_0^2 \vec{\psi}_T - 2i k n_0 \partial_Z \vec{\psi}_T = k^2 n^2 \vec{\psi}_T,  \label{transverse}
\end{eqnarray}
and the longitudinal equation 
\begin{eqnarray}
i k n_0 \nabla_T \cdot \vec{\psi}_T + \partial_Z \left( \nabla_T \cdot \vec{\psi}_T \right) - \nabla^2_T \psi_Z = k^2 n^2 \psi_Z. \label{longitudinal}
\end{eqnarray}
Equations \eqref{transverse} and \eqref{longitudinal} can be approximated and simplified when introducing a small parameter. In this problem we have three different scales, first the wavelength $\lambda$, second the characteristic size of the beam in the transverse direction  $x_0$ and finally a longitudinal distance $\ell$ defined as $\ell = n_0 k x_0^2$ known as diffraction length. We define our parameter as $\nu=x_0/\ell$. Introducing the scaled variables 
\begin{eqnarray}
x=X/x_0, \quad y=Y/x_0, \quad z=Z/2 \ell,
\end{eqnarray}
equations \eqref{transverse} and \eqref{longitudinal} take the form 
\begin{eqnarray}
\nabla_\perp \left( \nu \nabla_\perp \cdot \vec{\psi}_T+ i  \psi_Z + \frac{\nu^2}{2} \partial_z \psi_Z \right) - \nu\nabla^2_\perp \vec{\psi}_T - \frac{\nu^3}{4}\partial_z^2 \vec{\psi}_T - i \nu \partial_z \vec{\psi}_T &=& \nu (k x_0)^2 (n^2-n_0^2) \vec{\psi}_T,  \label{transverse2} \\
i \nu \nabla_\perp \cdot \vec{\psi}_T + \frac{\nu^3}{2}\partial_z \left( \nabla_\perp \cdot \vec{\psi}_T \right) - \nu^2 \nabla^2_\perp \psi_Z &=& \nu^2 (k n x_0)^2 \psi_Z, \label{longitudinal2}
\end{eqnarray} respectively, where the scaled differential operators are $\nabla_\perp = x_0 \nabla_T$ and $\partial_z= 2\ell \partial_Z$. To introduce non-linear effects of the media, let us consider the refractive  index as 
\begin{equation}
n^2=n_0^2+n_0 k g m, \label{index nonlinear}
\end{equation}
where $g$ is the signal gain per meter and $m$ is called homogeneous broadening. Substituting \eqref{index nonlinear} in \eqref{longitudinal2} we obtain:
\begin{eqnarray}
i \nu \nabla_\perp \cdot \vec{\psi}_T + \frac{\nu^3}{2}\partial_z \left( \nabla_\perp \cdot \vec{\psi}_T \right) - \nu^2 \nabla^2_\perp \psi_Z = \left(1+\nu^2 \ell ~g~m \right) \psi_Z. \label{longitudinal3}
\end{eqnarray}
If the parameter $\nu$ is small, $\nu << 1$, we can expand our functions $\psi_Z$ and $\vec{\psi}_T$ in powers of $\nu$, i. e. 
\begin{eqnarray}
\psi_Z(x,y,z)&=& \psi_Z^{(0)}+\nu \psi_Z^{(1)}+ \nu^2
\psi_Z^{(2)}+ \dots \\
\vec{\psi}_T(x,y,z)&=& \vec{\psi}_T^{(0)}+ \nu \vec{\psi}_T^{(1)}+ \nu^2 \vec{\psi}_T^{(2)}+ \dots.  
\end{eqnarray}
From the zeroth-order term in $\nu$ of \eqref{longitudinal3} we obtained $\psi_Z^{(0)}=0$ and from the first-order terms $i \nabla_\perp \cdot \vec{\psi}_T^{(0)}= \psi_Z^{(1)}$. Thus, the lowest order in $\nu$ of \eqref{transverse2} can be written as 
\begin{eqnarray}
 i \partial_z \vec{\psi}_T^{(0)} + \nabla_\perp^2 \vec{\psi}_T^{(0)} - k^2 x_0^2(n_0^2-n^2)\vec{\psi}_T^{(0)} = 0.\label{paraxial 1}
\end{eqnarray} 
Each  vector component in \eqref{paraxial 1} satisfies a time dependent Schr\"odinger equation:
\begin{eqnarray}
i\partial_t \psi + \partial_x^2 \psi - V \psi = 0,
\end{eqnarray}
where the $z$ variable plays the role of time parameter and the potential $V = k^2 x_0^2(n_0^2-n^2)$. Typical numbers in $\text{LiNbO}_3$ waveguides are \cite{Zhang,Chen}: refractive index varying from $n_0=2.217$ to $n_\text{max}=2.230$, wavelength of light $\lambda = 1064$nm and characteristic size of beam $x_0=10\mu$m. Then, diffraction length is $\ell=1.30919$mm, the parameter $\nu$ takes the value $\nu=0.00763829$ and the potential $V$ is zero where $n=n_0$ and $V=-201.598$ in regions where $n=n_{\text{max}}$.

%Typical numbers in step index optical fibers are \cite{Jackson}: index of refraction of core and cladding of $n_1=1.4$ and $n_0=1.39298$, respectively, core diameter $50~ \mu$m, wavelength $\lambda =  0.85~ \mu$m and characteristic size of beam $x_0 = 25~ \mu$m, then using as longitudinal scale the diffraction length $\ell = 4619~ \mu$m and the potential $V$ is a potential well being zero in the cladding and $V=-669.355$ in the core region.     

\subsection{Time-dependent Darboux transformation and $\mathcal{PT}$-symmetry} \label{SUSY and PT}

To our best knowledge, the Darboux transformation in the context of optical systems was employed in the analysis of effectively one-dimensional models. In the current article, we shall focus on settings where the Schr\"odinger equation cannot be reduced to an effectively one-dimensional equation as the fluctuations of the refractive index (both its real and imaginary part) are genuinely two dimensional. 

\subsubsection*{Standard 1D supersymmetric quantum mechanics}
Standard one-dimensional quantum mechanics is based on the factorization of the 1D Hermitian Hamiltonian $H_0$
\begin{equation}\label{factorization}H_0=-\partial_x^2+V(x)=L^{\dagger}L, \quad \mbox{where}\quad L=\partial_x+\mathcal{W}(x), \quad  \mathcal{W}(x)=-\partial_x\ln u,\end{equation}  
$\mathcal{W}(x)$ is called superpotential and $u$ solves $(H_0-E_0)u=0$. 
The factorization allows for the construction of a new operator $H_1$ that is intertwined with $H_0$ by either $L$ or $L^\dagger$, 
$$H_1L=LH_0,\quad L^\dagger H_1=H_0L^\dagger,\quad H_1=LL^\dagger=H_0-2\partial_x^2\ln u(x).$$ 
The intertwining relations imply that we can get solutions of $(H_1-E)\phi=0$ from the solutions of $(H_0-E)\psi=0$ by $\phi=L\psi$.   The function $u$ is annihilated by $L$. However, one can define the eigenstate of $H_1$ corresponding to $E_0$ as $u_m=u^{-1}$.  It satisfies $(H_1-E_0)u_m=0.$ Its definition suggests that it can be identified with the bound state of $H_1$ provided that $u$ is exponentially expanding and has no zeros. %It  does not need to be a physical () 
Then $E_0$ represents a discrete energy of $H_1$ but it does not belong to the energy spectrum of $H_0$. The function $u_m$ is called missing state. The intertwining operator $L$ can be utilized for mapping scattering states of $H_0$ onto scattering states of $H_1$.  When $\mathcal{W}(x)$  is asymptotically constant for large $|x|$, the action of $L$ on the scattering states just alter their phase.

\subsubsection*{Time-dependent Darboux transformation and the missing states}
Let us suppose that the following Schr\"odinger equation 
\begin{eqnarray}\label{S_0}
S_0\psi=i \partial_z \psi + \partial_x^2 \psi -V_0(x,z) \psi=0,\quad x\in\mathbb{R},\quad z\in\mathbb{R},  \label{SUSY SE}
\end{eqnarray}
is exactly solvable and its solutions are known. We suppose that $V_0(x,z)$ has no singularities in $\mathbb{R}^2$ and it is sufficiently smooth.  We will use the time-dependent Darboux transformation discussed in \cite{Samsonov1,Samsonov2,Contreras17,Zelaya17} to generate another exactly solvable equation with a different potential term. Let us present here the main steps of the construction. As the factorization of (\ref{S_0}) in the spirit of (\ref{factorization}) is not possible, the construction is based on the intertwining relation 
\begin{eqnarray}
S_1 \mathcal{L} = \mathcal{L} S_0  \label{SUSY Intertwining}
\end{eqnarray}
that guarantees that we can get solutions of the new equation $S_1\phi=0$, where $\phi$ is defined as $\phi=\mathcal{L}\psi$, provided that $S_0\psi=0$ and $\mathcal{L}$ maps the domain of $S_0$ into the domain of $S_1$.
The ansatz for the intertwining operator $\mathcal{L}$ is in the form a first order differential operator, $S_1$ is a Schr\"odinger operator with an altered potential term,
\begin{eqnarray}\label{S1L}
\mathcal{L}= L_1(z) \left[ \partial_x+\mathcal{W}(x,z)\right], \quad S_1= i \partial_z +\partial_x^2 -V_1(x,z), \quad  \mathcal{W}(x,z)=- \frac{\partial_x u(x,z)}{u(x,z)}. \label{S01L}
\end{eqnarray}
Here, $V_1(x,z)$, $u(x,z)$ and $L_1(z)$ are to be fixed such that the intertwining relation (\ref{SUSY Intertwining}) is satisfied. 
Substituting (\ref{S_0}) and (\ref{S1L}) into (\ref{SUSY Intertwining}), one can find that the latter relation can be satisfied as long as 
\begin{eqnarray}\label{V_1}
V_1(x,z)=V_0(x,z)+i \partial_z \ln L_1(z)-2 \partial_x^2 \ln u(x,z) \label{SUSY V1}
\end{eqnarray}
and 
\begin{equation}\label{uc}
S_0 u(x,z)= c(z) u(x,z),
\end{equation}
see \cite{Samsonov1} for details. As the function $c(z)$ affects just the phase of the solution\footnote{If $S_0\psi=0$ holds, then we can find solution of $(S_0-c(t))\tilde{\psi}=0$ that reads $\tilde{\psi}=\exp(-i\int c(t))\psi$.} but not the potential $V_1$, it can be set to zero, $c(z)=0$. In what follows, we will denote by $u$ the solution of $S_0u=0$ used in definition of the new potential (\ref{V_1}) and of the intertwining operator \eqref{S1L}. It will be called transformation function.

Relations (\ref{V_1}) and (\ref{uc}) are sufficient to establish the intertwining relation. In addition, the function $u(x,z)$ as well as $L_1(z)$ are also required to be nodeless, otherwise, the transformation would be singular and it would fail to provide the mapping between the domains of $S_0$ and $S_1$. When $\mathcal{L}$ as well as $V_1$ are regular, relation \eqref{SUSY Intertwining} guarantees that we can generate solutions of $S_1 \phi(x,z)=0$ from the solutions $\psi(x,z)$ of \eqref{SUSY SE} by $\mathcal{L}$, 
\begin{eqnarray}
\phi(x,z)= \mathcal{L} \psi(x,z). \label{SUSY states}
\end{eqnarray}

We can try to find the ``inverse" transformation $\mathcal{L}^\sharp$ such that it satisfies
\begin{equation}\label{inverseInt}
 S_0\mathcal{L}^\sharp=\mathcal{L}^\sharp S_1.
\end{equation}
%i.e. it is the ``inverse'' of $\mathcal{L}$. 
Using the general formulas (\ref{S1L}), we take $S_1$ as the initial system and we define $\mathcal{L}^\sharp=L_2(z)v(x,z)\partial_x \frac{1}{v(x,z)}$ where $v(x,z)$ solves $S_1v(x,z)=0$. Then it is granted that there holds $\mathcal{L}^\sharp S_1=S_2\mathcal{L}^\sharp$ for 
\begin{equation}\label{S2}
 S_2=i \partial_z +\partial_x^2 -V_0(x,z)-i\partial_z\ln L_1 L_2+2\partial_x^2\ln u v.
\end{equation}
In order to identify $S_2=S_0$, we have to eliminate the last two terms by setting $2\partial_x^2\ln u v=i\partial_z\ln L_1 L_2$. As the right-hand side of the latter equation is $x$-independent, we have to fix $v$ such that 
\begin{equation}\label{uuvv}
 \partial_x^3\ln u v=0
\end{equation}
and also we must fix $L_2=L_1^{-1}(z)\exp (-2i\int^z (\ln u v)'')$. Then the last two terms in (\ref{S2}) vanish  and $\mathcal{L}^\sharp$ represents the inverse 
intertwining operator. 

\subsubsection*{Finding a missing state and symmetry operators}
The operator $\mathcal{L}$ can map any solution of $S_0f=0$ to a nontrivial solution of $S_1g=0$ as $g=\mathcal{L}f$, except the case where $f\equiv u$ as it gets annihilated by $\mathcal{L}$, $\mathcal{L}u=0$. Hence, the image of $u$ is missing in the new system. We can try to find another solution of $S_1g=0$ given in terms of the function $u$. In the one-dimensional supersymmetric quantum mechanics, this missing state is defined as $u^{-1}$. This formula hints on the importance of the missing state; when $u$ is exponentially growing, the missing state is square integrable and represents a bound state of the new system. In \cite{Samsonov1}, similar formula was used for the time-dependent, Hermitian systems. Inspired by these results, let us make an ansatz for the missing state in the following form
\begin{equation}\label{um}
 u_m=\frac{1}{f(z)\mathcal{S}u},
\end{equation}
where $f(z)$ is a function and $\mathcal{S}$ is an operator whose properties are to be fixed such that the equation $S_1u_m=0$ is satisfied.  We shall compute $S_1u_m$. We have
\begin{equation}\label{28}
 S_1\frac{1}{f\mathcal{S}u}=\frac{1}{(\mathcal{S}u)^2f}\left(-i\dot{(\mathcal{S}u)}+(\mathcal{S}u)''-V_0 \mathcal{S}u+2\left(\ln \frac{u}{\mathcal{S}u}\right)''-i\dot{(\ln{(L_1 f)})}\mathcal{S}u\right).
\end{equation}
If the condition $\partial_x^3\ln \frac{u}{\mathcal{S}u}=0$ holds, then we can fix $f(z)=L_1^{-1}(z)\exp(-2i\int^z \left(\ln \frac{u}{\mathcal{S}u}\right)'')$ and  the last two terms in (\ref{28}) vanish. 
Hence, $u_m$ solves $S_1u_m=0$ provided that there holds 
\begin{equation}\label{Scond}
 [-\partial_x^2+V_0,\mathcal{S}]=0,\quad \{i\partial_z,\mathcal{S}\}=0,\quad\partial_x^3\ln \frac{u}{\mathcal{S}u}=0,
\end{equation}
and $f(z)$ is fixed as
\begin{equation}\label{f(z)}
 f(z)=L_1^{-1}(z)\exp\left(-2i\int^z \left(\ln \frac{u}{\mathcal{S}u}\right)''\right).
\end{equation}
We can see that the third relation in (\ref{Scond}) coincides with (\ref{uuvv}) for $v\equiv u_m$. Therefore, when (\ref{Scond}) are satisfied, there also exist the inverse operator,
\begin{equation}\label{tildeL}
 \mathcal{L}^\sharp=f(z)u_m\partial_x \frac{1}{u_m},
\end{equation}
where $u_m$ and $f(z)$ are defined in (\ref{um}), (\ref{Scond}) and (\ref{f(z)}).  

Existence of the inverse operator (\ref{tildeL}) implies another interesting fact; both $S_0$ and $S_1$ have symmetry operators
\begin{equation}
 [S_0,\mathcal{L}^\sharp\mathcal{L}]=0,\quad [S_1,\mathcal{L}\mathcal{L}^\sharp]=0,
\end{equation}
where
\begin{eqnarray}\label{tildeLL}
\mathcal{L}^\sharp \mathcal{L}= \exp\left(-2i\int^z \left(\ln u_m u\right)''\right) \left[  \partial_x^2-\left(\ln u_m u\right)'\partial_x+ (\ln u)' (\ln u_m)'-(\ln u)'' \right],\nonumber\\
\mathcal{L}\mathcal{L}^\sharp=\exp\left(-2i\int^z \left(\ln u_m u\right)''\right) \left[ \partial_x^2-\left(\ln u_m u\right)'\partial_x+ (\ln u)' (\ln u_m )'-(\ln u_m)'' \right].
\end{eqnarray}

\subsubsection*{$\mathcal{PT}$-symmetry}
Up to now, we did not make any assumption on the Hermiticity or $\mathcal{PT}$-symmetry of the new potential $V_1$. In \cite{Samsonov1}, both $V_0$ and $V_1$ were required to be real in order to preserve Hermiticity of $S_0$ and $S_1$. In the Hermitian case the operator $\mathcal{S}$ can be identified with $\mathcal{S}f(x,z)=\overline{f(x,z)}$. Then $V_1$ is real whenever $u$ satisfies $\partial_x^3\ln \frac{u}{\overline{u}}=0$ and $L_1$ is fixed as $L_1=\exp\left(-i\int\left(\ln \frac{u}{\overline{u}}\right)''dx\right)$. This ``Hermitian" definition of $\mathcal{S}$ complies with (\ref{Scond}).  

We are interested in the settings where $S_1$ ceases to be Hermitian but possesses an antilinear symmetry that we shall identify with the simultaneous action of the operators of time-reversal $\mathcal{T}$ and space inversion $\mathcal{P}$. 
Most of the $\mathcal{PT}$-symmetric systems discussed in the literature are effectively one-dimensional so that the space inversion $\mathcal{P}$ is defined unambiguously as $\mathcal{P}f(x)=f(-x)$. In two dimensions, we can define $\mathcal{P}$ as the reflection with respect to a fixed point or with respect to an axis,
\begin{equation}\label{Ps}
 \mathcal{P}_xf(x,z)=f(-x,z)\quad\mbox{or}\quad\mathcal{P}_2f(x,z)=f(-x,-z).
\end{equation}
The antilinear operator $\mathcal{T}$ is given as
\begin{equation}
 \mathcal{T}f(x,z)=\overline{f(x,z)}.
\end{equation}
We suppose that $V_0$ is $\mathcal{PT}$-symmetric and we require $V_1$ to be $\mathcal{PT}$-symmetric as well,
 \begin{equation}\label{PTV_1PT}
 \mathcal{PT}V_1(x,z)\mathcal{PT}=V_1(x,z).
\end{equation}
It will restrict the possible choice of $u(x,z)$ and $L_1$ in dependence on the actual definition of $\mathcal{P}$. 

First, let us consider $\mathcal{P}\equiv\mathcal{P}_x$.  Then $V_1$ is $\mathcal{P}_x\mathcal{T}$-symmetric provided that $u$ and $L_1$ satisfy
\begin{equation}\label{uPT1}
2\partial_x^2\ln\frac{u}{\overline{u(-x,z)}}=i\partial_z\ln |L_1(z)|^2.
\end{equation}
The $\mathcal{P}_x\mathcal{T}$ operator anticommutes with  $i\partial_z$ and it commutes with $V_0$, so that it fulfills the first two conditions in (\ref{Scond}). As the condition (\ref{uPT1}) is stronger than the third relation in (\ref{Scond}), we find that when the potential is $\mathcal{P}_x\mathcal{T}$-symmetric, then it also possesses missing state defined by (\ref{um}) (with $\mathcal{S}\equiv\mathcal{P}_x\mathcal{T}$), the inverse operator $\mathcal{L}^\sharp$ and the symmetry operators (\ref{tildeLL}).

If we set $\mathcal{P}=\mathcal{P}_2$, the requirement (\ref{PTV_1PT}) reduces to 
\begin{equation}\label{uPT2}
2\partial_x^2\ln\frac{u(x,z)}{\overline{u(-x,-z)}}=i\partial_z\ln\frac{L_1(z)}{\overline{L_1(-z)}}.
\end{equation}
The operator $P_2T$ commutes with $i\partial_z$, so that we cannot identify it with $\mathcal{S}$ in (\ref{Scond}). 
	
A few comments are in order. Having the transformation function $u$, we can define a whole family of systems that differ by the choice of $L_1$. When the function $u$ satisfies $\partial_x^3\ln\frac{u(x,z)}{\mathcal{P}_x\mathcal{T}u(x,z)}=0$, then we can identify $\mathcal{S}\equiv \mathcal{P}_x\mathcal{T}$ and the missing state is defined by (\ref{um}). In this family, we can set $L_1$ in accordance with (\ref{uPT1}) and the resulting system will be  $\mathcal{P}_x\mathcal{T}$-symmetric. When the function $u$ also satisfies  $\partial_x^3\ln\frac{u(x,z)}{\mathcal{P}_2\mathcal{T}u(x,z)}=0$, then we can find $\mathcal{P}_2\mathcal{T}$-symmetric systems in the family as one can define $L_1$ in coherence with (\ref{uPT2}). If $\partial_x^3\ln\frac{u(x,z)}{\mathcal{P}_2\mathcal{T}u(x,z)}=0$ holds, but  $\partial_x^3\ln\frac{u(x,z)}{\mathcal{P}_x\mathcal{T}u(x,z)}=0$ does not, there can be only $\mathcal{P}_2\mathcal{T}$ symmetric systems in the family and we cannot use the definition (\ref{um}) of the missing state.

\subsubsection*{Higher order (Crum)-Darboux transformations}
Let us make a few comments on the repeated use of the time-dependent Darboux transformation (\ref{V_1}). Having the Schr\"odinger operators $S_0$ and $S_1$ intertwined by $\mathcal{L}_1$,
\begin{equation}\label{ir1}
S_1\mathcal{L}_1=\mathcal{L}_1S_0, \quad S_1=S_0+2\partial_x^2\ln u_1-i\partial_z \ln L_1,\quad \mathcal{L}_1=L_1u_1\partial_xu_1^{-1},
\end{equation}
we can select a function $\check{u}_2$ such that $S_1\check{u}_2=0$ and use it to define the new intertwining operator $\mathcal{L}_2$ that satisfies 
\begin{equation}\label{ir2}
S_2\mathcal{L}_2=\mathcal{L}_2S_1, \quad S_2=
S_0+2\partial_x^2\ln u_1+2\partial_x^2\ln \check{u}_2-i \partial_z\ln L_1L_2,\quad \mathcal{L}_2=L_2\check{u}_2\partial_x\check{u}_2^{-1}.
\end{equation}
In this manner, a chain of the new solvable equations can be obtained.
Combining (\ref{ir1}) and (\ref{ir2}), we can see immediately the that the operators $S_0$ and $S_2$ are intertwined by the operator $\mathcal{L}_{12}=\mathcal{L}_2\mathcal{L}_1$. When we find a preimage ${u}_2$ of $\check{u}_2$ such that $\mathcal{L}_1u_2=\check{u}_2$ (the function $u_2$ does not need to be a solution of $S_0u_2=0$), we can rewrite both $\mathcal{L}_{12}$ and $S_2$ directly in terms of $u_1$ and $u_2$,
\begin{equation}\label{s2}
S_2\mathcal{L}_{12}=\mathcal{L}_{12}S_0,\quad S_2=S_0+2\partial_x^2\ln W(u_1,u_2)-i\partial_z\ln L_1L_2, 
\end{equation}
and
\begin{equation}\label{ir12}
\mathcal{L}_{12}=\frac{L_1L_2}{W(u_1,u_2)}\left|\begin{array}{ccc}u_1&u_2&1\\u_1'&u_2'&\partial_x\\u_1''&u_2''&\partial^2_x\end{array}\right|=L_2L_1\left(\partial_x^2+\frac{u_2u_1''-u_1u_2''}{W(u_1,u_2)}\partial_x+\frac{-u_2'u_1''+u_1'u_2''}{W(u_1,u_2)}\right),
\end{equation}
where $W(u_1,u_2)=u_1u_2'-u_1'u_2$. The formulas (\ref{ir12}) can be generalized for an arbitrary chain-length of time-dependent Darboux transformations, see \cite{Samsonov1}. The properties of the final system, existence and properties of the missing states in particular, can be deduced from the careful analysis of the intermediate models (e.g. a missing state of $S_1$ gets transformed into a missing state of $S_2$ by $\mathcal{L}_2$). However, it is worth noticing that despite $S_1$ can have singularities in the potential, the potential term of $S_2$ can be a regular function. It stems from the fact that despite $u_1$ can have zeros (which would introduce singularities into $S_1$), the Wronskian of $u_1$ and $u_2$ can be nodeless, keeping $S_2$ regular. %We will discuss such a case in the following text.

When $V_0$ is $z$-independent, we can write the solutions $S_0u=0$ in terms of the stationary states $u(x,z)=e^{-i\epsilon z}\psi_\epsilon(x)$, $(-\partial_x^2+V_0-\epsilon)\psi_\epsilon=0$. When $u_1$ and $\check{u}_2=\mathcal{L}_1u_2$ are stationary states of $S_0$ and $S_1$, respectively, then the potential term of $S_2$ is also $z$-independent. The transformation $\mathcal{L}_{12}$ can be identified as the $N=2$ (time-independent) Crum-Darboux transformation.

The stationary states $u_1$ and $u_2$ can be selected as two eigenstates corresponding to different energy levels. Alternatively, we can define the function $u_2$ in terms of $u_1$: taking $u_1=e^{-iE_m z}{\psi}_{m}(x)$, we can fix   
\begin{equation}\label{confluentu2}
u_2(x,z)=e^{-iE_mz}{\psi}_m(x)\left(\int_{x_0}^x\frac{1}{{\psi}_m^2}\left(\int_{s_0}^s{\psi}_m^2(r)dr+\alpha\right)ds+a\right),
\end{equation}
where $a$ and $\alpha$ are complex constants. The function $u_2$ satisfies $S_1\mathcal{L}_1u_2=0$, but $S_0u_2\neq 0$. Instead, it fulfills  $S_0^2u_2=0$, see \cite{Correa15}. The operator $\mathcal{L}_{12}$ is called  the confluent Crum-Darboux transformation in the literature, see e.g. \cite{MatveevSale,Correa15,Mielnik00,Salinas03,Schulze13,Contreras15,Plyushchay16} and references therein. The new Schr\"odinger operator $S_2$ can be written in terms of $\psi_m$ as
\begin{align}
& S_2=i\partial_z+\partial_x^2-V_2(x), \nonumber \\
&V_2(x)=V_0 -2\partial_x^2 \ln \left(\alpha+\int^x_0{\psi}^2_m(s)ds\right)= V_0 - 4\frac{{\psi}_m \partial_x{\psi}_m }{\alpha + \int_{0}^x {\psi}^2_m(s) ds} + 2 \frac{{\psi}_m^4}{\left( \alpha + \int_{0}^x {\psi}^2_m(s) ds \right)^2}. \label{Confluent SUSY V2}
\end{align}
When ${\psi}_m$ is a real function, the new potential will be free of singularities provided that $\alpha$ is a complex number with a non-vanishing imaginary part. The stationary states $f_n$ of $S_2$ for $n\neq m$ can be found by direct application of $\mathcal{L}_{12}$,
\begin{eqnarray}\label{Confluent SUSY Solutions}
f_n(x,z)&=&\mathcal{L}_{12}e^{-iE_nz}{\psi}_n(x)=L_1 L_2 \left(\partial_x-\frac{\partial_x\check{u}_2}{\check{u}_2}\right)\left(\partial_x-\frac{\partial_x u_1}{u_1}\right){\psi}_n(x)e^{-iE_nz}.
\end{eqnarray}
For $n=m$, we can find the following solution (see e.g. \cite{Correa15}) that represents the missing state of $S_2$,
\begin{eqnarray}\label{confluentmissingstate}
f_m(x,z)= \frac{{{\psi}}_m(x)}{\alpha + \int_{0}^x {{\psi}}_m^2(s) ds}e^{-iE_mz}. \label{Confluent missing state}
\end{eqnarray}
It is worth comparing the confluent transformation $\mathcal{L}_{12}$ with the first order (time-independent) Darboux transformation $\mathcal{L}$. Both transformations are defined in terms of a single function $\psi_m$ which also determines the form of the missing state; it is (\ref{confluentmissingstate}) for the confluent transformation whereas $\sim\psi_m^{-1}$ for the first order transformation. One can see from (\ref{confluentmissingstate}) that $f_m$ can be square integrable even in the case when ${\psi}_m$ is a bounded function \cite{Correa15}. This result cannot be obtained with the first order transformation.

%%%%%%%%%%%%%%%%%%%%%%%%%%%%%%%%%%%%%%%%%%%%%%%5
%%%%%%%%%%%%%%%%%%%%%%%%%%%%%%%%%%%%%%%%%%%%%%%
%%%%%%%%%%%%%%%%%%%%%%%%%%%%%%%%%%%%%%%%%%%%%%%%%

%%%%%%%%%%%%%%%%%%%%%%%%%%%%%%%%%%%%%%%%%%%%%%%5
%%%%%%%%%%%%%%%%%%%%%%%%%%%%%%%%%%%%%%%%%%%%%%%
%%%%%%%%%%%%%%%%%%%%%%%%%%%%%%%%%%%%%%%%%%%%%%%%%

\subsection{Prelude to the next sections}
	In the forthcoming text, we will focus on two different scenarios:
	\begin{itemize}
		\item  a localized defect of the refractive index 
		\item a straight wave guide with a periodically modulated profile
	\end{itemize}
In the explicit construction of the solvable models, we will depart from the free particle system described by the equation
\begin{equation}\label{FPEq}
S_0f=(i\partial_z+\partial_x^2)f=0.
\end{equation}
This choice will help us to keep the illustrative examples simple enough and provide straightforward analysis of the missing states. It is worth mentioning that in the literature, see e.g. \cite{Miri, MiriPRL, Mathias1, Mathias2}, the supersymmetric techniques are usually utilized to annihilate a given localized mode (ground state) so that it is no longer present in the new, superpartner system. We intend to go the opposite way; the new systems should posses additional localized solutions that have no preimage in the original one. 
	
The transformation function $u$, $S_0u=0$, determines the properties of the new system to a large extend.
There are the two notoriously known types of solutions of (\ref{FPEq}), the plane waves
\begin{equation}\label{planewave}
\Phi_{k,x_0,z_0,v_0}=e^{\pm i k (x-x_0+v_0z)-\frac{iv_0}{4}(2x+v_0 z)-i k^2(z-z_0)}
\end{equation}	
and the wave packets
\begin{equation}\label{wavepacket}
\Psi_{x_0,z_0,v_0,\sigma}=\frac{1}{\sqrt{i(z-z_0)+\sigma}}e^{-\frac{(x-x_0+v_0(z-z_0))^2}{4(i(z-z_0)+\sigma)}-\frac{i}{4}v_0(2x+v_0z)},
\end{equation}
where $k$, $x_0$, $z_0$, $v_0$ and $\sigma$ are real parameters. The wave packet (\ref{wavepacket}) can be written as an infinite linear combination of the plane waves. Let us consider properties of the intertwining operator for different choices of $u$. 

Neither (\ref{planewave}) nor (\ref{wavepacket}) are optimal for direct identification with the transformation function $u$; the finite combination $u$ of $\Phi_{k,x_0,z_0,v_0}$ does not satisfy $\partial_x^3\ln\frac{u(x,z)}{\overline{u(-x,z)}}=0$, i.e. the condition (\ref{Scond}) is not satisfied and the formula (\ref{um}) for the missing state cannot be used\footnote{Here we identify $\mathcal{S}\sim \mathcal{P}_x\mathcal{T}$.}. As we shall see, identification of $u$ with the wave packets $\Psi_{x_0,z_0,v_0,\sigma}$ does not lead to the system with required properties of the refractive index. 
We will circumvent both these difficulties: in the first case, we will provide an alternative way for construction of the missing states. In the second case, we will construct other wave-packet-like solutions via a transformation that relates the free particle system with the one of the harmonic oscillator. This mapping consists of a specific change of coordinates and a gauge-like transformation. It can be written as
\begin{equation}
e^{-if(x,z)}S_{HO}(y(x,z),t(z))e^{if(x,z)}=g(z)S_{0}(x,z),
\end{equation}
where $S_{HO}(y,t)=i\partial_t+\partial_y^2-y^2/4$. The functions $y=y(x,z)$, $t=t(z)$, $f(x,z)$, and $g(z)$ are to be specified in section \ref{tres}.

The formula (\ref{V_1}) for the potential term of $S_1$ 
suggests that when $u$ is a (finite) linear combination of the plane waves (\ref{planewave}), the potential term will  be non-vanishing and oscillating  along the $z$-axis and, hence, it could form a wave guide. The wave packet solutions will be the candidates for the construction of the localized defects of $n(x,z)$. Hence, the following two sections will be distinguished by these two different choices of $u$,
\begin{align}
u=\begin{cases}\mbox{wave-packet-like solutions} \longrightarrow \mbox{localized defects of $n(x,z),$}\\\mbox{finite combination of stationary solutions}\longrightarrow \mbox{straight wave guides}.\end{cases}
\end{align}

 The action of the intertwining operator $\mathcal{L}$ can dramatically change the profile of the transformed function. 
If we select $u$ as a finite linear combination of the plane waves (\ref{planewave}) (that has no zeros), the superpotential $\mathcal{W}(x,z)=-\partial_x\ln u(x,z)$ is bounded both for large $|z|$ and $|x|$. It resembles the one-dimensional superpotential $\mathcal{W}(x)$ that is usually fixed such that it is asymptotically constant for large $|x|$. 
When we identify $u$ with the wave packet that has $x^2$ term in the exponential, the superpotential then behaves as $\mathcal{W}(x,y)=-\partial_x\ln u(x,y)=O(x)$ for large $|x|$ and constant $z$. When we apply such $\mathcal{L}$ on the plane waves (\ref{planewave}), the resulting function will be an unbounded function of $x$. However, when we apply it on another wave packet (\ref{wavepacket}), we get a function that is still vanishing rapidly for large $|x|$. It is promising, as we would like to transform a generic wave packets $\Psi_{x_0,z_0,v_0,\sigma}$ into functions that have also bounded amplitude for all $x$ and $z$. The behavior of $\mathcal{L}\Psi_{x_0,z_0,v_0,\sigma}$ along the $z$-axis is largely affected by the choice of $L_1(z)$. We introduce an additional requirement  that should guarantee boundedness of the transformed wave packets, 
	\begin{equation}\label{wavepacketpreservation}
	\mathcal{L}\Psi_{x_0,z_0,v_0,\sigma}=G(x,z)\Psi_{x_0,z_0,v_0,\sigma} \quad \mbox{where} \quad  |G(x,z)|\leq C<\infty,\quad \forall x,y\in\mathbb{R}.%\quad \mbox{for} \quad |z|\rightarrow\infty\quad\mbox{and constant $x$},
	\end{equation}
The function $G(x,z)$ reflects how the wave packet gets changed by the transformation and its asymptotic properties depend on the explicit form of both $u$ and $L_1(z)$. Hence, the relation (\ref{wavepacketpreservation}) will impose additional restriction (besides (\ref{uPT1}) or (\ref{uPT2}) imposed by $\mathcal{PT}$-symmetry) on the possible choice of $L_1(z)$ and will be used to fix this function in the explicit models.  
		
\section{Systems with localized defects of refractive index\label{tres}}

By selecting different wave-packet-like solutions as transformation function $u$, we will construct systems with localized defects of refractive index in this section.

\subsection{Straight wave guide divided symmetrically by gain and loss regions}
In our seek for systems with localized defects of refractive index, let us start with a simple choice of $u$ 
\begin{equation}\label{gaussian1}
u(x,z)=\frac{(2 \pi)^{1/4}}{\sqrt{1-iz}}\exp\left( \frac{x^2}{4(1-iz)}\right).
\end{equation}
This function expands exponentially for large $|x|$ and it can be obtained from the  Gaussian wave packet by the substitution $z\rightarrow-z$, $x\rightarrow ix$.
It satisfies the relation (\ref{Scond})  
and it is nodeless. Therefore, the missing state (\ref{um}) and  the symmetry operators (\ref{tildeLL}) are well defined. 

\begin{figure}[t!]
	\begin{center}
	\begin{subfigure}[b]{0.3\textwidth}
        \includegraphics[width=\textwidth]{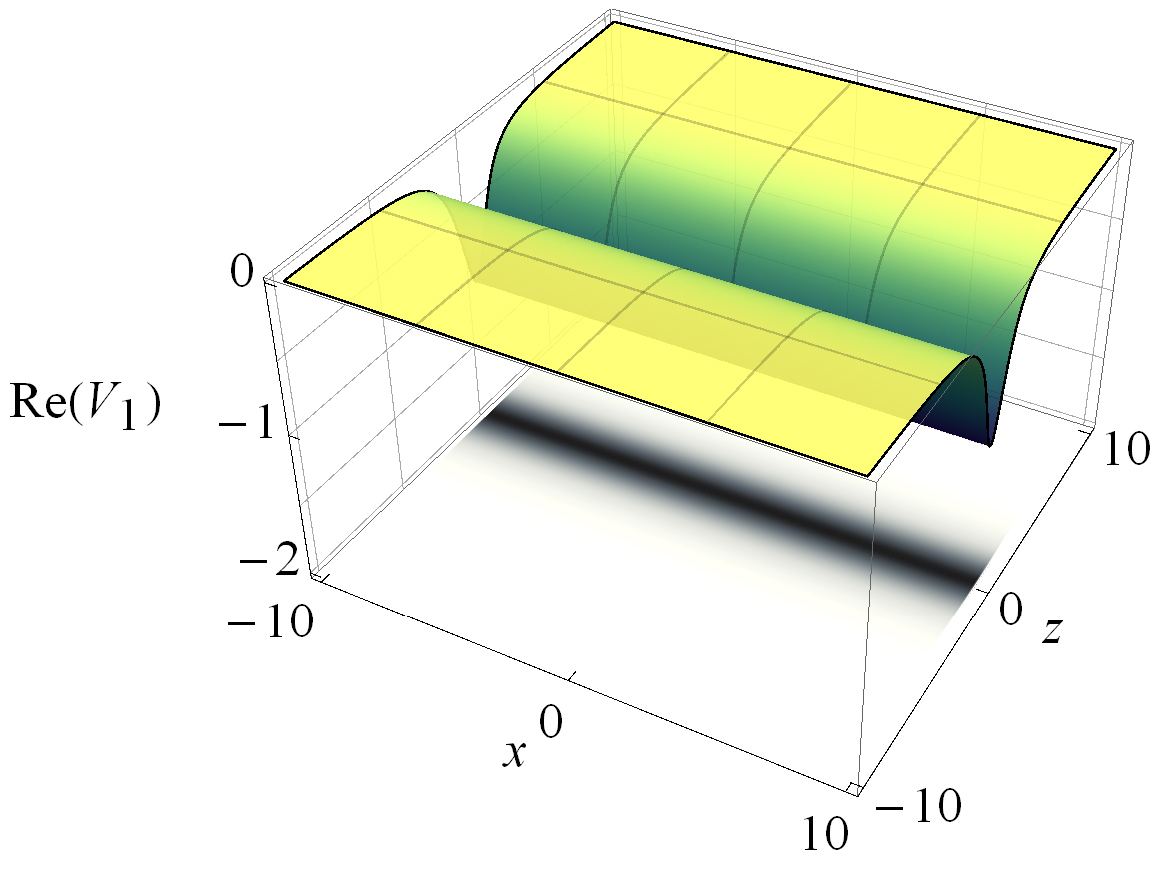}
        \caption{}
    \end{subfigure}
\begin{subfigure}[b]{0.3\textwidth}
        \includegraphics[width=\textwidth]{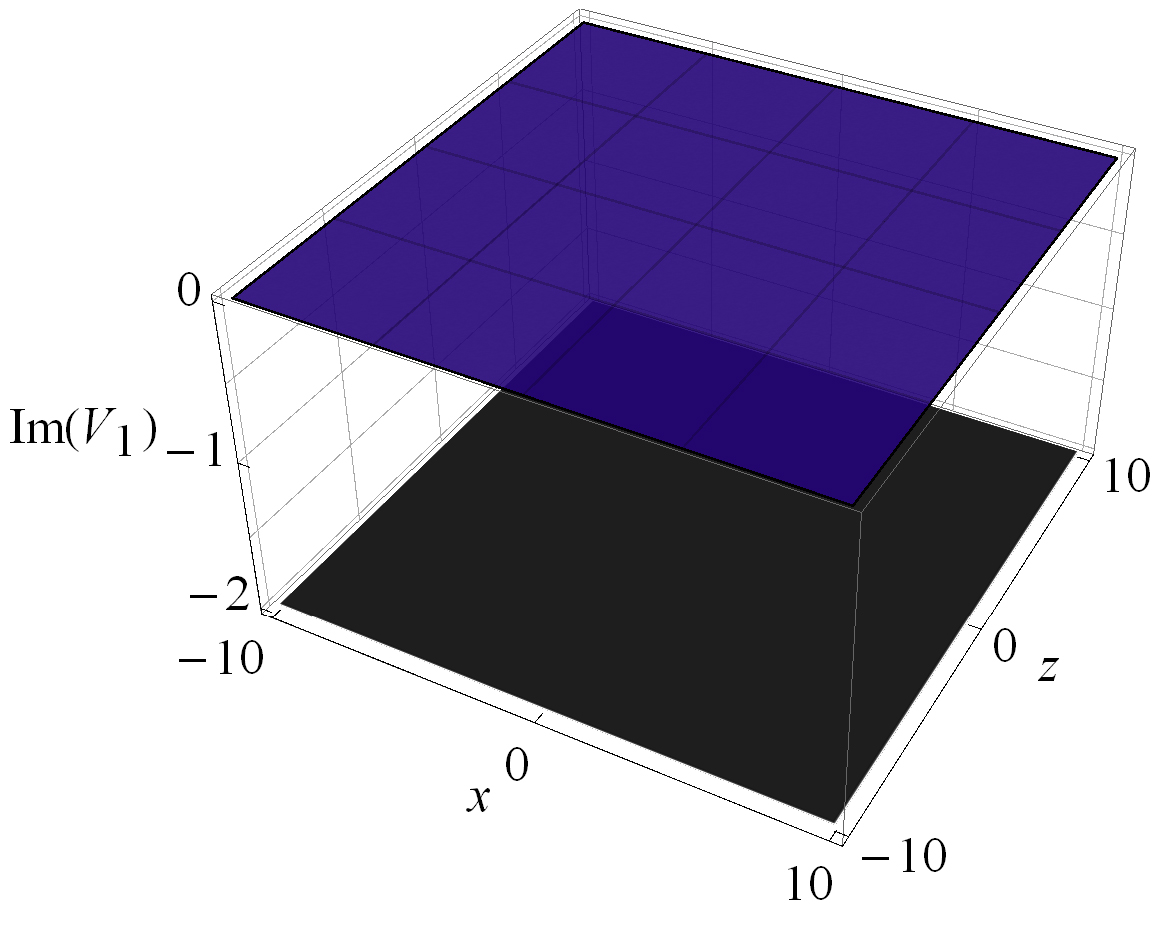}
        \caption{}
    \end{subfigure}
    \begin{subfigure}[b]{0.3\textwidth}
        \includegraphics[width=\textwidth]{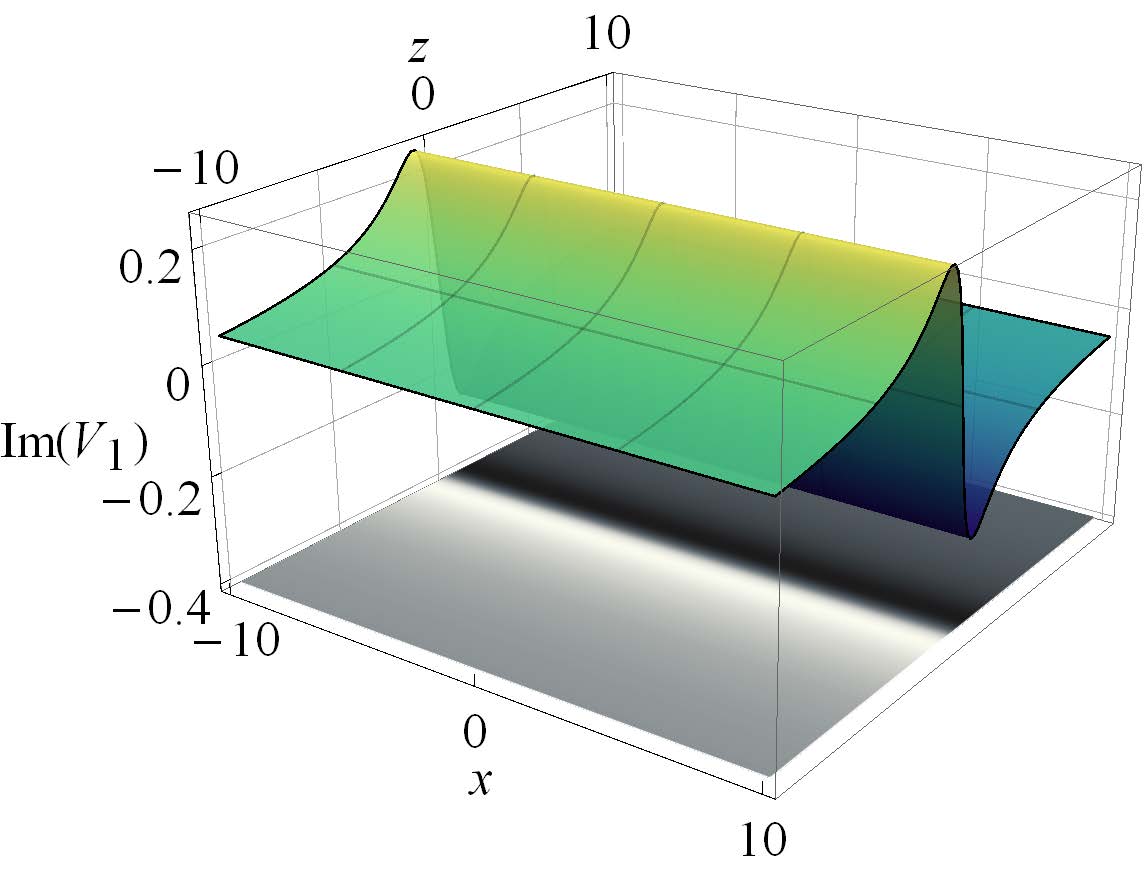}
        \caption{}
    \end{subfigure}\\
\begin{subfigure}[b]{0.3\textwidth}
        \includegraphics[width=\textwidth]{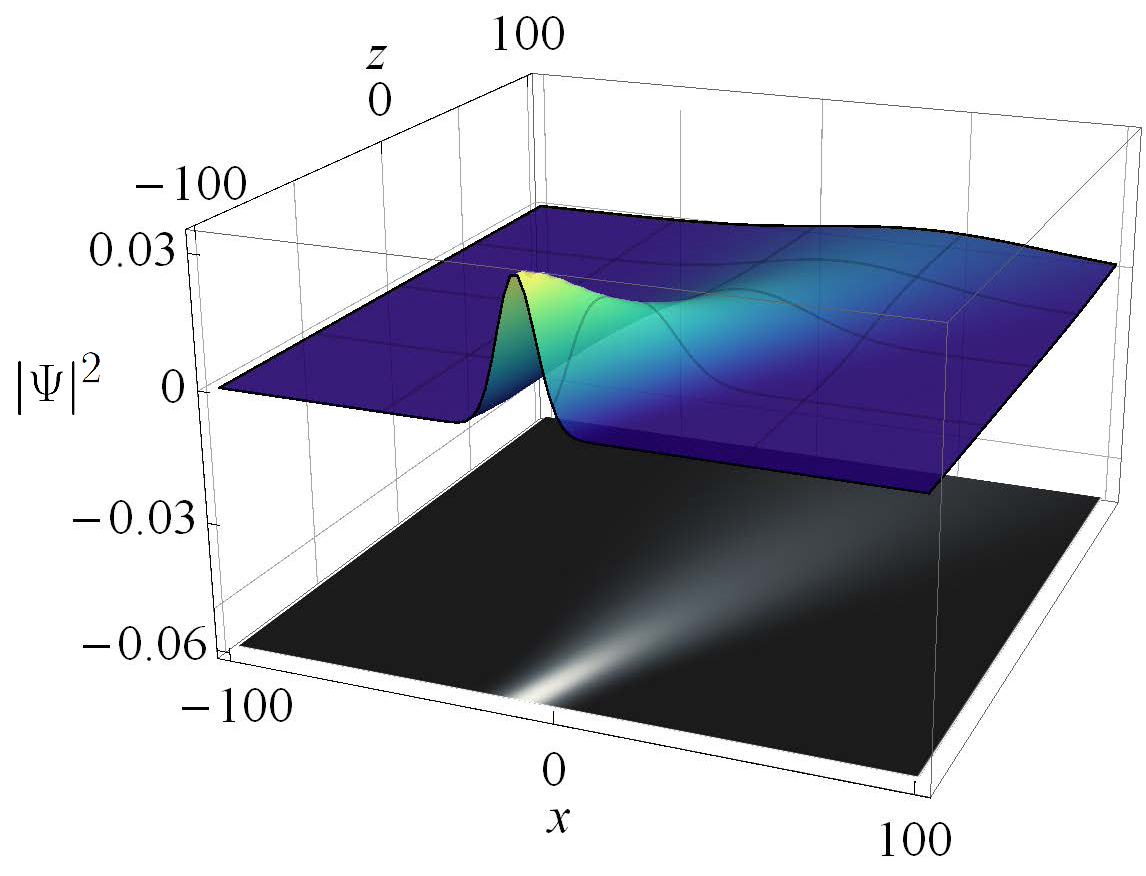}
        \caption{}
    \end{subfigure}
\begin{subfigure}[b]{0.3\textwidth}
        \includegraphics[width=\textwidth]{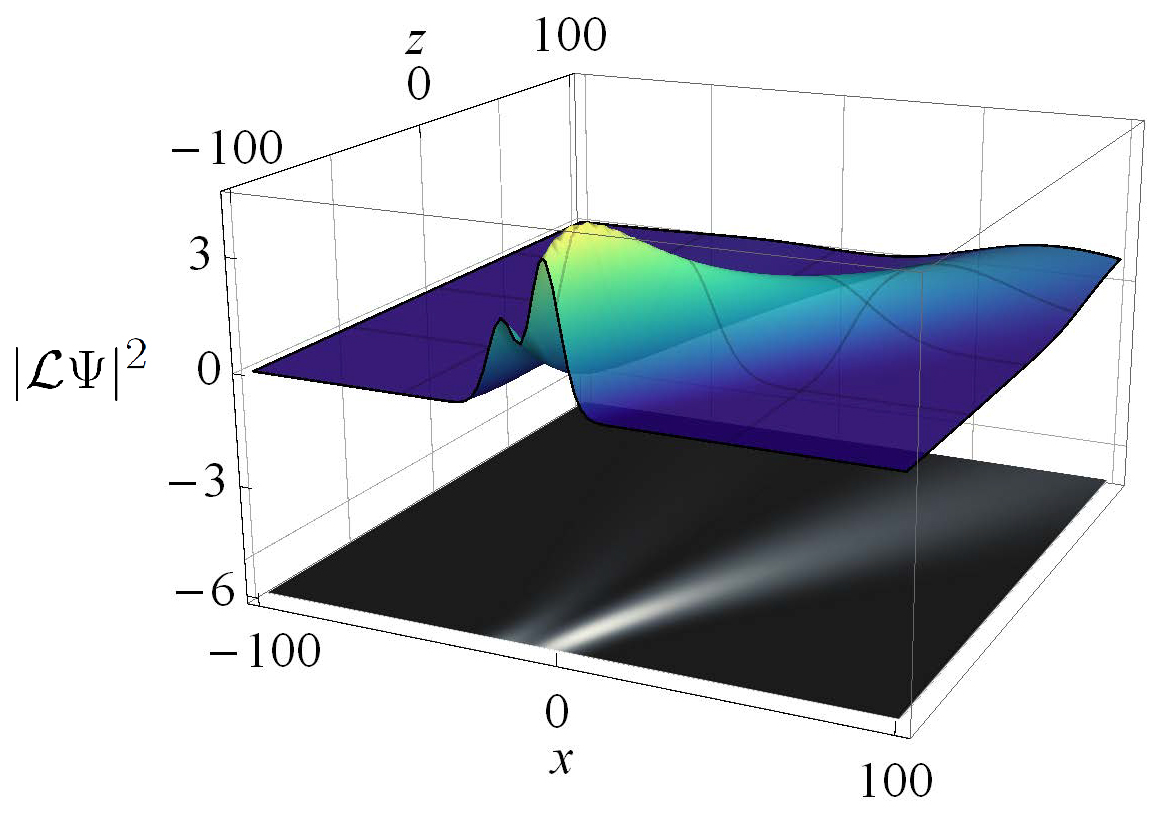}
        \caption{}
    \end{subfigure}
\begin{subfigure}[b]{0.3\textwidth}
        \includegraphics[width=\textwidth]{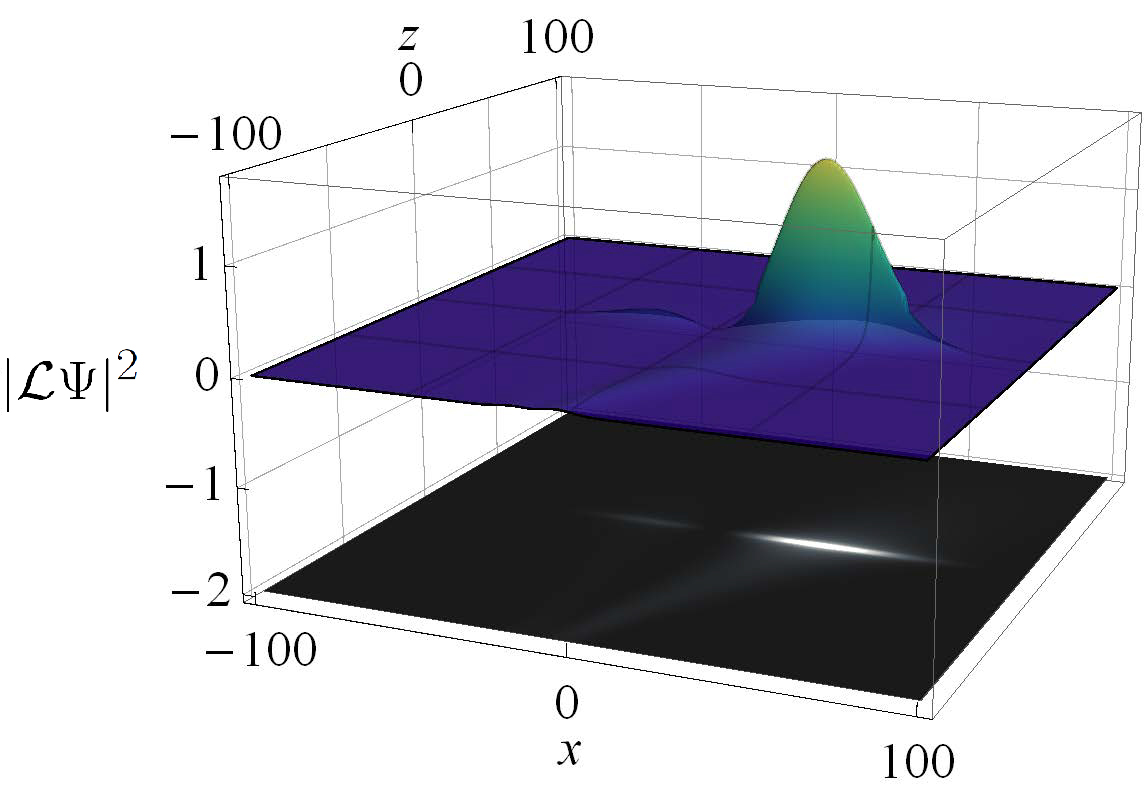}
        \caption{}
    \end{subfigure}\\
\begin{subfigure}[b]{0.3\textwidth}
        \includegraphics[width=\textwidth]{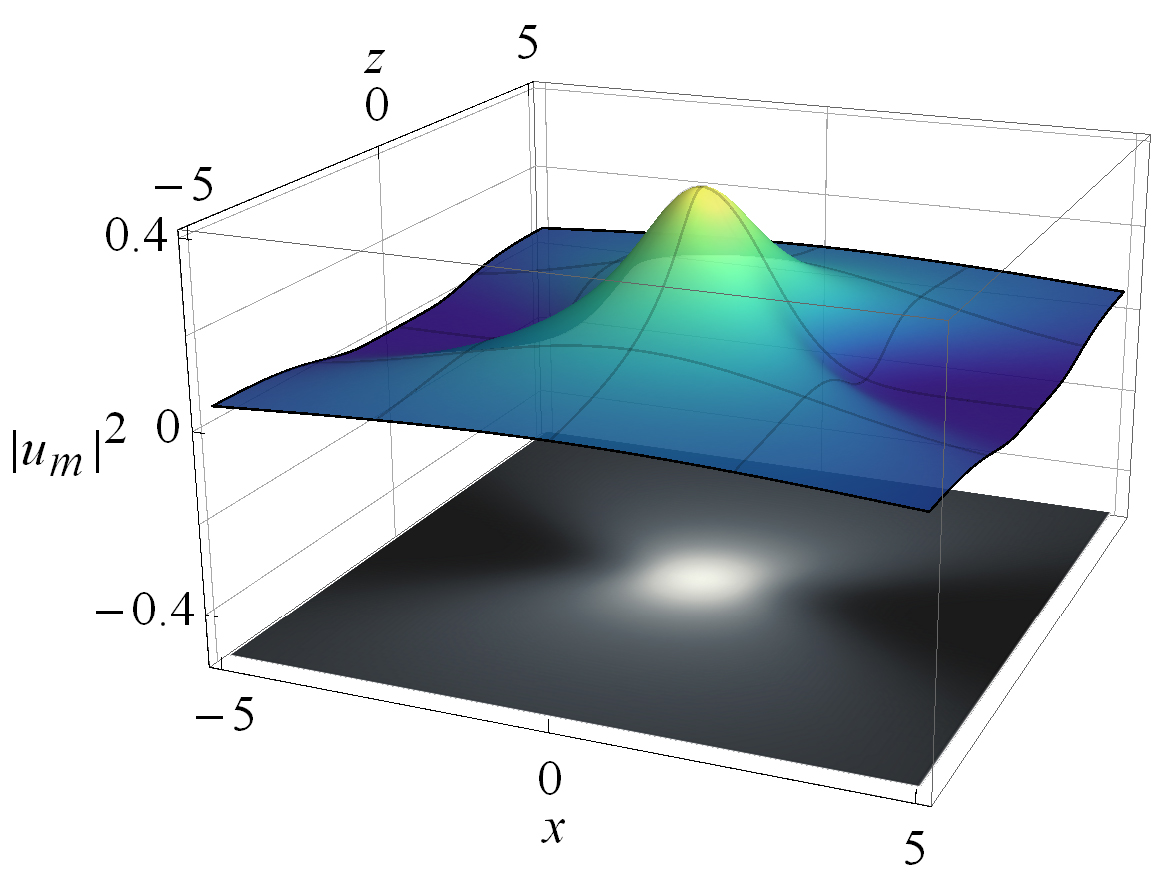}
        \caption{}
    \end{subfigure}
\begin{subfigure}[b]{0.3\textwidth}
        \includegraphics[width=\textwidth]{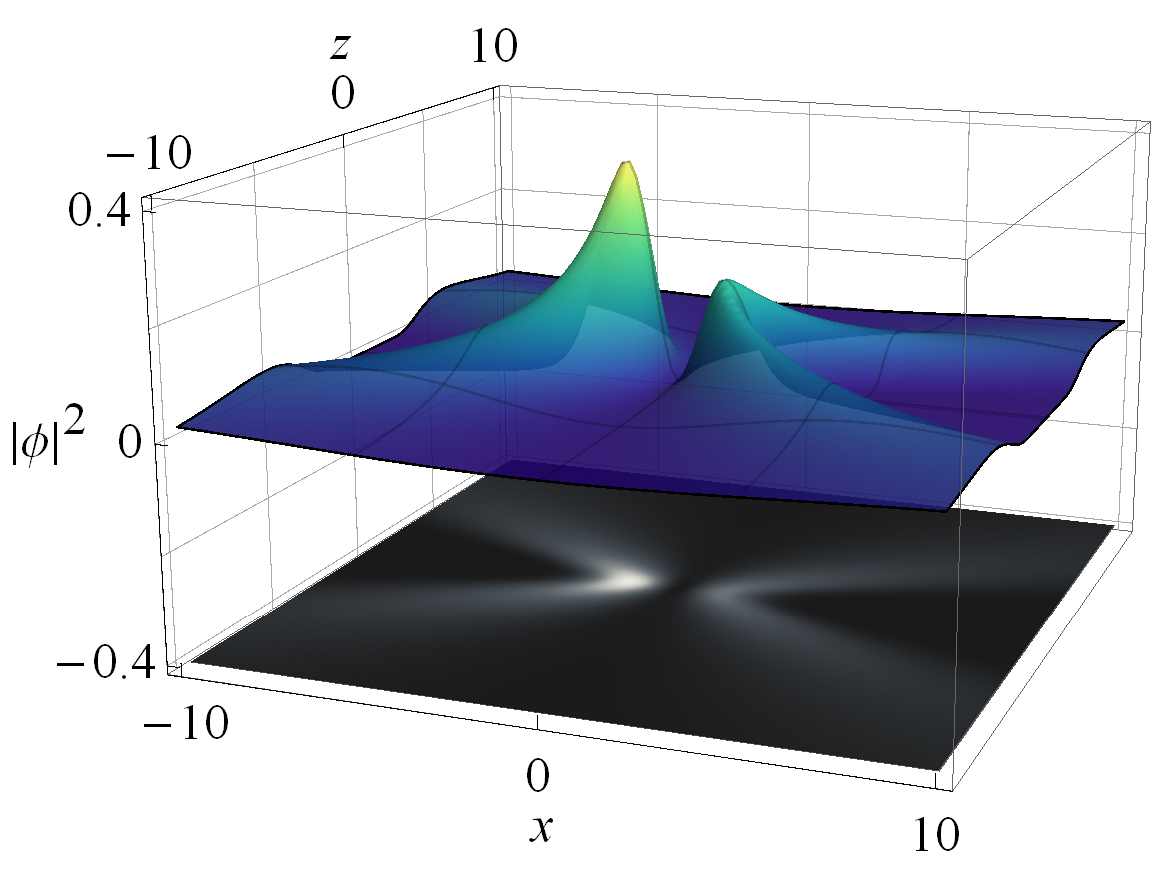}
        \caption{}
    \end{subfigure}
\begin{subfigure}[b]{0.3\textwidth}
        \includegraphics[width=\textwidth]{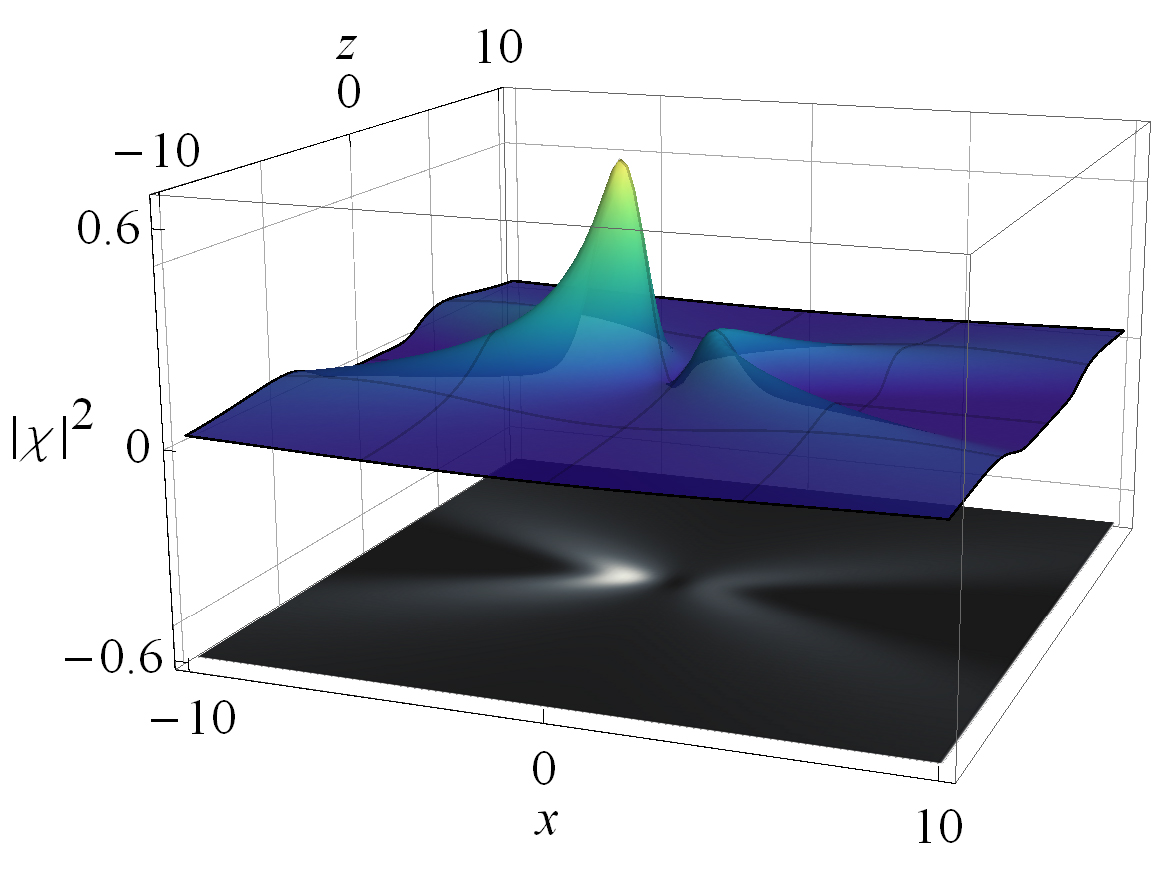}
        \caption{}
    \end{subfigure}
		\caption{A straight wave guide divided symmetrically by gain and loss regions. Real (a) and imaginary (b) parts of the potential term $V_1=-\frac{1}{1+z^2}$. In (c) the imaginary part for the potential $V_1=-\frac{1}{2(1-iz)}$. The intensity density $|\Psi_{x_0,z_0,v_0,\sigma}|^2$ is shown in (d).   In (e) we show $|\mathcal{L}\Psi_{x_0,z_0,v_0,\sigma}|^2$ where $L_1(z)=\sqrt{1+z^2}$. In (f) we show $|\mathcal{L}\Psi_{x_0,z_0,v_0,\sigma}|^2$ where $L_1(z)=\sqrt{1-iz}$. The parameters used in (d)-(f) are: $x_0=-10$, $z_0=-100$, $v_0=-0.25$, $\sigma=30$. Three different solutions of the time dependent Schr\"odinger equation for the potential $V_1=-\frac{1}{1+z^2}$, see (\ref{fpum}), (\ref{fpphi}) and (\ref{fpchi}), $|u_m|^2$ (g), $|\phi|^2$ (h) and $|\chi|^2$ (i) are also graphed.} \label{spontaneously pot}
	\end{center}
\end{figure}

To make the definition of the intertwining operator and the new system unambiguous, we have to fix the function $L_1$. The intertwining operator is required to preserve the amplitude of the wave packets, see (\ref{wavepacketpreservation}). We have
\begin{equation}\label{GaussWP}
\mathcal{L}\Psi_{x_0,z_0,v_0,\sigma}=G(x,z)\Psi_{x_0,z_0,v_0,\sigma},\quad G(x,z)=\frac{L_1(z)}{2}\left(-\frac{x}{1-iz}-\frac{x-x_0-iv_0\sigma}{\sigma+i(z-z_0)}\right).
\end{equation}
We should select $L_1$ such that $G(x,z)$ is a bounded function of $z$. We also require the new potential to be $\mathcal{PT}$-symmetric. The requirement (\ref{uPT1}) tells us that the new system will be $\mathcal{P}_x{T}$-symmetric provided that $|L_1|=\sqrt{1+z^2}$. It suggests $L_1=\sqrt{1+z^2}$ or $L_1=1\pm i z$ as the viable candidates. The requirement of $\mathcal{P}_2\mathcal{T}$-symmetry is less restrictive. Substituting $u$ into (\ref{uPT2}), we find that $V_1$ is $\mathcal{P}_2\mathcal{T}$-symmetric provided that $L_1(z)=\overline{L_1(-z)}$. We fix\footnote{Fixing $L_1=1-iz$ gives $V_1=0$ and $L_1=1+iz$ gives $V_1=-\frac{2}{1+z^2}$. } $L_1=\sqrt{1+z^2}$. Then we get
\begin{equation}\label{fpS1}
V_1=-\frac{1}{1+z^2},\quad \mathcal{L}=\sqrt{1+z^2}~\left( \partial_x - \frac{x}{2(1-iz)}\right).
\end{equation}
The potential term is $x$-independent. It has the form of a straight wave guide along $x$-axis, Fig. \ref{spontaneously pot} (a) and (b).

The intertwining operator $\mathcal{L}$ alters the profile of the wave packets and keeps them bounded for large $|z|$. Alternatively, we can define $\mathcal{L}$ such that the wave packets are mapped into the localized states of the new system. Taking $L_1=\sqrt{1-iz}$, the new potential $V_1$ is no longer $\mathcal{P}_x \mathcal{T}$-symmetric but it possesses $\mathcal{P}_2 \mathcal{T}$-symmetry, $V_1=-\frac{1}{2(1-iz)}$ and the transformed wave packets $\mathcal{L}\Psi_{x_0,z_0,v_0,\sigma}$ are strongly suppressed for large $|z|$, see Fig. \ref{spontaneously pot} (c)-(f). However, we prefer to fix $L_1$ such that it preserves the amplitude of the wave packets and we take $L_1=\sqrt{1+z^2}$.

The missing state (\ref{um}) reads
\begin{eqnarray}\label{fpum}
u_m(x,z)=\frac{1}{(2 \pi)^{1/4} (1-iz)^{1/2}} \exp\left(- \frac{x^2}{4(1+iz)} \right),
\end{eqnarray}
where $u_m$ fulfills $S_1 u_m =0$. The solution is well localized in the wave guide; it vanishes exponentially along $x$-axis while it has $\sim z^{-1/2}$ decay along the $z$-axis, see Fig. \ref{spontaneously pot} (g).
%%%%%%%%%%%%%%%%%%%%%%
%%%%%%%%%%%%%%%%%%%%%%
%%%%%%%%%%%%%%%%%%%%%%
%%%%%%%%%%%%%%%%%%%%%%

Symmetry operators (\ref{tildeLL}) can be constructed as well, 
\begin{eqnarray}\label{FPLL}
\mathcal{L}^\sharp \mathcal{L}&=& (1+z^2) \partial_x^2 -i x z\partial_x - \frac{1}{4}(x^2+2 i z +2), \quad [S_0,\mathcal{L}^{\sharp}\mathcal{L}]=0, \nonumber \\
\mathcal{L}\mathcal{L}^\sharp&=& (1+z^2) \partial_x^2 -i x z\partial_x - \frac{1}{4}(x^2+2 i z -2),   \quad [S_1,\mathcal{L}\mathcal{L}^\sharp]=0.
\end{eqnarray}
By construction, $\mathcal{L}^\sharp u_m=0$. In order to illustrate the action of the symmetry operator $ \mathcal{L}\mathcal{L}^\sharp$, let us consider the wave packet $\psi=(2 \pi)^{-1/4}(1+iz)^{-1/2} \exp(-(x+1/2)^2/4(1+iz) )$ solving $S_0 \psi = 0$.
We transform it into a solution of $S_1 \phi=0$ by the application of the intertwining operator, 
\begin{eqnarray}\label{fpphi}
\phi= \mathcal{L} \psi = - \frac{4x -iz+1}{4(2 \pi)^{1/4}(1+iz)\sqrt{1-iz}} \exp\left(- \frac{(x+\frac{1}{2})^2}{4(1+iz)}  \right),\quad S_1\phi=0.
\end{eqnarray}
Through the successive applications of the symmetry operator $ \mathcal{L}\mathcal{L}^\sharp$ we can obtain a whole family of  solutions. After the first iteration, we get  
\begin{eqnarray}\label{fpchi}
\chi= \mathcal{L}\mathcal{L}^\sharp \phi= \frac{-16x^2 + 56x +72 i xz +2iz +33 z^2+31}{64 (2\pi)^{1/4}  (1+iz)^2 \sqrt{1-iz}} \exp\left(- \frac{(x+\frac{1}{2})^2}{4(1+iz)}  \right),\quad S_1\chi=0.
\end{eqnarray}
The solutions $u_m$, $\phi$ and $\chi$ are illustrated in the Fig. \ref{spontaneously pot} (g)-(i). It is worth mentioning that as the system possesses translational invariance with respect to $x$-axis, all the presented solutions (\ref{GaussWP}), (\ref{fpum}), (\ref{fpphi}) and (\ref{fpchi}) can be modified by $x\rightarrow x+a$, $a\in\mathbb{R}$, without compromising validity of the Schr\"odinger equation.

As we can see, identification of $u$ with the Gaussian wave packet (\ref{gaussian1}) resulted in the construction of the new system with a  localized missing state. However, the new Schr\"odinger operator has separable potential (\ref{fpS1}). The refractive index possesses translational invariance and forms a barrier in propagation of the light beam. It is not quite satisfactory result as we seek for a non-separable, two-dimensional system with localized defects of the refractive index.
We shall find alternative transformation function $u$ that would serve better in construction of the models with desired properties. 

\subsection{Free particle solutions via harmonic oscillator} \label{Point T subsection}

The free particle and harmonic oscillator systems are related through a specific point transformation, see  \cite{Abraham80,Ray82,Bluman83,Finkel99,Guerrero11,Schulze14}. It allows to map the solutions of one system into the solutions of the other system. 

Let us consider the Schr\"odinger equation of the harmonic oscillator given in terms of the variables $y$ and $t$, 
\begin{eqnarray}\label{SchHO}
S_{HO}\widetilde{\psi}(y,t)=\left(i\partial_t+\partial_y^2-\frac{1}{4}y^2\right)\widetilde{\psi}(y,t)=0,   \label{Finkel TISE}
\end{eqnarray}
Now, let $y$ and $t$ be defined in terms of the new variables $z$ and $x$ as
\begin{eqnarray}
y(x,z)= \frac{x}{\sqrt{1+z^2}},\quad t=\arctan z.
\label{FinkelVariable} 
\end{eqnarray}
Then the Schr\"odinger operator $S_{HO}$ of the Harmonic oscillator can be transformed into the Schr\"odinger operator of the free particle multiplied by a $z$-dependent function,
\begin{equation}\label{FinkelFunction}
U^{-1}S_{HO}U=(1+z^2)(i\partial_z+\partial_x^2)=(1+z^2)S_0,\quad U=%\frac{e^{\frac{ix^2z}{4(1+z^2)}}}{(1+z^2)^{1/4}}=
e^{-\frac{ix^2z}{4(1+z^2)}}(1+z^2)^{1/4}.
\end{equation}
This transformation allows us to transform the solutions $S_{HO}\widetilde{f}(y,t)=0$ into the solutions of $S_{0}f(x,z)=0 $,
\begin{equation}\label{pointtPsi}
f(x,z)=U^{-1}\widetilde{f}(y(x,z),t(z)).
\end{equation}

The stationary solutions
$\widetilde{u}_{I,n}$ and $\widetilde{u}_{II,n}$ of (\ref{SchHO}) are, see \cite{Flugge},
\begin{eqnarray}
\widetilde{u}_{I,n}(y,t)&=&~_1F_1 \left(-\frac{n}{2},\frac{1}{2};\frac{1}{2}y^2 \right) \exp \left(-\frac{1}{4} y^2 \right) e^{-itE_n}, \nonumber\\
\widetilde{u}_{II,n}(y,t)&=& ~y~_1F_1\left(\frac{1-n}{2},\frac{3}{2};\frac{1}{2}y^2 \right) \exp\left(-\frac{1}{4} y^2 \right)e^{-itE_n}. \label{HOgensols}
\end{eqnarray}
Here, $_1F_1(a,b;z)$ is a confluent hypergeometric function \cite{Abramowitz,Bateman}. This functions satisfy $\widetilde{u}_{I,n}(y,t)=\widetilde{u}_{I,n}(-y,t)$ and $\widetilde{u}_{II,n}(-y,t)=-\widetilde{u}_{II,n}(y,t)$, i.e. they are even and odd functions in $y$, respectively. It implies that all the functions $\widetilde{u}_{II,n}$ share at least one zero at $y=0$,  $\widetilde{u}_{II,n}(0)=0$.  The Wronskian
of the two solutions for fixed $t$ is constant, $W(\widetilde{u}_{I,n}, \widetilde{u}_{II,n})|_{t=0}=1$. In the special case of $n$ being a non-negative integer, one of (\ref{HOgensols}) reduces to a square integrable function as the confluent hypergeometric function is truncated  to a Hermite polynomial. 

  The point transformation (\ref{pointtPsi}) maps the solutions (\ref{HOgensols}) into  
\begin{align}
u_{I,n}(x,z)=&  \frac{1}{(1+z^2)^{1/4}} \exp\left\{\frac{i}{4} \left[\frac{x^2}{z-i}  - 4 E_n \arctan(z)    \right]  \right\}~ _1F_1 \left(-\frac{n}{2},\frac{1}{2};\frac{x^2}{2(z^2+1)} \right)  \label{u even} , \\
u_{II,n}(x,z)=&  \frac{x}{(1+z^2)^{3/4}} \exp\left\{\frac{i}{4} \left[\frac{x^2}{z-i} - 4E_n \arctan(z)    \right]  \right\} ~_1F_1\left(\frac{1-n}{2},\frac{3}{2};\frac{x^2}{2(z^2+1)} \right). \label{u odd} 
\end{align}
They satisfy
\begin{equation}
 S_0u_{I,n}=S_0u_{II,n}=0.
\end{equation}

Let us fix $u$ as the following linear combination of $u_{I(II),n}$,
\begin{eqnarray}\label{uu0}
u(x,z)= \sum_{j=1}^N \left( \alpha_{I,n_j} u_{I,n_j} + i~ \alpha_{II,n_j} u_{II,n_j} \right),\quad \alpha_{I(II),n_j}\in\mathbb{R},\quad n_j\in\mathbb{R}. \label{Superposition}
\end{eqnarray}
We can see that it satisfies\footnote{The functions $u_{I(II),n}$ fulfill  $\mathcal{P}_2\mathcal{T}u_{I,n}=u_{I,n}$ and $\mathcal{P}_2\mathcal{T}u_{II,n}=-u_{II,n}$.} $\mathcal{P}_2\mathcal{T}u=\epsilon u$, where $\epsilon\in\{-1,1\}$. 
Considering the other definition of the $\mathcal{P}$ operator, then
\begin{equation}\label{PxTu}
 \mathcal{P}_x\mathcal{T}u_{I(II),n}=u_{I(II),n}\exp\left(2i E_n\arctan(z)+i\frac{z}{2(z^2+1)}x^2\right).
\end{equation}
The function $u$ complies with (\ref{Scond}) provided that it is a linear combination of the solutions associated with the same energy. It can be written as
\begin{equation}\label{uu1}
 u(x,z)=\alpha_{I,n}u_{I,n}+i\alpha_{II,n}u_{II,n},\quad \alpha_{I(II),n}\in\mathbb{R}.
\end{equation}
We will use this function to construct the new system of required properties.

\subsection{Optical wave guide with a localized defect}

We identify $u$ with \eqref{uu1}. It can be written as 
\begin{eqnarray}\label{uHO}
u(x,z)&=&\frac{1}{(1+z^2)^{1/4}} \exp\left\{\frac{i}{4} \left[\frac{x^2}{z-i} - 4\left(n+\frac{1}{2}\right) \arctan(z)    \right]  \right\} \nonumber \\
& ~& \times  \left[\alpha_{I,n} ~_1F_1 \left(-\frac{n}{2},\frac{1}{2};\frac{x^2}{2(z^2+1)} \right)+i ~ \alpha_{II,n}\frac{x}{(1+z^2)^{1/2}}~ ~_1F_1\left(\frac{1-n}{2},\frac{3}{2};\frac{x^2}{2(z^2+1)} \right) \right]. \label{u no missing state}
\end{eqnarray} 
Let us analyze its zeros. When $\alpha_{I,n}\alpha_{II,n}\neq0$, then $u_{I,n}$ and $u_{II,n}$ cannot vanish in the same points as we have $W(u_{I,n},u_{II,n})|_{z=const}\neq 0$. Hence, the function $u$ is nodeless in this case.  
When $\alpha_{II,n}=0$ and $n \leq 0$, $u\equiv u_{I,n}$ is nodeless by the oscillation theorem. 
As we discussed above, $u$ satisfies (\ref{Scond}) %for $\mathcal{S}\equiv \mathcal{P}_x\mathcal{T}$ 
so that the missing state $u_m$ can be constructed as in (\ref{um}) (in the definition of $u_m$, the function $f(z)$ reads $f(z)= L_1^{-1}(z)(z^2+1)$, see  \eqref{f(z)}). 

As we require the missing state to be vanishing for large $|x|$ and $|z|$, we take $n=-2$ that corresponds to the exponentially expanding solutions for large $|x|$ (it is associated with non-physical stationary state of the Harmonic oscillator) and also $\alpha_{I,-2}=1$, $\alpha_{II,-2}=\alpha$. Then (\ref{uHO}) can be simplified to  
\begin{eqnarray}
u(x,z)&=&\frac{1}{\delta^{1/4}} \exp\left\{\frac{i}{4} \left[\frac{x^2}{z-i}  + 6 \arctan(z)    \right]  \right\}   \left[ 1+\sqrt{\frac{\pi}{2 \delta}} x ~\text{erf}\left(\frac{x}{\sqrt{2 \delta}}\right) \exp\left(\frac{x^2}{2 \delta} \right) + i\frac{\alpha ~x}{\sqrt{\delta}} \exp\left(\frac{x^2}{2 \delta} \right)  \right], \label{u guide defect}
\end{eqnarray} 
where we used the abbreviation $\delta=z^2+1$, and $\text{erf}(\cdot)$ is the error function \cite{Abramowitz}.

\begin{figure}[t!]
	\begin{center}
	\begin{subfigure}[b]{0.3\textwidth}
        \includegraphics[width=\textwidth]{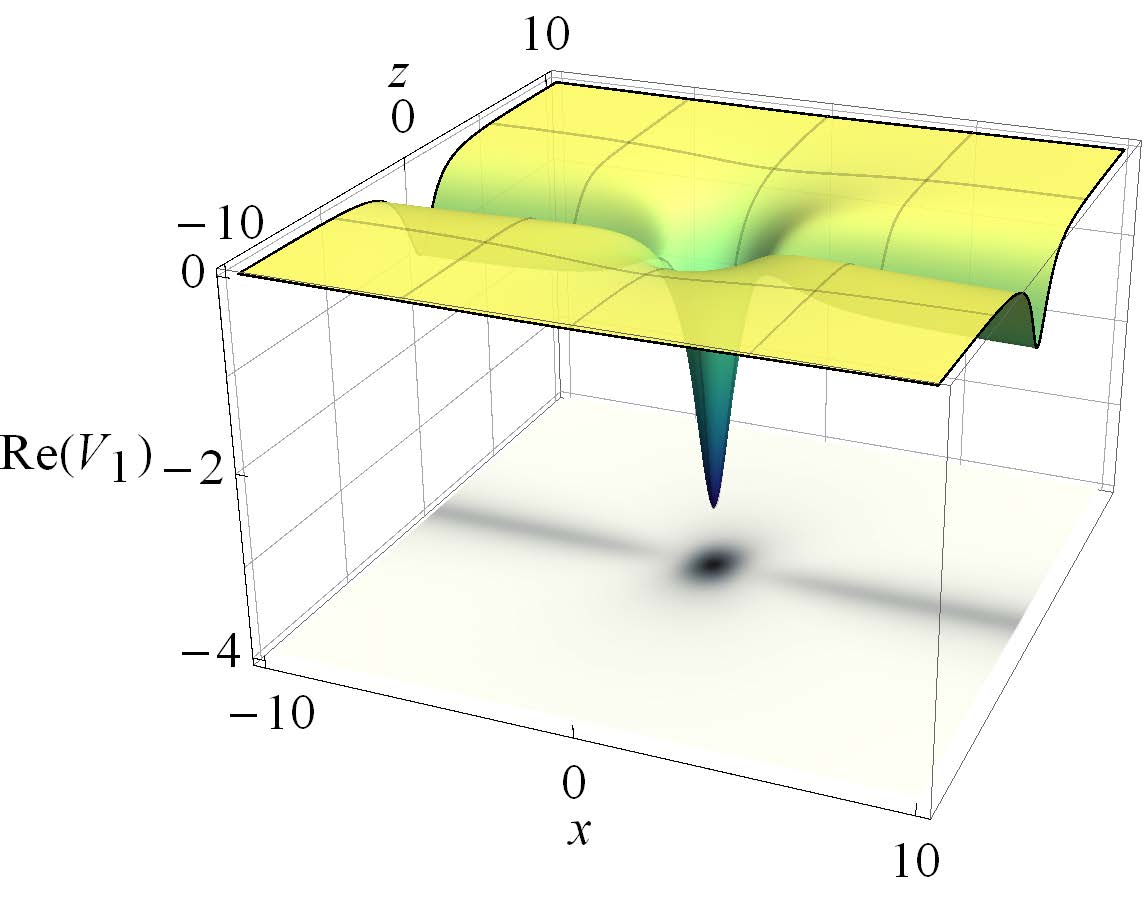}
        \caption{}
    \end{subfigure}
    \begin{subfigure}[b]{0.3\textwidth}
        \includegraphics[width=\textwidth]{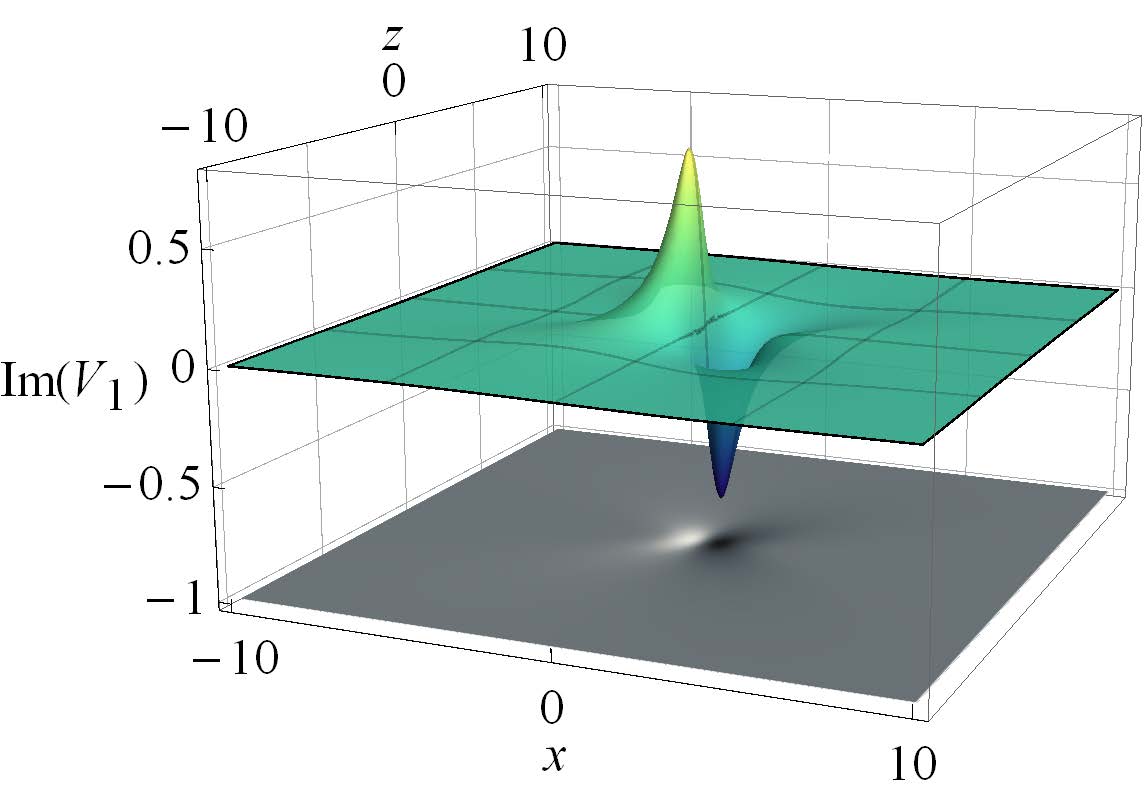}
        \caption{}
    \end{subfigure}
    \begin{subfigure}[b]{0.38\textwidth}
        \includegraphics[width=\textwidth]{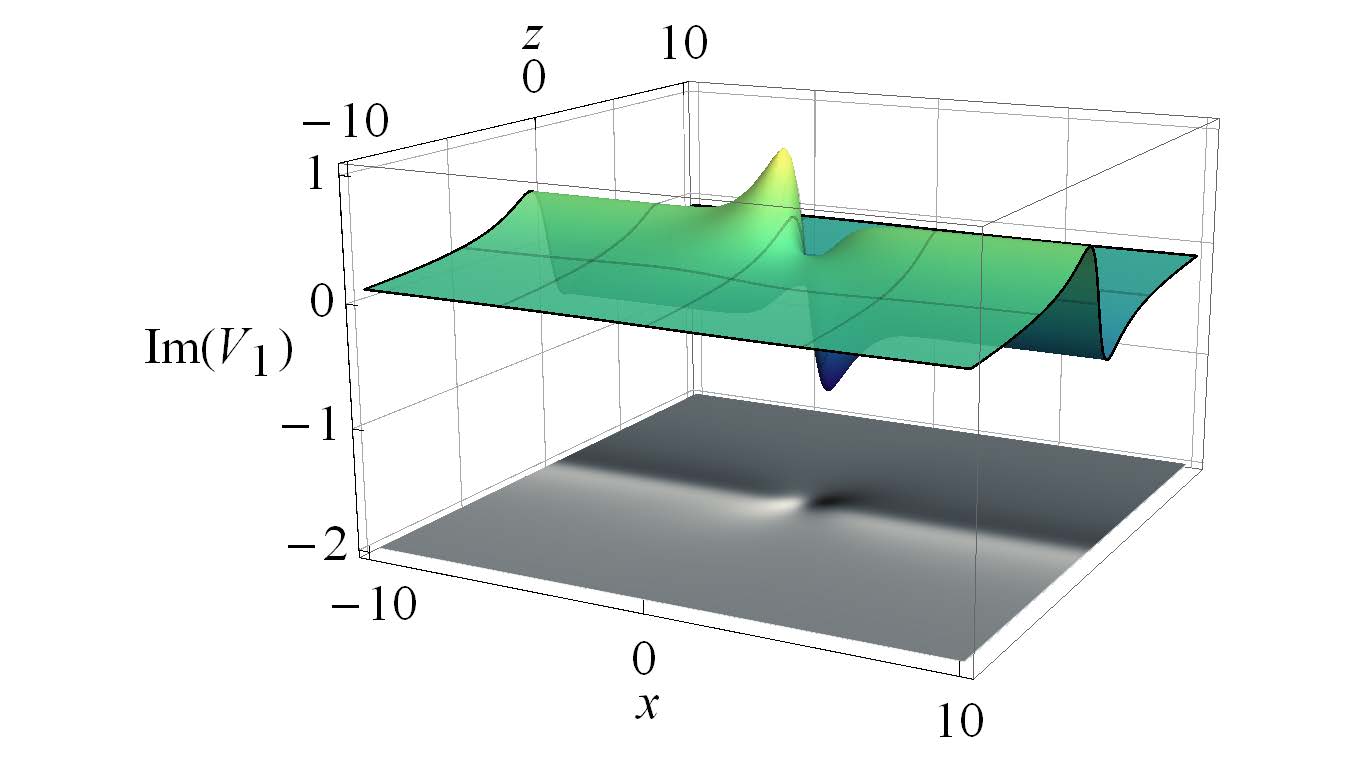}
        \caption{}
    \end{subfigure}\\
    \begin{subfigure}[b]{0.3\textwidth}
        \includegraphics[width=\textwidth]{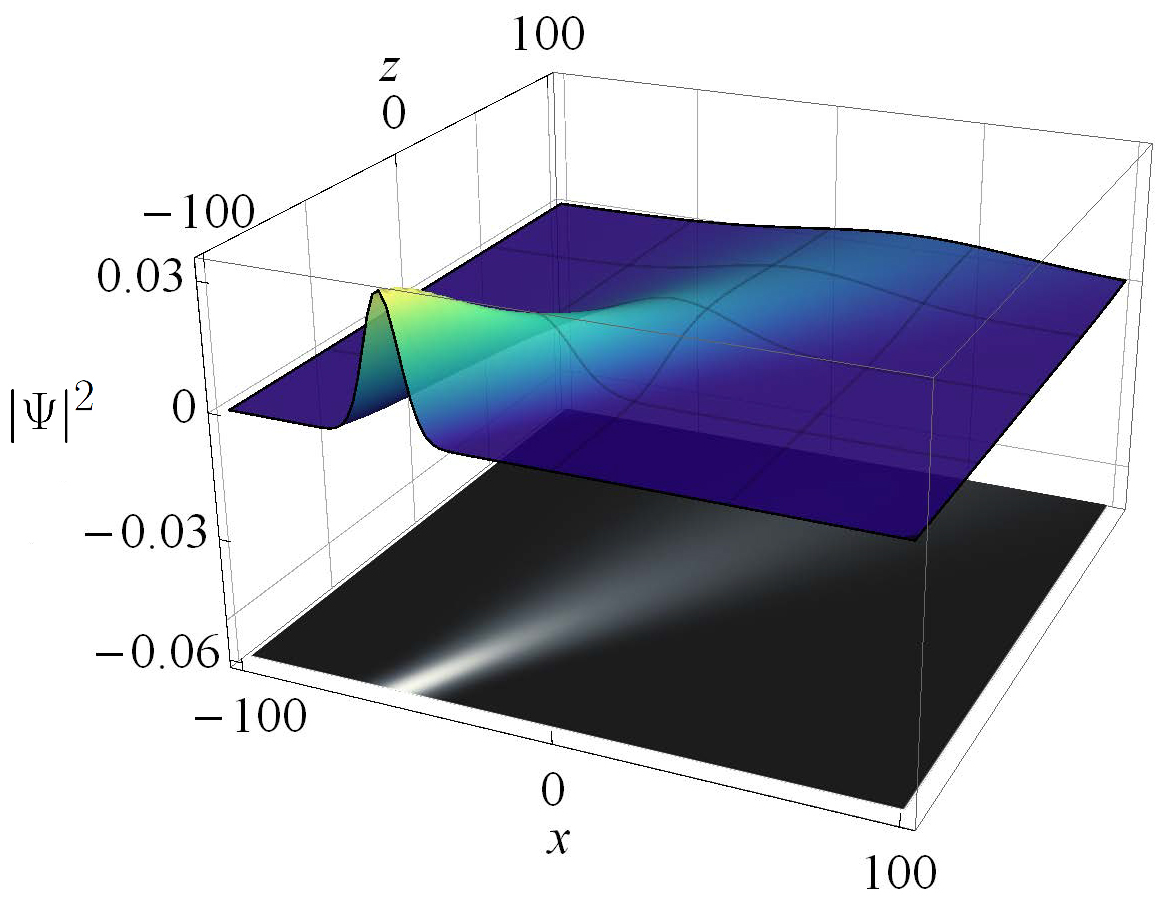}
        \caption{}
    \end{subfigure}
    \begin{subfigure}[b]{0.3\textwidth}
        \includegraphics[width=\textwidth]{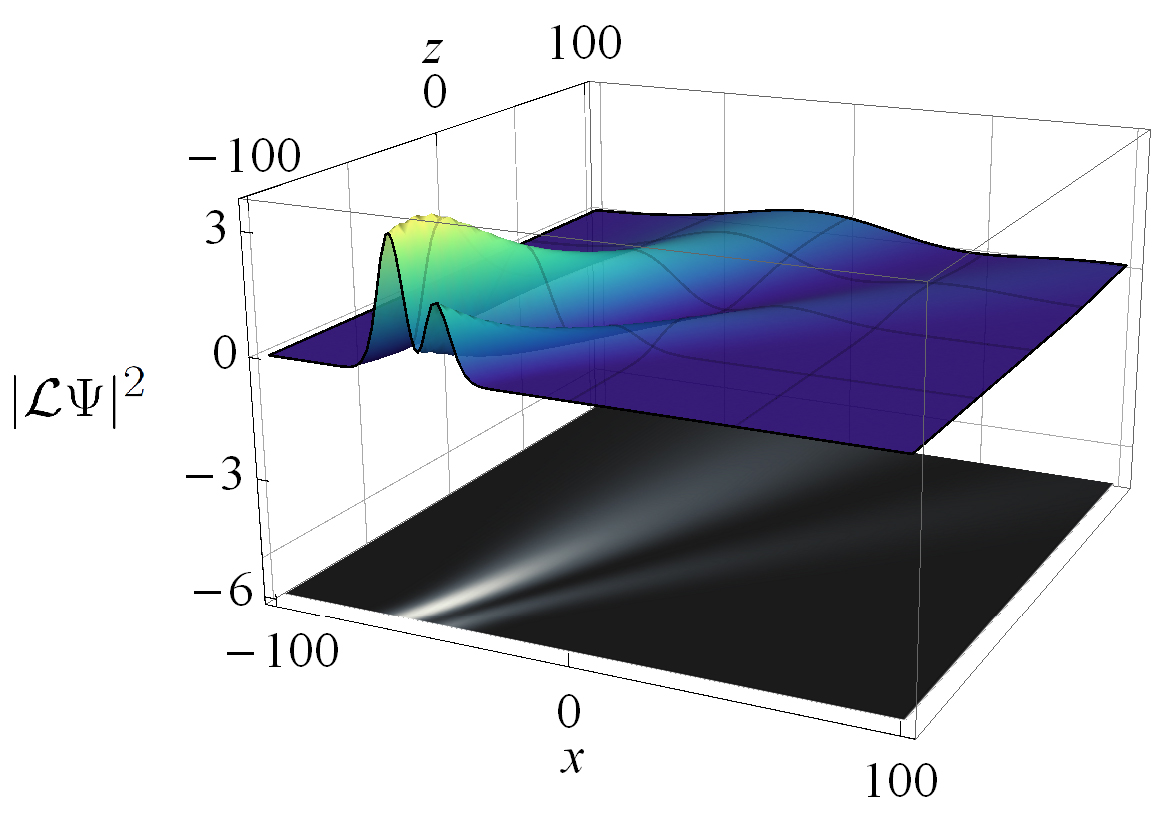}
        \caption{}
    \end{subfigure}
    \begin{subfigure}[b]{0.3\textwidth}
        \includegraphics[width=\textwidth]{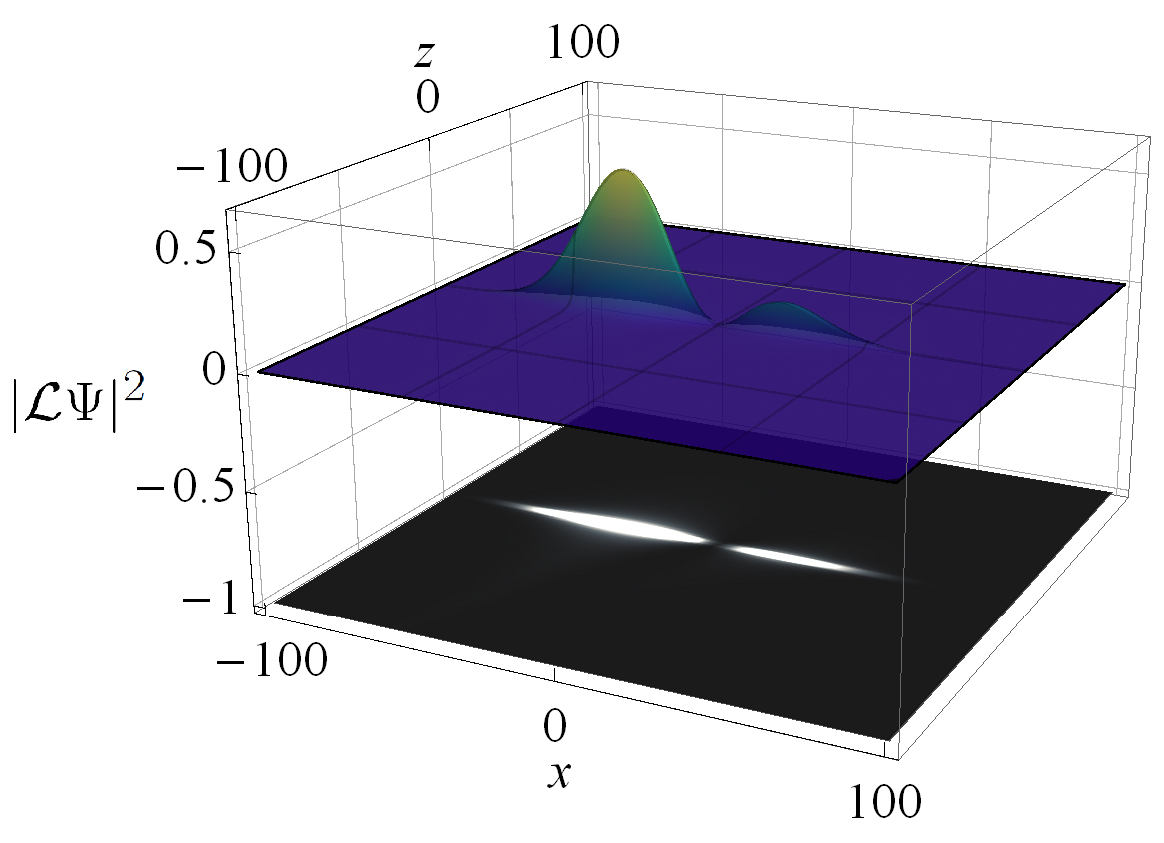}
        \caption{}
    \end{subfigure}\\
    \begin{subfigure}[b]{0.3\textwidth}
        \includegraphics[width=\textwidth]{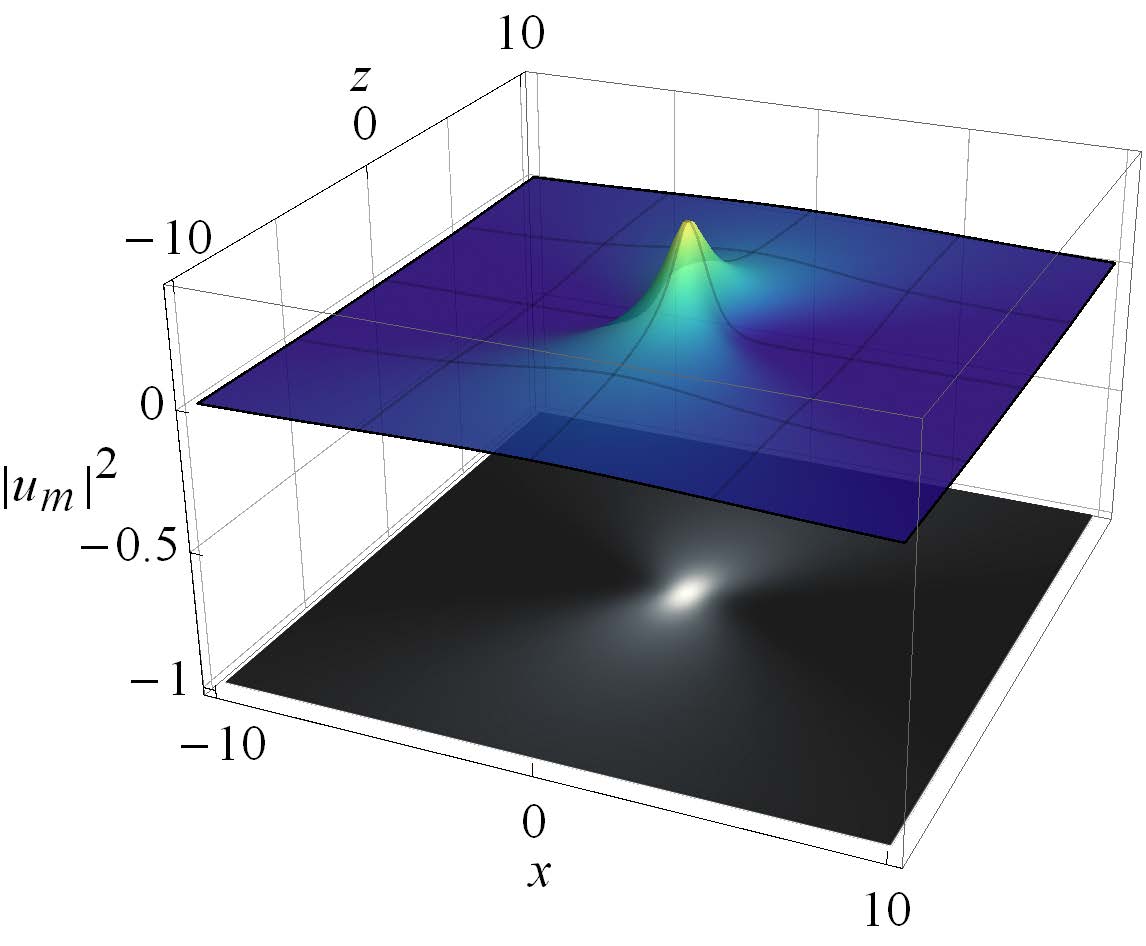}
        \caption{}
    \end{subfigure}
    \begin{subfigure}[b]{0.3\textwidth}
        \includegraphics[width=\textwidth]{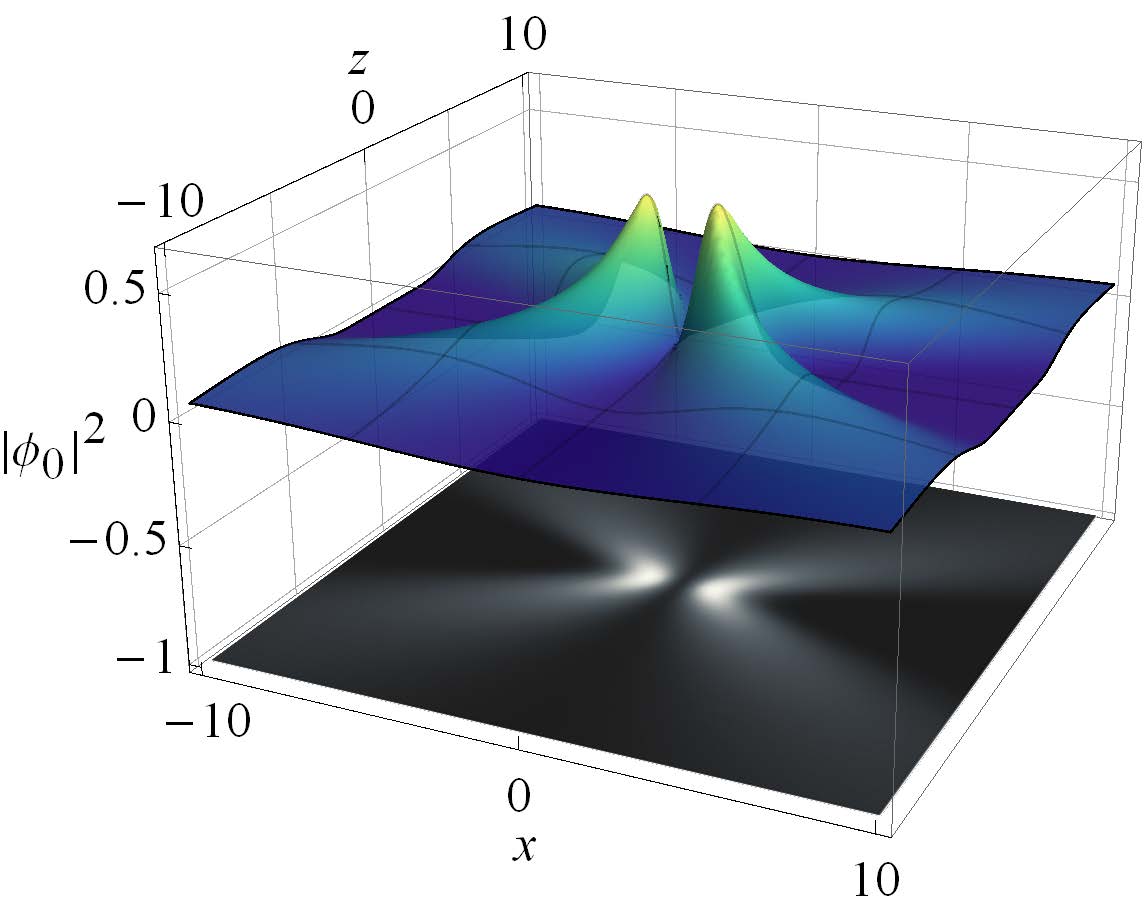}
        \caption{}
    \end{subfigure}
    \begin{subfigure}[b]{0.3\textwidth}
        \includegraphics[width=\textwidth]{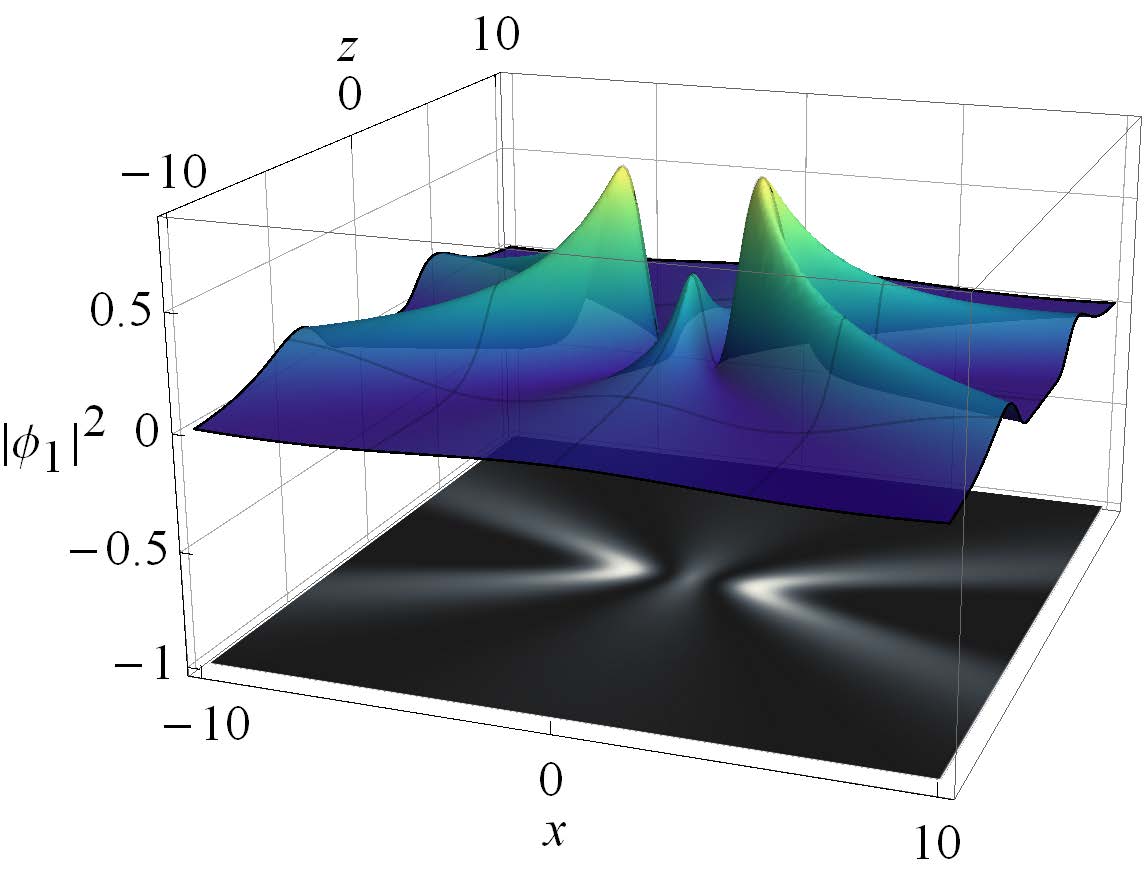}
        \caption{}
    \end{subfigure}
 	\caption{Optical wave guide with a localized defect. Plots of the real (a) and imaginary (b) parts of $V_1$, see \eqref{V1HO}, when $L_1=\sqrt{1+z^2}$. In (c) the imaginary part of $V_1$ when $L_1=1$, see \eqref{SUSY V1} and \eqref{u guide defect}. The intensity density $|\Psi_{x_0,z_0,v_0,\sigma}|^2$ is shown in (d).  In (e) we show $|\mathcal{L}\Psi_{x_0,z_0,v_0,\sigma}|^2$ where $L_1(z)=\sqrt{1-iz}$. In (f) we show $|\mathcal{L}\Psi_{x_0,z_0,v_0,\sigma}|^2$ where $L_1(z)=1$. The parameters used in (d)-(f) are: $x_0=-50$, $z_0=-100$, $v_0=-0.4$, $\sigma=30$. Three different solutions for the corresponding time dependent Schr\"odinger equation for $V_1$ as in \eqref{V1HO} are shown: first $|u_m|^2$ (g), see \eqref{um missing}, then $|\phi_0|^2=|\mathcal{L}\psi_0|^2$ (h) and finally $|\phi_1|^2=|\mathcal{L}\psi_1|^2$ (i). In all plots $\alpha=-(2\pi)^{-1/2}$.  }
\label{Missing pot}.
	\end{center}
\end{figure}

Let us consider how the intertwining operator $\mathcal{L}$ transforms the wave packet $\Psi_{x_0,z_0,v_0,\sigma}$, see (\ref{wavepacket}). We get
\begin{eqnarray}\label{Lpsi}
\mathcal{L}\Psi_{x_0,z_0,v_0,\sigma}&=&G(x,z)\Psi_{x_0,z_0,v_0,\sigma},
\end{eqnarray}
where
\begin{eqnarray}\label{Lpsi2}
G(x,z)&=&\frac{iL_1(z)}{2}\left(\frac{x-x_0+iv_0\sigma}{z-z_0-i\sigma}-\frac{x}{i+z}-\frac{2}{ix+\frac{2e^{-\frac{x^2}{2(1+z^2)}}\sqrt{1+z^2}}{2\alpha-i\sqrt{2\pi}\mbox{erf}\left(\frac{x}{\sqrt{2}\sqrt{1+z^2}}\right)}}\right)
\end{eqnarray}
The function $L_1$ should satisfy either (\ref{uPT1}) or (\ref{uPT2}) in order to have a $\mathcal{PT}$-symmetric potential. The relation (\ref{uPT1}) results in $|L_1|=1+z^2$ while the conditions (\ref{uPT2}) gives $L_1(z)=\overline{L_1(-z)}$. Additionally, we require the function $G(x,z)$ to be bounded (\ref{wavepacketpreservation}). These requirements (i.e. the potential is both $\mathcal{P}_x\mathcal{T}$-symmetric and $\mathcal{P}_2\mathcal{T}$-symmetric and $G(x,z)$ is bounded) can be met by fixing 
	$$L_1=\sqrt{1+z^2}.$$
With this selection of $L_1$, the new potential \eqref{V_1} reads
\begin{eqnarray}\label{V1HO}
V_1(x,z)=\frac{2 \left\{ \left(x^2-2\delta\right) \left[\sqrt{2}\alpha- i \sqrt{\pi}\text{erf}\left(\frac{x}{\sqrt{2\delta} }\right) \right]^2-4 \sqrt{2 \pi  \delta} x  e^{-\frac{x^2}{2\delta }} \text{erf}\left(\frac{x}{\sqrt{2 \delta} }\right)-8 i \alpha  x \sqrt{\delta} e^{-\frac{x^2}{2\delta }}-6\delta e^{-\frac{x^2}{\delta}} \right\}}{\delta \left[2 \sqrt{\delta}e^{-\frac{x^2}{2 \delta}}+\sqrt{2}ix  \left(\sqrt{2} \alpha-i\sqrt{\pi } \text{erf}\left(\frac{x}{\sqrt{2 \delta} }\right) \right)\right]^2}.
\end{eqnarray}
The expression (\ref{V1HO}) represents a one-parameter family of potentials where $\alpha$ can acquire any real value. For $\alpha=0$, the potential is real function and $S_1$ is Hermitian. The potential behaves asymptotically as
\begin{eqnarray}
V_1(x,z)&=&\frac{1}{\delta}+o(1),\quad (|x|\rightarrow\infty),\\
V_1(x,z)&=&O\left(\frac{1}{\delta}\right),\quad (|z|\rightarrow\infty),\quad \delta=1+z^2.
\end{eqnarray}
Hence, it represents a real wave guide with a localized $\mathcal{PT}$-symmetric defect, see Fig. \ref{Missing pot} (a) and (b).   

The intertwining operator $\mathcal{L}$ changes profile of the wave packet $\Psi_{x_0,z_0,v_0,\sigma}$. It gets divided it into two beams that pass around the origin from both sides. If we had fixed $L_1=1$ that still respects (\ref{wavepacketpreservation}), the potential (\ref{V1HO}) would acquire an additional term $i\partial_z\ln(1+z^2)$, adding a gain-loss profile to the potential barrier, see Fig. \ref{Missing pot} (c). The transformed wave packets would change their form radically; they would be concentrated in region with non-vanishing gain and loss, see Fig.\ref{Missing pot} (d)-(f).
%%%%%%%%%%%%%%%%%%%%%%
%%%%%%%%%%%%%%%%%%%%%%
%%%%%%%%%%%%%%%%%%%%%%
%%%%%%%%%%%%%%%%%%%%%%

The missing state $u_m$ for the system with potential term as in \eqref{V1HO} can be written as
\begin{eqnarray} \label{um missing}
 u_m(x,z) & = & \frac{1}{\delta^{1/4}} \exp\left\{\frac{i}{4} \left[\frac{x^2}{i+z}  + 6 \arctan(z)    \right]  \right\} \left[ 1+\sqrt{\frac{\pi}{2 \delta}} x ~\text{erf}\left(\frac{x}{\sqrt{2 \delta}}\right) \exp\left(\frac{x^2}{2 \delta} \right) + i\frac{\alpha ~x}{\sqrt{\delta}} \exp\left(\frac{x^2}{2 \delta} \right)  \right]^{-1}. 
\end{eqnarray}
It is vanishing for large values of $|x|$ and $|z|$, i.e. it represents a light dot. It decreases exponentially for large $|x|$ and fixed $z$, whereas it behaves as $(1+z^2)^{-1/4}$ for large $|z|$ and  fixed $x$, see Fig. \ref{Missing pot} (g).

The symmetry operators (\ref{tildeLL}) can be found explicitly as 
\begin{eqnarray}\label{LLHO}
\mathcal{L}^\sharp \mathcal{L}&=& (1+z^2) \partial_x^2 - i z x \partial_x - \frac{1}{4}(x^2+2iz+6), \\
\mathcal{L}\mathcal{L}^\sharp & = & (1+z^2) \partial_x^2 - i z x \partial_x- \frac{1}{4}(x^2+2iz+6) -(z^2+1) V_1, 
\end{eqnarray}
where $[S_0,\mathcal{L}^\sharp \mathcal{L}]=[S_1,\mathcal{L}\mathcal{L}^\sharp ]=0$. Notice that $\mathcal{L}\mathcal{L}^\sharp-\mathcal{L}^\sharp \mathcal{L}=-L_1(z)^2 V_1$.

The point transformation can be used to get other localized solutions that are based on the bound states of the harmonic oscillator. It is convenient to introduce the following notation 
\begin{equation}\label{psi_n}
 \psi_n(x,z)=\begin{cases}\frac{1}{\sqrt{\sqrt{2 \pi } 2^n n!}}u_{I,n}(x,z),&\quad n\ \mbox{is even},\ n\geq 0,\\
         \frac{1}{\sqrt{\sqrt{2 \pi } 2^n n!}}u_{II,n}(x,z),&\quad n\ \mbox{is odd},\ n\geq 0,
	     \end{cases}
\end{equation}
where $\psi_n$ are square integrable functions for fixed $z$ that are obtained from the bound states of the harmonic oscillator by the point transformation.  Then $\phi_n\equiv \mathcal{L}\psi_n$  represent light dots in the current system as they vanish both for large $x$ and $z$, see Fig. \ref{Missing pot} (h)-(i) for illustration. 
The action of $\mathcal{L} \mathcal{L}^\sharp$ on $\phi_n$ can produce new solutions of $S_1\mathcal{L} \mathcal{L}^\sharp\phi_n=0$.

In principle, one can define a Darboux transformation $\mathcal{L}$ by identifying $u$ with an arbitrary linear combination of the wave packets (\ref{psi_n}). Such $u$ will not comply with (\ref{Scond}) in general and the formula for the missing state (\ref{um}) will not be applicable. Despite it is not clear how to construct a missing state in terms of $u$ in this case, there are localized solutions, the light dots, associated with $\phi_n=\mathcal{L}\psi_n$. See Appendix where we illustrated such construction on an explicit example.

\subsection{Localized defects in a homogeneous crystal}

Let us construct now the system where uniformity of the refractive index is violated by a single localized defect. We shall use the stationary confluent Crum-Darboux transformation together with the point transformation. The former transformation allows us to get a localized (time-independent) deformation of the harmonic oscillator with the use of  the ground state \footnote{Let us notice that when we fix the transformation function $u$ as the ground state of the HO for the first order transformation, the resulting new system coincides with the original one up to an additive constant. It follows from the so called shape invariance of the Harmonic oscillator, the property that underlies its exact solvability \cite{Cooper}.}. 
Then, the point transformation will transform the deformed Harmonic oscillator into the system with asymptotically vanishing potential that has a localized defect. 
	
The confluent transformation is defined in terms of two functions, $u_1$ and $u_2$. However, the two functions are not independent, $u_2$ can be written in terms of $u_1$. Hence, $u_1$ defines the transformation together with some constant parameters, see  (\ref{confluentu2}).
Let us select $u_1$ as the stationary solution $S_{HO}\tilde{u}_1=0$,
$$
\tilde{u}_1(y,t)={\tilde{\psi}}_m(y)e^{-iE_mt},\quad\mbox{where}\quad \tilde{\psi}_m(y)=H_{m}\left(\frac{y}{\sqrt{2}}\right)e^{-\frac{1}{4}y^2},\quad E_m=m+\frac{1}{2},
$$
where $H_m(y)$ is a Hermite polynomial. As the second function $u_2$, we take  
$$
\tilde{u}_2(y,t)=e^{-iE_mt}\tilde{\psi}_m\left(\int_{y_0}^y\frac{1}{\tilde{\psi}_m^2}\left(\int_{s_0}^s\tilde{\psi}_m^2(r)dr+\alpha\right)ds+a\right).
$$
where $a$ is a constant.
 Then the Schr\"odinger operator of harmonic oscillator is intertwined with the new one by the operator $\tilde{\mathcal{L}}_{12}$, see (\ref{Confluent SUSY V2}), % $u_1(x,z)=U^{-1}u_1(y(x,t),t(z))$.....
	\begin{equation}
\tilde{S}_{HO}\tilde{\mathcal L}_{12}=\tilde{\mathcal{L}}_{12}S_{HO}, \quad \tilde{S}_{HO}=S_{HO}+2\partial_y^2\ln \left(\alpha+\int^y_0\tilde{\psi}_1^2(s)ds\right).
	\end{equation}
The formula (\ref{Confluent missing state}) gives us the missing state for the new system,
\begin{eqnarray}\label{confluum}
\tilde{f}_m(y,t)= \frac{{\tilde{\psi}}_m(y)}{\alpha + \int_{0}^y {\tilde{\psi}}_m^2(s) ds}e^{-iE_m t}.
\end{eqnarray}
There are also other stationary solutions of $\tilde{S}_{HO} \tilde{f}=0$ that can be obtained from the square integrable eigenstates of the Harmonic oscillator $\psi_n$ when $n\neq m$. Let us denote
\begin{eqnarray}\label{Confluent SUSY Solutions}
\tilde{f}_n(y,t)&=&\tilde{\mathcal{L}}_{12}\tilde{\psi}_n(y) e^{-i E_n t}= \left(\partial_y-\frac{\partial_y\check{u}_2}{\check{u}_2}\right)\left(\partial_y-\frac{\partial_y \tilde{u}_1}{\tilde{u}_1}\right)\tilde{\psi}_n(y) e^{-i E_n t},\quad \check{u}_2=\tilde{u}_1\partial_y\frac{\tilde{u}_2}{\tilde{u}_1}.
\end{eqnarray}	
After the point transformation (\ref{FinkelFunction}) of the Schr\"odinger operator $\tilde{S}_{HO}$, we get
\begin{equation}
S_2=(1+z^2)^{-1}U^{-1}\tilde{S}_{HO}(y(x,z),t(z))U=S_{0}+2\partial_x^2\ln \left(\alpha+\int^{y(x,z)}_0u_1^2(s)ds\right).
\end{equation}

\begin{figure}[t!] 
	\centering
	\begin{subfigure}[b]{0.3\textwidth}
        \includegraphics[width=\textwidth]{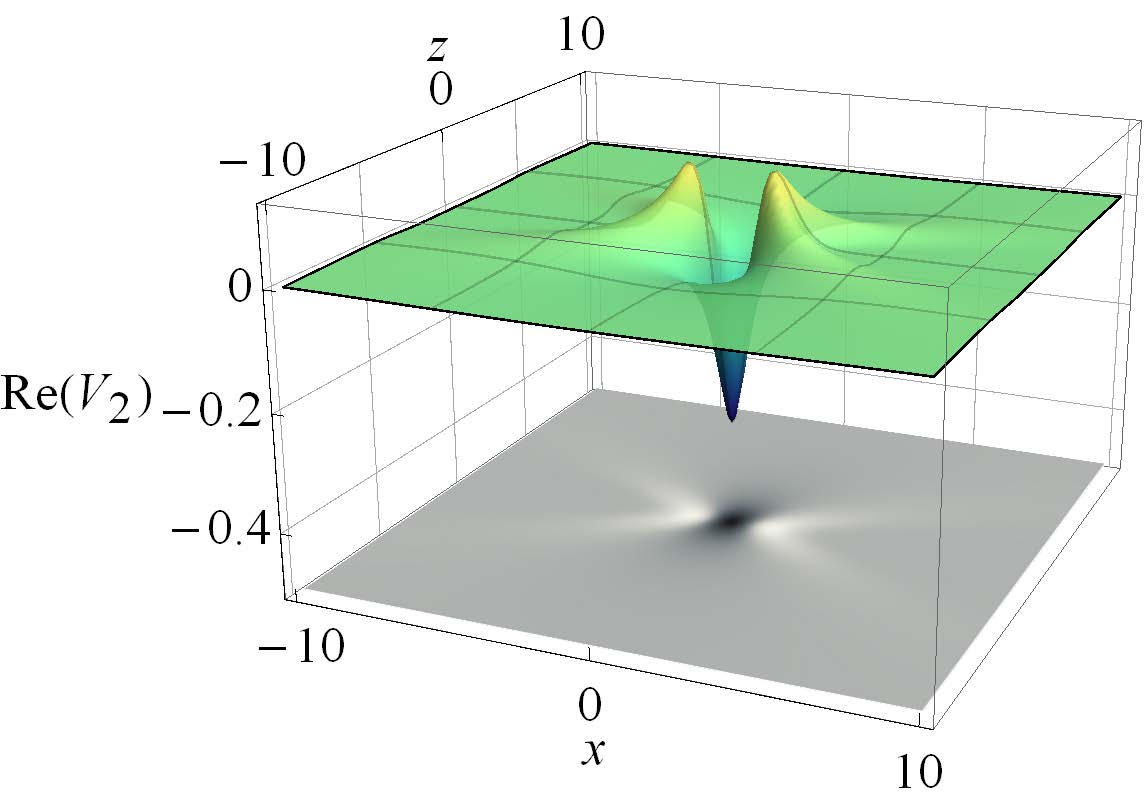}
        \caption{}
    \end{subfigure}
    \begin{subfigure}[b]{0.3\textwidth}
        \includegraphics[width=\textwidth]{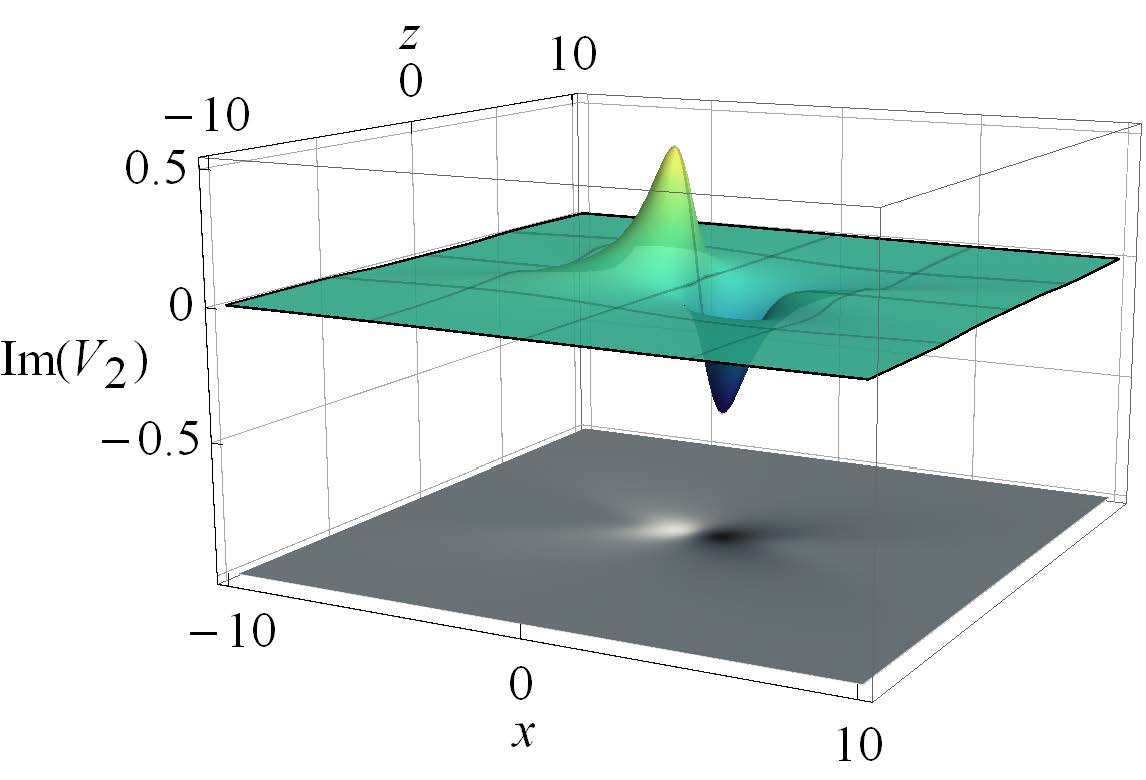}
        \caption{}
    \end{subfigure}
    \begin{subfigure}[b]{0.3\textwidth}
        \includegraphics[width=\textwidth]{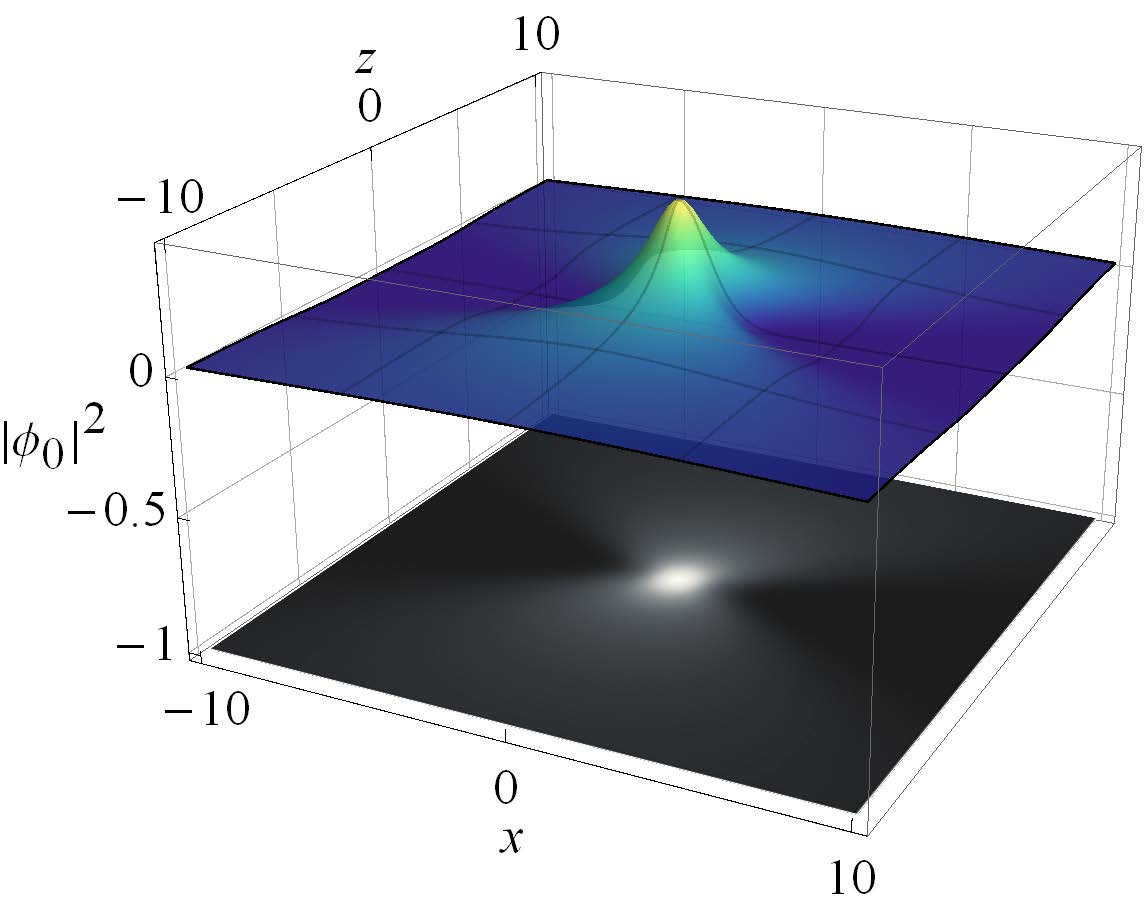}
        \caption{}
    \end{subfigure} \\
    \begin{subfigure}[b]{0.3\textwidth}
        \includegraphics[width=\textwidth]{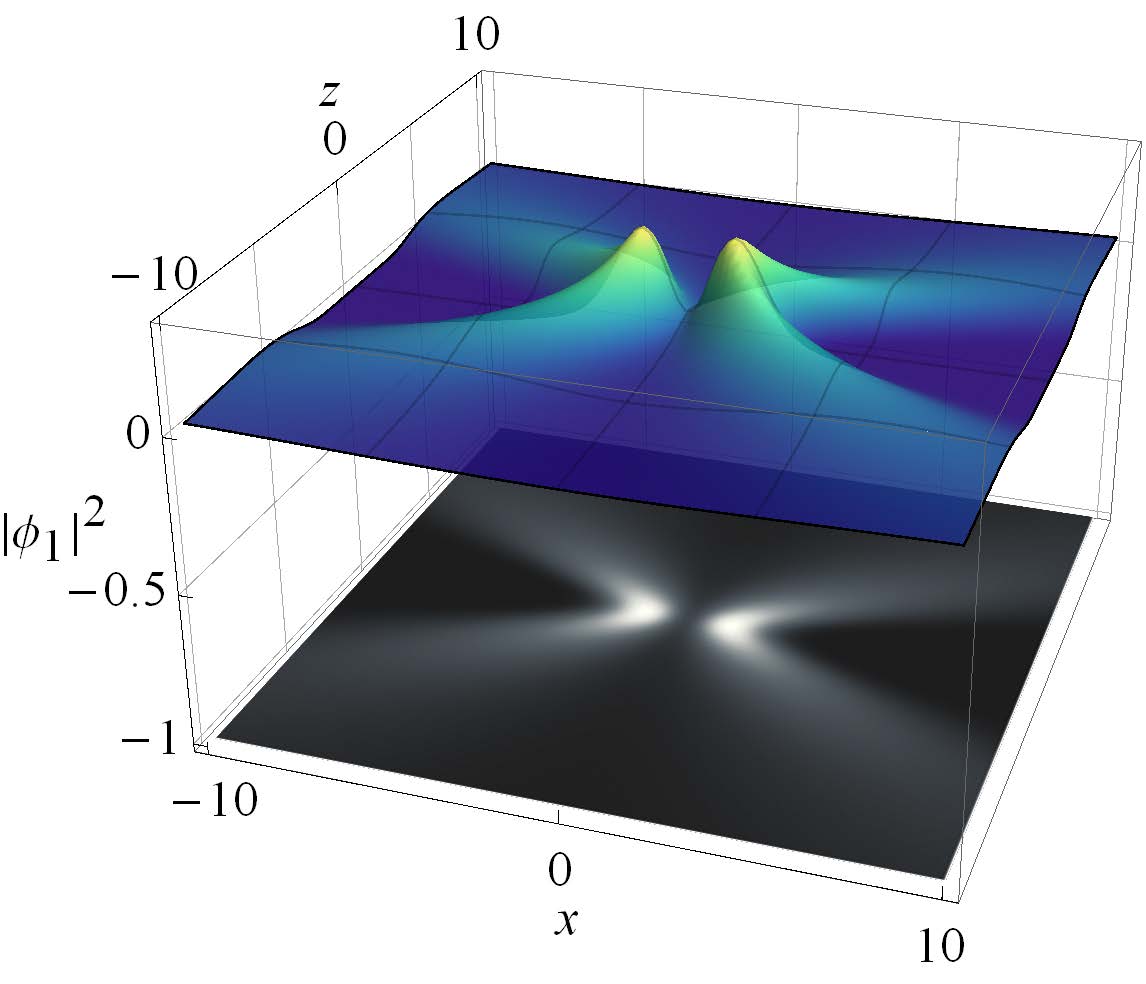}
        \caption{}
    \end{subfigure}
    \begin{subfigure}[b]{0.3\textwidth}
        \includegraphics[width=\textwidth]{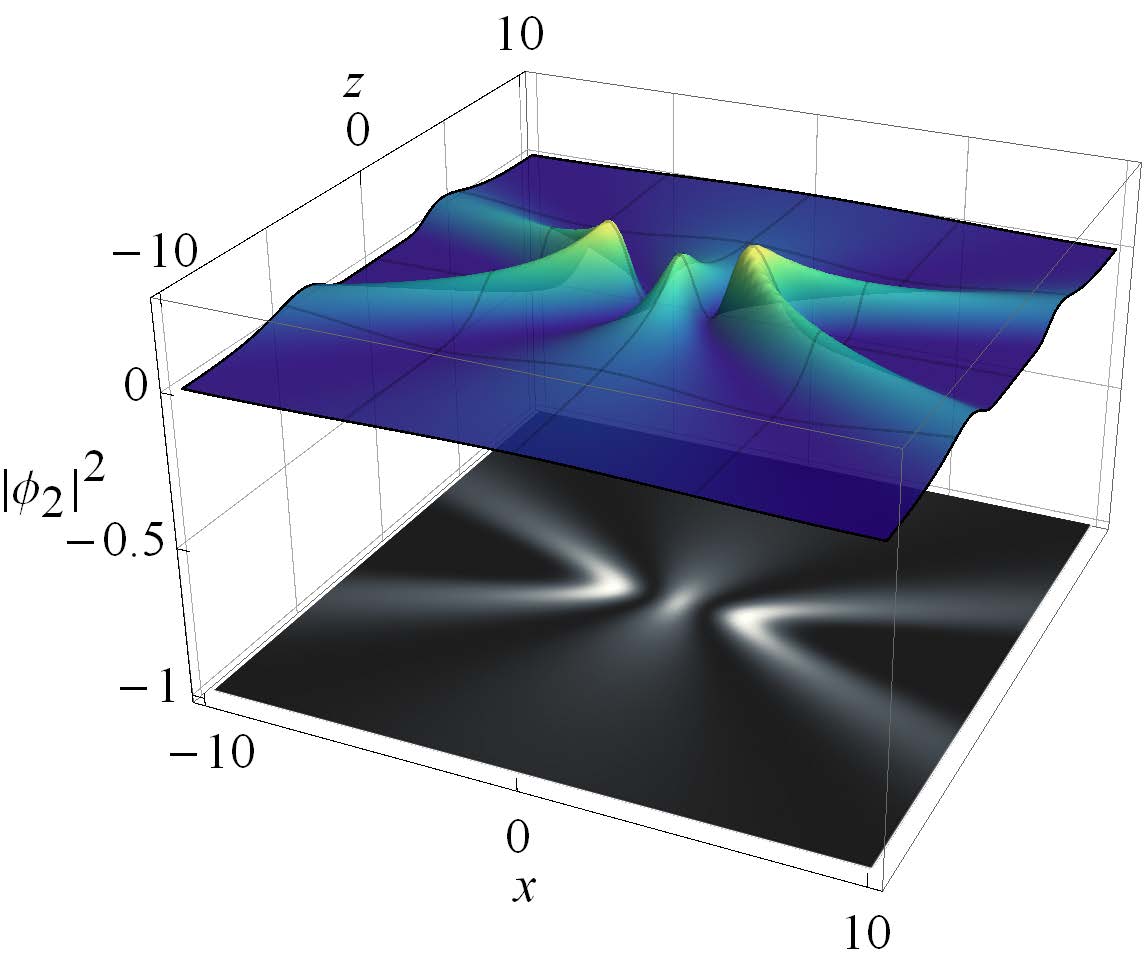}
        \caption{}
    \end{subfigure}
	\caption{Localized defect in a homogeneous crystal. Plots of the  real (a) and imaginary (b) parts of $V_2$ when $\alpha=i$, see \eqref{PTHO Example Potential}. Moreover,  the absolute value squared of three solutions are plotted: $|\phi_0|^2$ (c), see \eqref{phi0}, $|\phi_1|^2$ (d) and $|\phi_2|^2$ (e), see \eqref{phin}.   } \label{FigPTHOPotential}
\end{figure}

For explicit illustration, let us fix $m=0$. It means that the ground state of the harmonic oscillator has been selected as the transformation function to perform the confluent SUSY transformation and $E_m\equiv E_0 = 1/2$. Then the potential term (\ref{Confluent SUSY V2}) acquires the following explicit form
\begin{eqnarray}
V_2(x,z)=
\frac{4}{(1+z^2)^{3/2}}\frac{2\sqrt{1+z^2}e^{-\frac{x^2}{1+z^2}}+x e^{-\frac{x^2}{2(1+z^2)}}\left(2\alpha+\sqrt{2\pi}\mbox{erf}\left(\frac{x}{\sqrt{2(1+z^2)}}\right)\right)}{\left(2\alpha+\sqrt{2\pi}\mbox{erf}\left(\frac{x}{\sqrt{2(1+z^2)}}\right)\right)^2}.
\label{PTHO Example Potential}
\end{eqnarray}
Choosing $\alpha$ as a pure imaginary parameter and since $\text{erf}(-x)=- \text{erf}(x)$, one can check that the new potential $V_2$ is invariant with respect to both $\mathcal{P}_x \mathcal{T}$ and $\mathcal{P}_2 \mathcal{T}$. 
The potential $V_2$ represents a well localized defect of a uniform refractive index. Indeed, the potential behaves as $V_2(x,z)\sim O\left(x e^{-\frac{x^2}{2 \left(z^2+1\right)}}\right)$ for large $|x|$ and fixed $z$, whereas for fixed $x$ and large $|z|$, it behaves as $V_2(x,z)= O\left(\frac{e^{-\frac{x^2}{2 \left(z^2+1\right)}}}{z^2}\right)$.
In Fig. \ref{FigPTHOPotential} plots of the real (a) and imaginary (b) parts of $V_2$ for $\alpha=i$ are shown. 
%%%%%%%%%%%%%%%%%%%%%%
%%%%%%%%%%%%%%%%%%%%%%
%%%%%%%%%%%%%%%%%%%%%%
%%%%%%%%%%%%%%%%%%%%%%

\begin{figure}[t!]
	\centering
 \begin{subfigure}[b]{0.3\textwidth}
        \includegraphics[width=\textwidth]{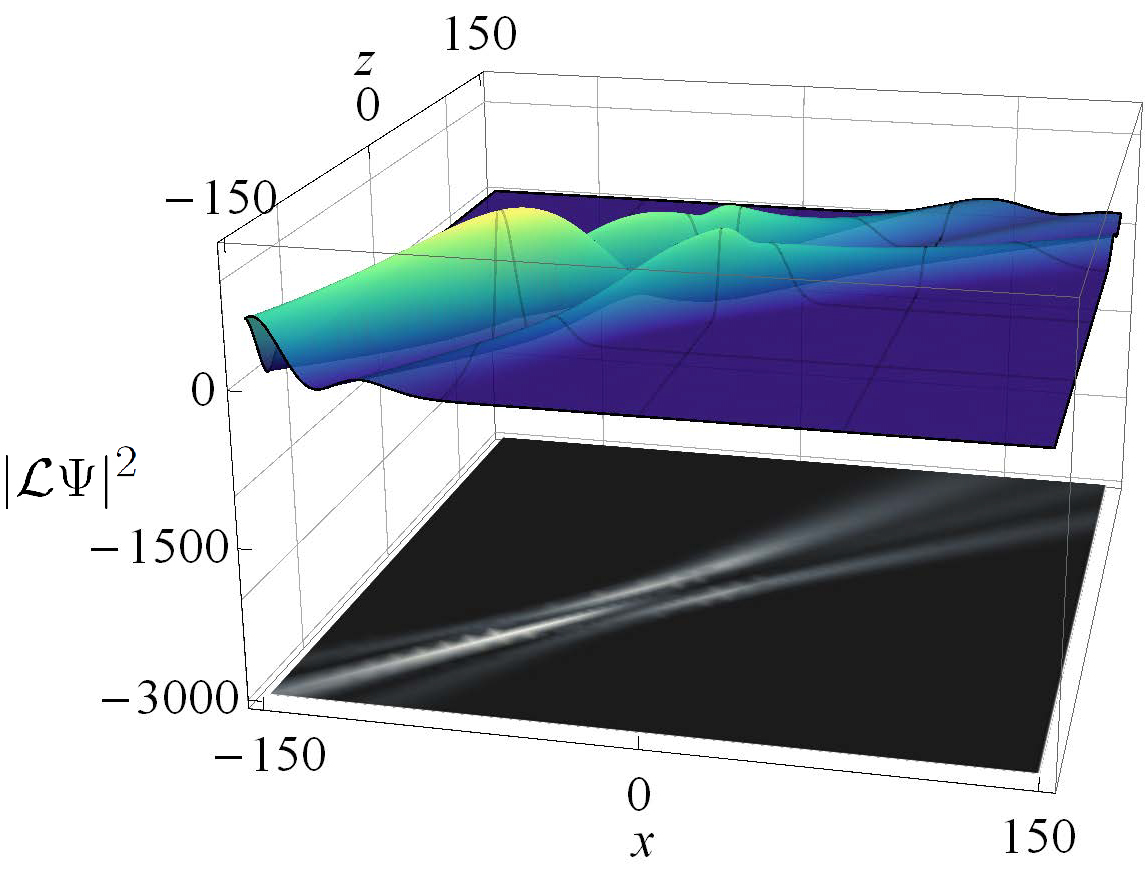}
        \caption{}
    \end{subfigure}
     \begin{subfigure}[b]{0.3\textwidth}
        \includegraphics[width=\textwidth]{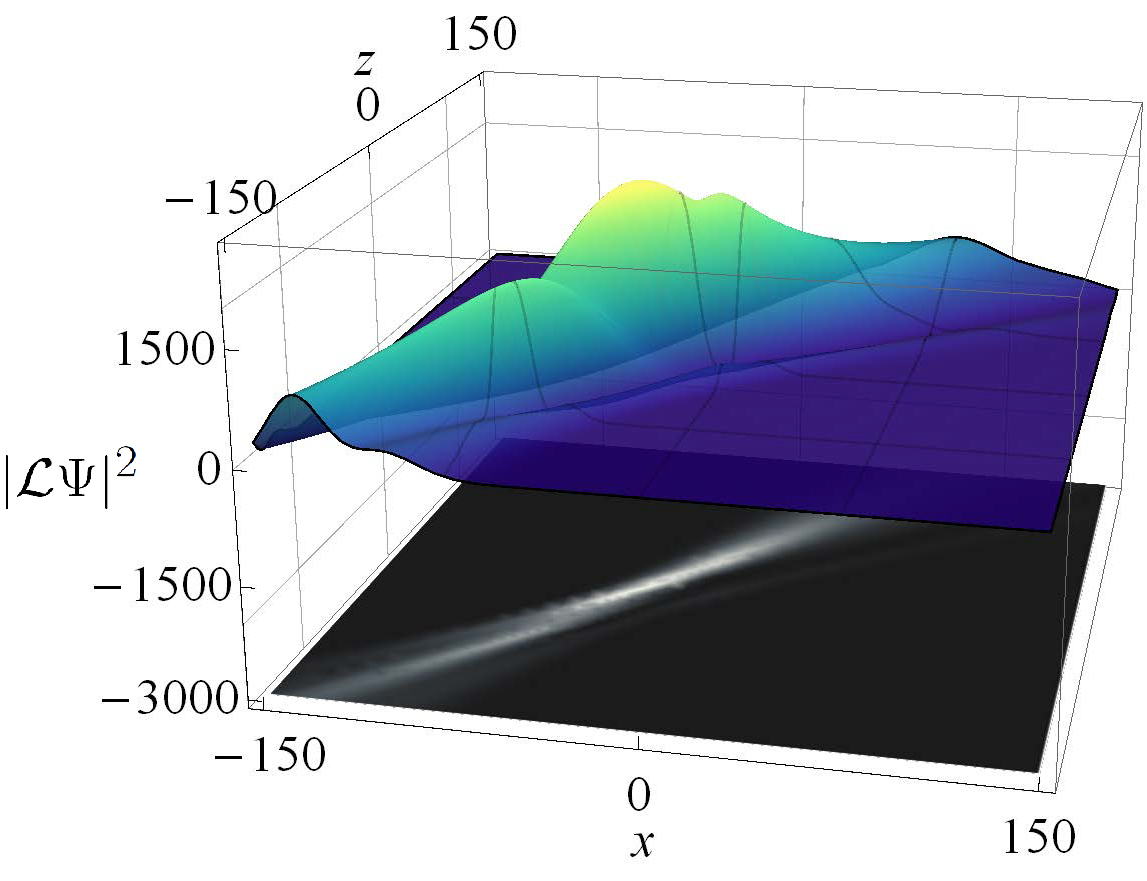}
        \caption{}
    \end{subfigure}
     \begin{subfigure}[b]{0.3\textwidth}
        \includegraphics[width=\textwidth]{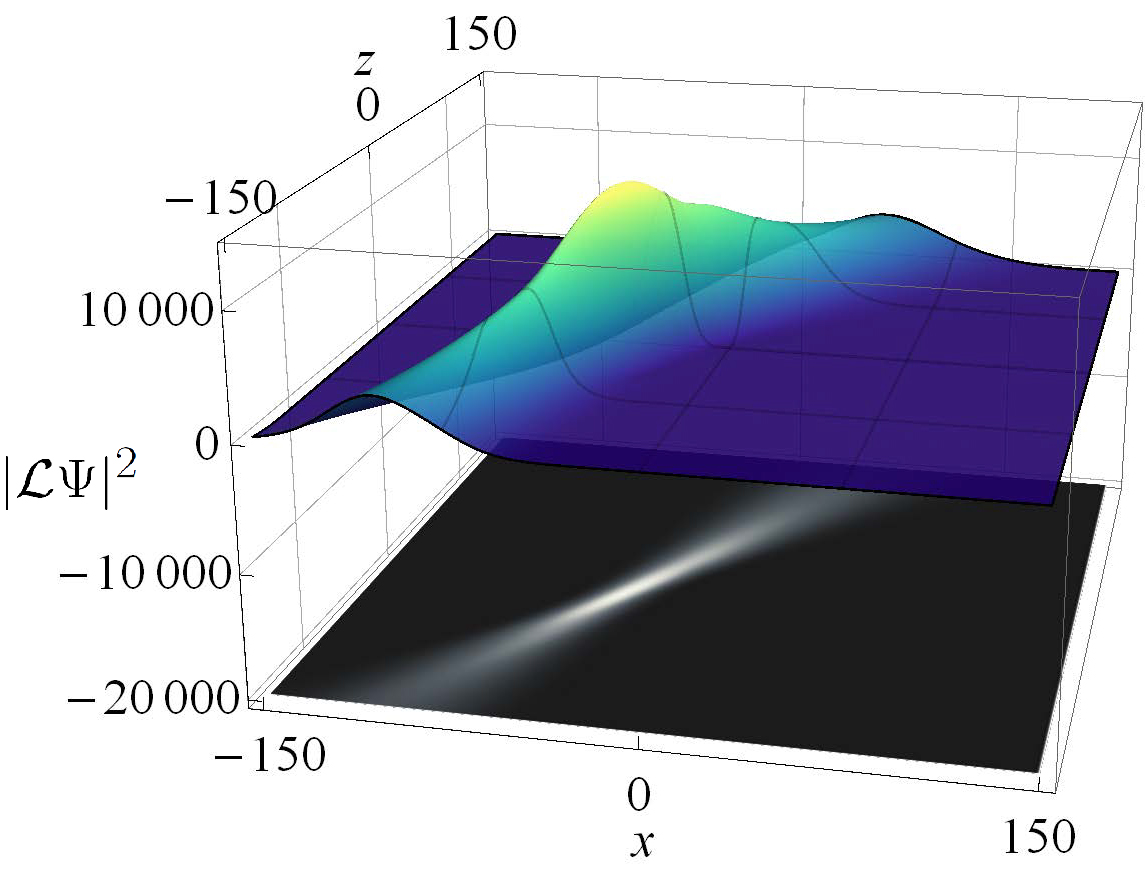}
        \caption{}
    \end{subfigure}	
	\caption{Propagation of three different wave packets of the form $\mathcal{L}_{12}\Psi_{x_0, z_0, v_0,\sigma} $, see \eqref{propagation LPsi}. The parameters used are $x_0=-40$, $z_0=-40$, $\sigma=10$, $\alpha=i$ (see \eqref{PTHO Example Potential}) and $v_0=0.98$ in (a), $v_0=0.85$ in (b) and $v_0=0.60$ in (c).    } \label{beamPD}
\end{figure}

The use of the point transformation (\ref{FinkelFunction}) allowed us to work with the less complicated form of the wave functions. However, we could make a direct, second order transformation to the free particle system that would transform it into $S_2$,
\begin{equation}
\mathcal{L}_{12}S_{0}={S}_{2}\mathcal{L}_{12}
\end{equation}
where %\begin{equation}
$\mathcal{L}_{12}=U\tilde{\mathcal{L}}_{12}(y(x,z),t(z))U^{-1}$. The explicit form of the intertwining operator is
\begin{equation}
\mathcal{L}_{12}=(1+z^2)\partial_x^2-\left(-ixz+2\frac{\sqrt{1+z^2}}{B(x,z)}\right)\partial_x+\left(\frac{1-\frac{1}{2}x^2-i z}{2}-\frac{x(1-iz)}{(1+z^2)^{1/2}B(x,z)}\right).
\end{equation}
We abbreviated $B(x,z)=e^{\frac{x^2}{2(1+z^2)}}\left(2\alpha+\sqrt{2\pi}\,\mbox{erf}\left(\frac{x}{\sqrt{2(1+z^2)}}\right)\right)$. One can check that it satisfies (\ref{wavepacketpreservation}) when applied on the wave packets $\Psi_{x_0,z_0,v_0,\sigma}$
\begin{equation} \label{propagation LPsi}
\mathcal{L}_{12}\Psi_{x_0,z_0,v_0,\sigma}=G(x,z)\Psi_{x_0,z_0,v_0,\sigma},\quad  
%\end{equation}
%where 
%\begin{equation}
G(x,z)=\begin{cases}O(1),\quad |z|\rightarrow \infty, \ x\ \mbox{constant},\\O(x^2),\quad |x|\rightarrow \infty, \ z\ \mbox{constant}.\end{cases}
\end{equation}
The function $G(x,z)$ changes the profile of the wave packet; it gets divided such that it flows around the defect in two beams, see Fig.\ref{beamPD} for illustration.
%%%%%%%%%%%%%%%%%%%%%%
%%%%%%%%%%%%%%%%%%%%%%
%%%%%%%%%%%%%%%%%%%%%%
%%%%%%%%%%%%%%%%%%%%%%

The missing state (\ref{confluum}), after the point transformation, reads explicitly
\begin{eqnarray} \label{phi0}
\phi_0(x,z)=U^{-1} \tilde{f}_0 \left(y(x,t),t(z) \right)=  \frac{e^{-\frac{x^2}{4(1+i z)}-\frac{i}{4}\arctan z}}{(1+z^2)^{1/4}\left(2\alpha+\sqrt{2\pi}\mbox{erf}\left(\frac{x}{\sqrt{2(1+z^2)}}\right)\right)}.
\end{eqnarray} 
Additionally, there are other localized solutions of $S_2 f=0$, the light dots, that are associated with (\ref{Confluent SUSY Solutions}). They are
\begin{align}\label{phin}
\phi_n(x,z) = \frac{1}{(1+z^2)^{1/4}}  \exp\left( \frac{i z}{4(1+z^2)} x^2    \right) \tilde{f}_n\left(\frac{x}{\sqrt{1+z^2}}, \arctan(z)\right). 
\end{align}
These functions satisfy $\mathcal{P}_x \mathcal{T}\phi_n(x,z)=(-1)^n\phi_n(x,z)$. We can see that for  large $|x|$ and fixed $z$, the functions vanish like an exponential multiplied by a polynomial. Along the curves $x= c \sqrt{1+z^2}$ where the argument of $\tilde{f}_n(y(x,t),t(z))$ is constant, the wave functions behave as $|\phi_n(c \sqrt{1+z^2},z)|= O\left((1+z^2)^{-1/4}\right)$. In Fig. \ref{FigPTHOPotential} (c)-(e), the intensity densities of three solutions are plotted: $|\phi_0|^2$, $|\phi_1|^2$ and $|\phi_2|^2$, respectively, see \eqref{phi0} and \eqref{phin}.

\begin{figure}[t!]
	\centering
    \begin{subfigure}[b]{0.3\textwidth}
        \includegraphics[width=\textwidth]{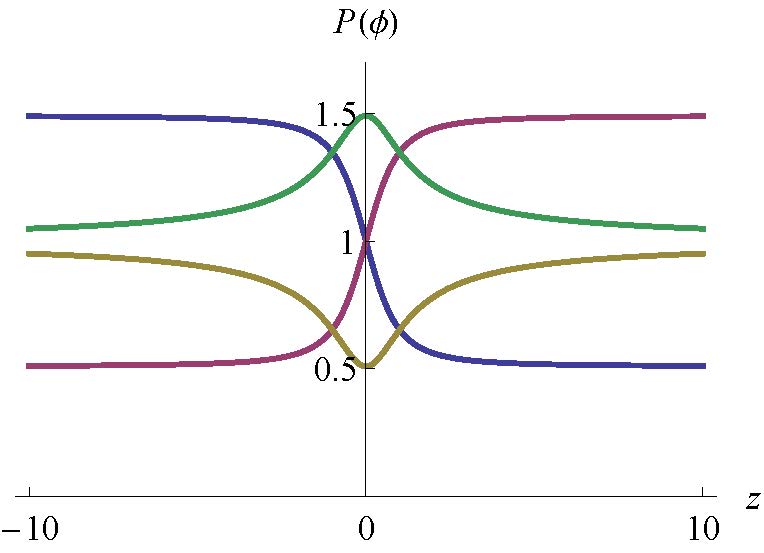}
        \caption{}
    \end{subfigure}
    \begin{subfigure}[b]{0.3\textwidth}
        \includegraphics[width=\textwidth]{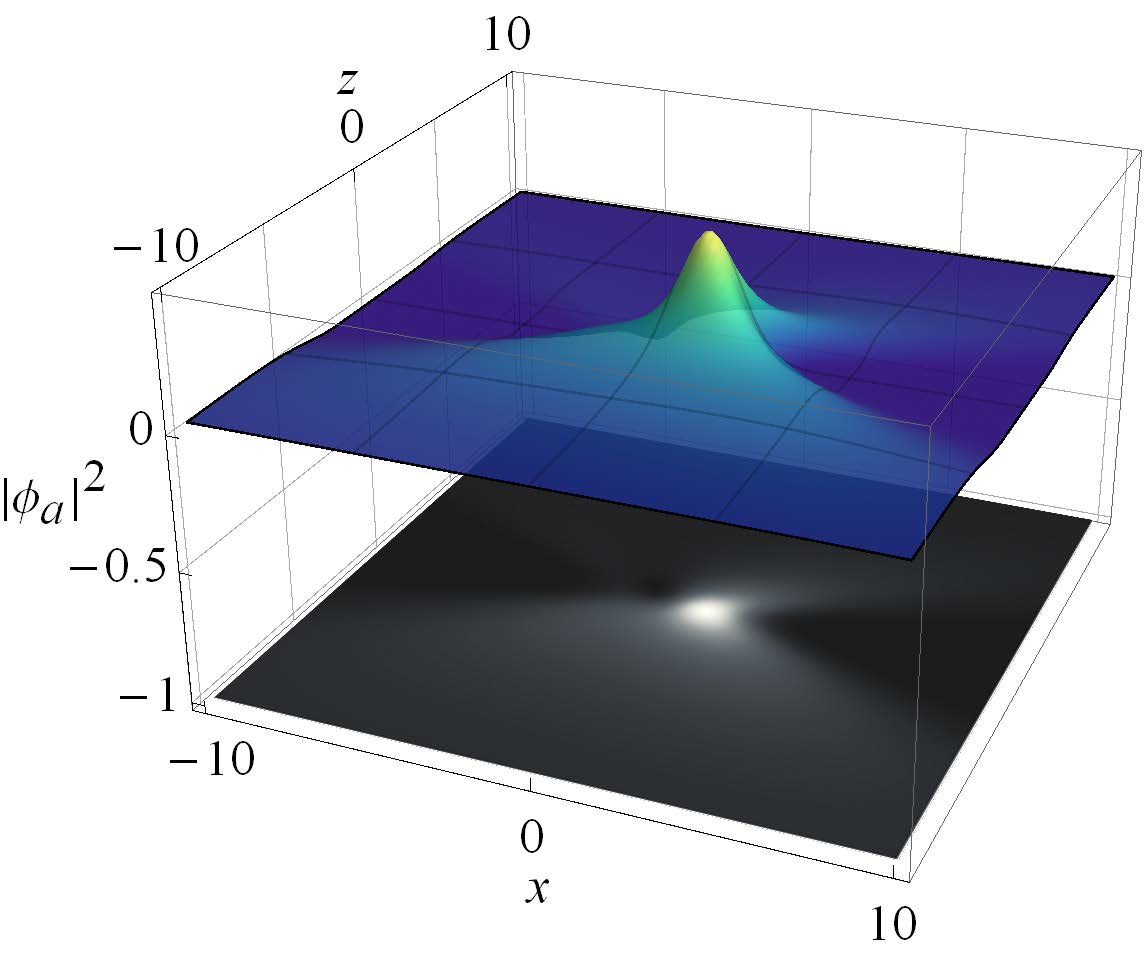}
        \caption{}
    \end{subfigure}
    \begin{subfigure}[b]{0.3\textwidth}
        \includegraphics[width=\textwidth]{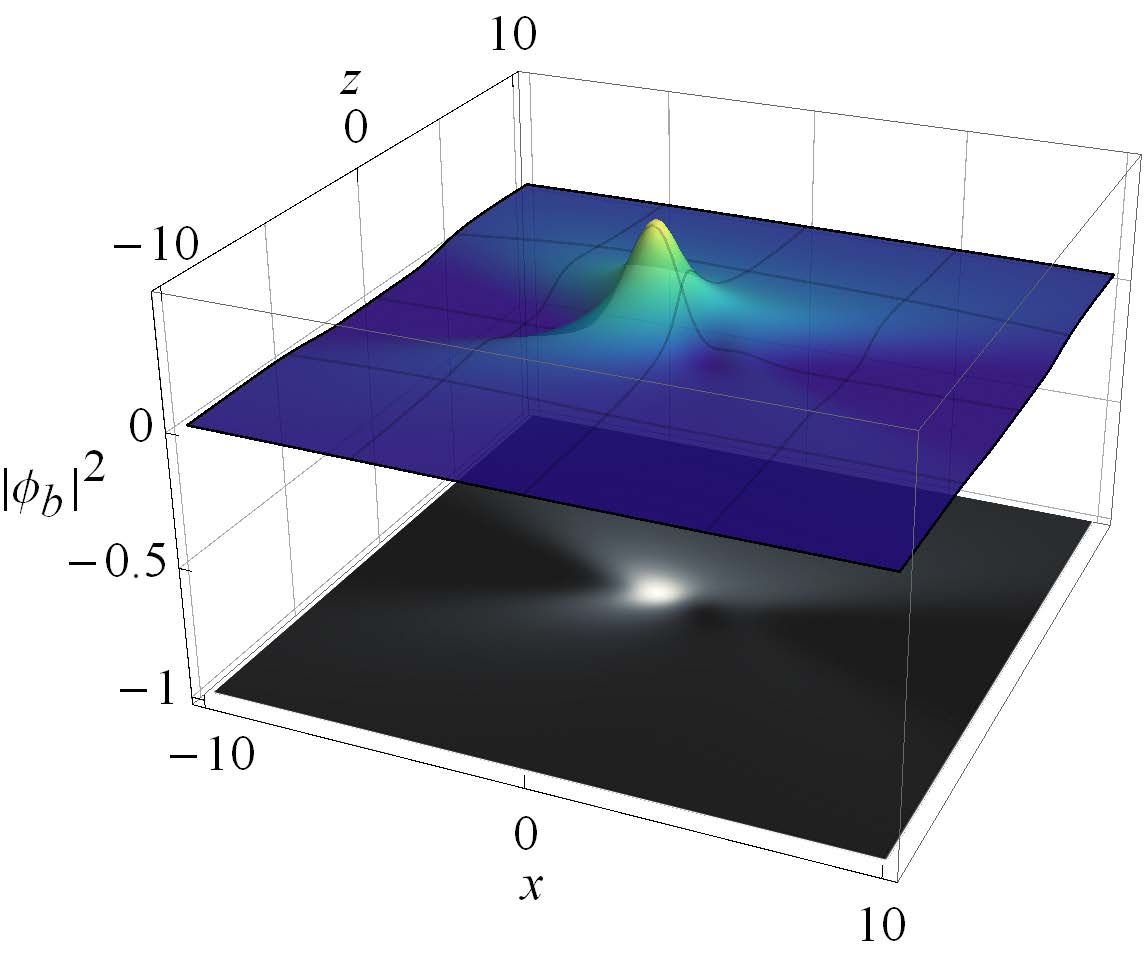}
        \caption{}
    \end{subfigure}\\
    \begin{subfigure}[b]{0.3\textwidth}
        \includegraphics[width=\textwidth]{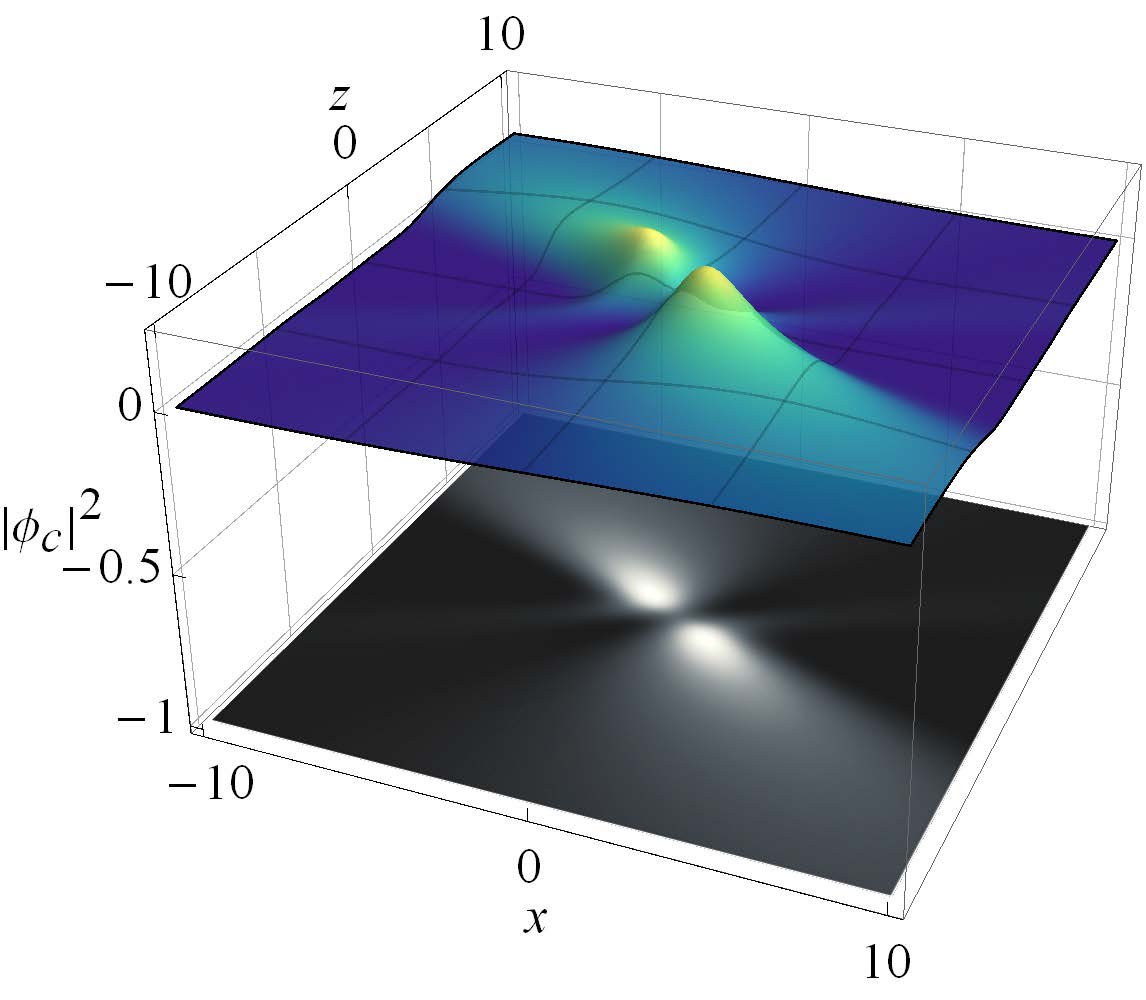}
        \caption{}
    \end{subfigure}
    \begin{subfigure}[b]{0.3\textwidth}
        \includegraphics[width=\textwidth]{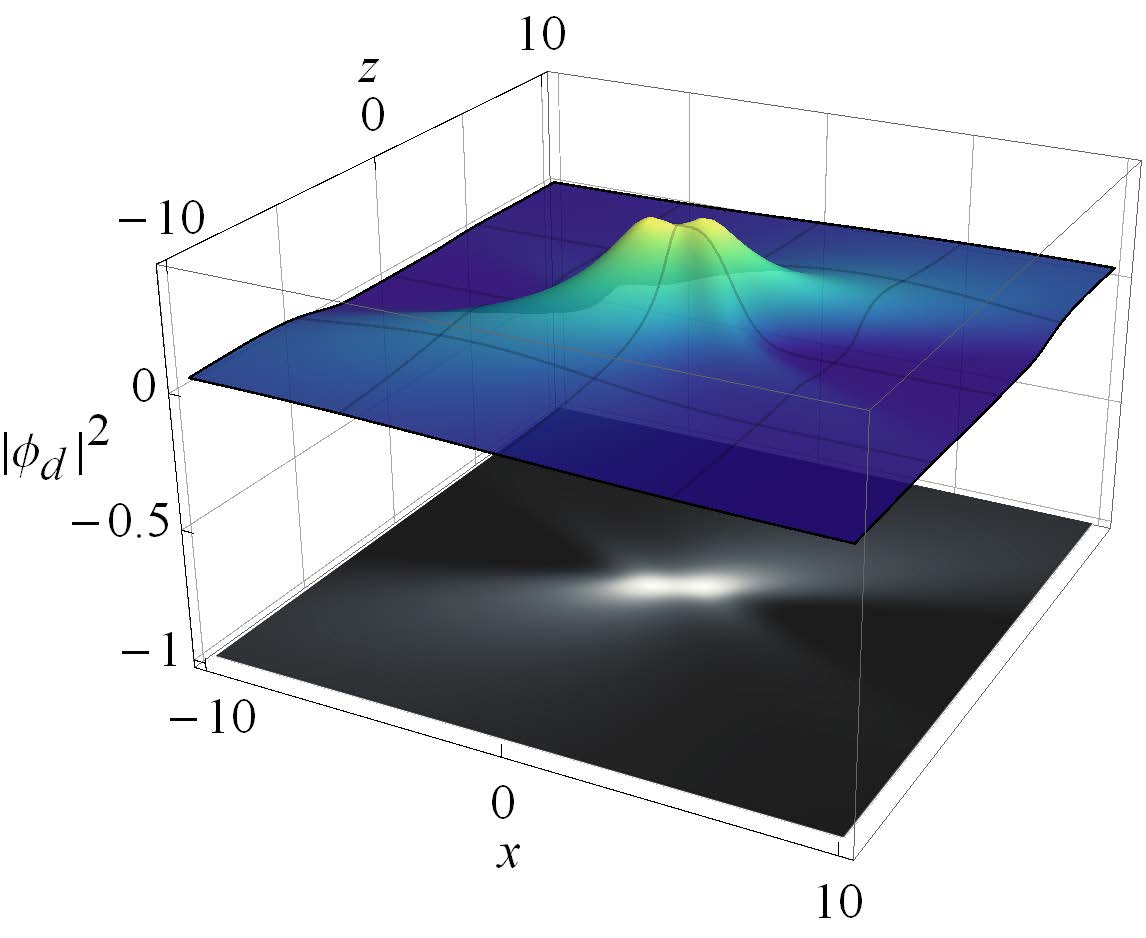}
        \caption{}
    \end{subfigure}
	\caption{(a): Plots of the power of four different superposition states for the localized defect in a homogeneous crystal system. In blue $P(\phi_a)$, in purple $P(\phi_b)$, in yellow $P(\phi_c)$ and in green $P(\phi_d)$, the superpositions are defined in \eqref{Superpositions}. Furthermore, the intensity densities of the states $\phi_a$ (b), $\phi_b$ (c), $\phi_c$ (d) and $\phi_d$ (e) are plotted.  } \label{Fig PTHO Power}
\end{figure}

It follows directly from the formulas (\ref{phin}) that the power of the light beam defined as
$
P(\phi_n)= \int_{-\infty}^{\infty} |\phi_n(x,z)|^2 dx
$
is constant, i.e. it does not depend on $z$. However, different superpositions of states $\phi_n$  would not have a constant power. Using the same example ($m=0, \alpha=i)$, consider the following four superpositions:
\begin{eqnarray} \label{Superpositions}
\phi_a= \frac{1}{\sqrt{2}}\left(N_0 \phi_0 + i N_1 \phi_1   \right), \quad \phi_b= \overline{\phi_a}, \quad \phi_c= \frac{1}{\sqrt{2}}\left(N_0\phi_0 + N_1\phi_1   \right),\quad 
\phi_d= \frac{1}{\sqrt{2}}\left(N_0\phi_0 - N_1\phi_1   \right)
\end{eqnarray}
where $N_0,~N_1$ are (positive) normalization constants. For such states, the power is a nontrivial function of $z$. In Fig. \ref{Fig PTHO Power} (a), the power of these states is plotted. First in blue the power of the first superposition $P(\phi_a)$ was plotted. It can be seen that power decreases near $z=0$,  the interaction of the light with the defect results in its absorption. In purple, we have $P(\phi_b)$ representing the opposite case, power increases after the interaction zone. In yellow, $P(\phi_c)$ has a minimum around zero whereas $P(\phi_d)$ in green has a maximum.  In Fig. \ref{Fig PTHO Power} (b)-(e)  the functions $|\phi_a|^2$, $|\phi_b|^2$, $|\phi_c|^2$, $|\phi_d|^2$ were plotted, respectively. 
%%%%%%%%%%%%%%%%%%%%%%
%%%%%%%%%%%%%%%%%%%%%%
%%%%%%%%%%%%%%%%%%%%%%
%%%%%%%%%%%%%%%%%%%%%%

\section{Guided modes in optical waveguides \label{cuatro} }
In this section, we will focus on models of optical wave guides. By an asymptotic analysis of the involved quantities, we will obtain general results for a large class of initial potentials that are just required to be integrable and to possess translational symmetry. In particular, we will show how to find solutions of the associated Schr\"odinger equation that are vanishing exponentially in the transverse direction to the wave guide,  and hence, represent guided modes.

\subsection{On the construction of guided modes\label{4.1}}
Let us suppose that the initial potential $V_0$ does not depend on $z$.
It is known (see e.g. Th. 4.1, p.70 in \cite{BS}) that when the absolute value of the potential term is integrable, $\int_{\mathbb{R}}|V_0(x)|dx<\infty$, then the stationary equation $(-\partial_x^2+V_0(x))\psi=\lambda^2\psi$ has two solutions $\psi^+(\lambda,x)$ and $\phi^+(\lambda,x)$  for a complex $\lambda\neq 0$ satisfying
\begin{equation}\label{set1}
 \psi^+(\lambda,x)=e^{i\lambda x}(1+o(1)),\quad \phi^+(\lambda,x)=e^{-i\lambda x}(1+o(1))\quad\mbox{as}\quad x\rightarrow+\infty.
\end{equation}
As the integrability of $V_0(x)$ is invariant with respect to $x\rightarrow -x$, there are also two solutions $\psi^-(\lambda,x)$ and $\phi^-(\lambda,x)$ satisfying
\begin{equation}\label{set2}
 \psi^-(\lambda,x)=e^{-i\lambda x}(1+o(1)),\quad \phi^-(\lambda,x)=e^{i\lambda x}(1+o(1))\quad\mbox{as}\quad x\rightarrow-\infty.
\end{equation}
The two sets (\ref{set1}) and (\ref{set2}) represent two possible choices of the fundamental solutions of the stationary equation, i.e. a function from one set can be written as a linear combination of the functions from the other set.  Let us mention that a finite square well is a simple example of the system where the potential term satisfies the condition of integrability. In this case, the small terms $o(1)$ are identically zero in the wave functions (\ref{set1}) and (\ref{set2}).

In order to perform a Darboux transformation to the potential $V_0$ we need to choose an adequate function $u$. Moreover, we will select a preimage $v$, satisfying $S_0 v = 0$, such that $\mathcal{L}v$ is a guided mode fulfilling $S_1 \mathcal{L}v=0$. First, let us assume that $\lambda^2 \in \mathbb{R}$, then $\lambda$ can be written either as $\lambda  = -i k, ~k>0,$ or as $\lambda = r, ~r>0$, depending on the sign of $\lambda^2$. Now, let us consider $N+M$ stationary solutions, $N$ of them with eigenvalues $\lambda_j=-i k_j$, $j=1,\dots,N$, where $k_1>k_2>\dots>k_N>0$; and $M$ solutions such that $\lambda_{N+\ell}=r_\ell>0$, $\ell=1,\dots,M$. We denote the corresponding functions (\ref{set1}) or (\ref{set2}) as $\psi^{\pm}_{k_j}(x)\equiv \psi^\pm(-ik_j,x)$,  $\phi^{\pm}_{k_j}(x)\equiv \phi^\pm(-ik_j,x)$, $\psi^{\pm}_{r_\ell}(x)\equiv \psi^\pm(r_\ell,x)$ and $\phi^{\pm}_{r_\ell}(x)\equiv \phi^\pm(r_\ell,x)$. Hence, we selected $N$ solutions that increase (decrease) exponentially for large $|x|$ and $M$ solutions that are asymptotically bounded and oscillating. We suppose that their derivatives satisfy
\begin{equation}
 (\psi_{k_j}^{\pm})'=\pm k_je^{\pm {k_j}x}(1+o(1)),\quad (\phi_{k_j}^{\pm})'=\mp k_je^{\mp {k_j}x}(1+o(1)),\quad  (x\rightarrow\pm\infty)
\end{equation}
and $(\psi_{r_\ell}^{\pm})'$ and $(\phi_{r_\ell}^{\pm})'$ are bounded functions. These requirements can be met provided that the functions do not have asymptotically small but rapidly oscillating terms.

Let us compose the following functions
\begin{eqnarray}\label{uv}
 &&u=\sum_{j=1}^{N}F_je^{ik_j^2z}+\sum_{\ell=0}^MG_\ell e^{-ir_\ell^2z},\quad v=\sum_{j=1}^{N}\tilde{F}_je^{ik_j^2z}+\sum_{\ell=0}^M\tilde{G}_\ell e^{-ir_\ell^2z},
\end{eqnarray}
where $F_j$, $\tilde{F}_j$, $G_\ell$, $\tilde{G}_\ell$, $j\in\{1,\dots,N\}$, $\ell\in\{1,\dots,M\}$ are generic linear combinations
\begin{eqnarray}
 &&F_j=a_j^{\pm}\psi_{k_j}^{\pm}+b_j^{\pm}\phi_{k_j}^{\pm}
,\quad
G_\ell=c_\ell^{\pm}\psi_{r_\ell}^{\pm}+d_\ell^{\pm}\phi_{r_\ell}^{\pm},\nonumber\\
&&%\begin{equation}
\tilde{F}_j=\tilde{a}_j^{\pm}\psi_{k_j}^{\pm}+\tilde{b}_j^{\pm}\phi_{k_j}^{\pm}
,\quad
\tilde{G}_\ell=\tilde{c}_\ell^{\pm}\psi_{r_\ell}^{\pm}+\tilde{d}_\ell^{\pm}\phi_{r_\ell}^{\pm},\label{FG}
\end{eqnarray}
and $G_0=\tilde{G}_0=0$.
The functions $G_\ell$ and $\tilde{G}_\ell$ are asymptotically oscillating and bounded. 
The functions $F_j$ and $\tilde{F}_j$ are exponentially expanding, 
\begin{align}\label{tildeFG}
 F_j=a_j^{\pm}e^{\pm k_jx}+o(e^{\pm k_jx}),\ (x\rightarrow\pm\infty),\nonumber\\ \tilde{F}_j=\tilde{a}_j^{\pm}e^{\pm k_jx}+o(e^{\pm k_jx}),\ (x\rightarrow\pm\infty),\nonumber\\
 (F_j)'=\pm k_ja_j^{\pm}e^{\pm k_jx}+o(e^{\pm k_jx}),\ (x\rightarrow\pm\infty),\nonumber\\ (\tilde{F}_j)'=\pm k_j\tilde{a}_j^{\pm}e^{\pm k_jx}+o(e^{\pm k_jx}),\ (x\rightarrow\pm\infty).
\end{align}

We define the Darboux transformation 
\begin{equation}\label{wgL}\mathcal{L}=L_1(z)\left(\partial_x-\frac{u'}{u}\right).\end{equation}
Contrary to the cases discussed in the previous section, the superpotential $\mathcal{W}(x,z)=-\partial_x\ln u$ is asymptotically constant for large $|x|$. Indeed, substituting the explicit form (\ref{uv}) of $u$ into (\ref{wgL}) and using (\ref{tildeFG}), one finds that $\mathcal{W}(x,z)=\mp k_1+o(1)$ for $x\rightarrow\pm\infty$. Hence, the action of $\mathcal{L}$ on a plane wave (\ref{planewave}) preserves its amplitude for large $|x|$, $\mathcal{L}\Phi_{k,x_0,z_0,v_0}=L_1(z)(ik\mp k_{1}+o(1))\Phi_{k,x_0,z_0,v_0}$ for $x\rightarrow\pm\infty$. When acting on the wave packets (\ref{wavepacket}), we get
\begin{equation}\label{LPWG}
\mathcal{L}\Psi_{x_0,z_0,v_0,\sigma}=G(x,z)\Psi_{x_0,z_0,v_0,\sigma},\quad G(x,z)=L_1(z)\left(\frac{i(x-x_0)-v_0\sigma}{2(z-z_0-i \sigma)}\mp k_1+g(x,z)\right).
\end{equation}
where $g(x,z)$ is asymptotically bounded and oscillating function in $z$. Hence, the requirement (\ref{wavepacketpreservation}) then suggest to fix the function $L_1(z)$ to be bounded (and non-vanishing) for all real $z$.

\subsubsection*{The potential of wave guide}
The function $u$ fixes the new potential $V_1=V_0+\delta V_1+i\partial_z\ln L_1(z)$, where $\delta V_1=-2\frac{u''}{u}+2\frac{(u')^2}{u^2}$ is asymptotically vanishing for large $|x|$. Indeed, we have   
\begin{eqnarray}
 \delta V_1&=&\frac{\sum\limits_{j,s=1}^N\left[(V_0+k_j^2)F_jF_s-F'_jF'_s\right]e^{i(k_j^2+k_s^2)z}}{u^2}+\frac{\sum\limits_{\{j,\ell\}=\{1,0\}}^{N,M}\left[(2V_0-r_\ell^2+k_j^2)F_jG_{\ell}-2F_j'G'_\ell\right]e^{i(k_j^2-r_\ell^2)z}}{u^2}\nonumber\\
&&+\frac{\sum\limits_{\ell,s=0}^M\left[(V_0-r_\ell^2)G_\ell G_s-G_\ell'G_s'\right]e^{-i(r_\ell^2+r_s^2)z}}{u^2}.
\end{eqnarray}
As we have $u^2= \left(a_1^{\pm}\right)^2e^{\pm 2k_1x}(1+o(1))$ for $x\rightarrow\pm \infty$, the last two terms vanish for large $|x|$ as their denominators increase much faster then their numerators. The first term will vanish as well provided that $(V_0+k_1^2)F_1^2-(F'_1)^2\rightarrow 0$ for $|x|\rightarrow \infty$. But taking into account (\ref{tildeFG}) and integrability of $V_0$ that implies $V_0=o(1)$ for $|x|\rightarrow\infty$, we can see that this is indeed the case. We get 
\begin{equation}
 \delta V_1=O(e^{\pm(-k_1+k_2)x})\quad \mbox{for}\quad x\rightarrow\pm\infty.
\end{equation}
The term $\delta V_1$ is regular provided that $u$ has no zeros. It is rather nontrivial to guarantee in general. We will discuss this point in the explicit examples in the end of this section.

\subsubsection*{Guided modes - the alternative construction of the missing state}
Let us inspect the asymptotic behavior of the function $\mathcal{L}v$. We shall fix the coefficients $\tilde{a}^{+}_k$ and $\tilde{b}^+_k$ such that $\mathcal{L}v$ is asymptotically vanishing for large $|x|$ or it is bounded at least. We have
\begin{eqnarray}\label{Lv}
 L_1^{-1}\mathcal{L}v&=&\frac{W(u,v)}{u}=\frac{\sum\limits_{j,s=1}^{N}W(F_j,\tilde{F}_s)e^{i(k_j^2+k_s^2)z}}{u}+\frac{\sum\limits_{\{j,\ell\}=\{1,0\}}^{N,M}\left(W(G_\ell,\tilde{F_j})+W(F_j,\tilde{G}_\ell)\right)e^{i(k_j^2-r_\ell^2)z}}{u}\nonumber\\&&+\frac{\sum\limits_{\ell,s=0}^MW(G_\ell,\tilde{G}_s)e^{-i(r_\ell^2+r^2_s)z}}{u},
\end{eqnarray}
where $W(f,g)=fg'-f'g$ is the Wronskian of two functions.
In the numerator of the first sum, there are terms of order $e^{\alpha x}$ with $\alpha\geq k_1$. We would like to fix $\tilde{a}^{\pm}$ and $\tilde{b}^{\pm}$ such that these terms vanish. Using (\ref{tildeFG}), we get
\begin{eqnarray}
\label{WFF}
W(F_j,\tilde{F}_s)+W(F_s,\tilde{F}_j)&=&e^{\pm (k_j+k_s)x}\left(\pm k_j(a_j^{\pm}\tilde{a}_s^{\pm}-a_s^{\pm}\tilde{a}_j^\pm)\pm k_s(a_s^{\pm}\tilde{a}_j^{\pm}-a_j^{\pm}\tilde{a}_s^\pm)\right)\nonumber\\&&+{o(e^{\pm (k_j+k_s)x})},\quad x\rightarrow \pm\infty.
\end{eqnarray}
We can make the term proportional to $e^{\pm (k_j+k_s)x}$ vanish by fixing
\begin{equation}\label{aa}
\tilde{a}_j^{+}=c^{+}a_j^{+},\quad \tilde{a}_j^{-}=c^{-}a_j^{-},
\end{equation}
where $c^{\pm}$ are constants.  However, the condition (\ref{aa}) cannot guarantee that $\mathcal{L}v$ will vanish exponentially; the condition (\ref{aa}) does not nullify the term $o(e^{\pm(k_j+k_s)x})$ in (\ref{WFF}) in general, so that it only forces the first term in (\ref{Lv}) to behave as $o(e^{\pm 2k_1x})$ for $x\rightarrow\pm\infty$. However, (\ref{aa}) can serve well as the guiding relation in the construction of guided modes in the explicit examples discussed later on.

Let us focus on the second term in (\ref{Lv}). It vanishes asymptotically provided that the function at  the term $e^{\pm k_1x}$ is vanishing. Using (\ref{aa}), we get
\begin{eqnarray}\label{FG}
 F_1\tilde{G}_{\ell}'-F_1'\tilde{G}_\ell+G_{\ell}\tilde{F}_1'-G_\ell'\tilde{F}_1=\left(a_1^{\pm}(\tilde{G}_\ell'-c^{\pm}G_\ell')\pm k_1a_1^{\pm}(c^{\pm}G_\ell-\tilde{G}_{\ell})\right)e^{\pm k_1 x}+o(e^{\pm k_1 x}),\quad x\rightarrow\pm\infty.\end{eqnarray}
The leading term above can be made zero provided that
\begin{equation}\label{G}
 \tilde{G}_\ell=c^{\pm}G_\ell,
\end{equation}
where $c^{\pm}$ are the two constants introduced in (\ref{aa}). When $c^{+}\neq c^-$, the only way how to make the second term of (\ref{Lv}) asymptotically vanishing for both $x\rightarrow \pm\infty$ is to make it identically zero by setting $G_\ell=\tilde{G}_\ell=0$. Fixing either $\tilde{G}_\ell=c^{+}G_\ell$ or $\tilde{G}_\ell=c^{-}G_\ell$ will make the second term vanishing either at $x\rightarrow+\infty$ or $x\rightarrow-\infty$.  Now, let us see what happens when $c^+=c^-$. Fixing $\tilde{G}_j=c^+G_j$ for all $j=1,\dots,M$, the first term on the right side of (\ref{FG}) is vanishing. We also have $\tilde{a}_j^{\pm}=c^+a_{j}^{\pm}$. Then, in general, the function $v$ can differ from $u$ only in the functions that vanish asymptotically for $|x|\rightarrow\infty$, i.e.  $u-c^+v$ is a linear combination of bound states of the initial system. In that case, it is straightforward to see that $\mathcal{L}v$ is an exponentially vanishing function for large $|x|$. When there are no bound states in the initial system, the choice $c^+=c^-$ would imply $v=c^+u$ and $\mathcal{L}v=0$ identically. Taking $c^+\neq c^-$ and $M>0$, the wave function $\mathcal{L}v$ cannot be exponentially vanishing on both sides of the wave guide. While decreasing rapidly to zero on one side, it has bounded and non vanishing oscillations on the other side. In this case, we call the wave guides weakly confining as the guided mode $v$ is ``leaking" from the wave guide on one side. We will illustrate this situation on the explicit examples below.

Both the new potential term $V_1$ and $\mathcal{L}v$ are periodic in $z$ provided that $k_1,\dots,k_N$ and $r_1,\dots,r_M$ are commensurable. For $M=0$, $V_1$ offers a strong confinement of the guided mode as $\mathcal{L}v$ vanishes outside exponentially. When $M\neq0$, the potential $V_1$ offers rather weak confinement as the guided mode ``leaks'' out of the wave guide, performing non-vanishing oscillations in $|x|\rightarrow\infty$.

\subsubsection*{$\mathcal{P}_2\mathcal{T}$-symmetry}
Up to now, we did not make any assumption on the $\mathcal{PT}$-symmetry of $V_1$. As we do not see how the function $u$ in (\ref{uv}) could satisfy (\ref{uPT1}), we look for $\mathcal{P}_2\mathcal{T}$-symmetry of the new potential. It is sufficient to fix the function $u$ such that it has definite $\mathcal{P}_2\mathcal{T}$-parity. It can be granted by taking the coefficients in (\ref{FG}) such that
\begin{equation}\label{PTFG}
 \mathcal{P}_2\mathcal{T}F_j=\epsilon\, F_j,\quad \mathcal{P}_2\mathcal{T}G_\ell=\epsilon\, G_\ell,\quad \epsilon\in\{-1,1\}.
\end{equation}
These relations imply $a^{-}_j=\epsilon\, \overline{a^{+}_j}$.

The following examples will differ by the choice of the transformation function $u$; when it consists of exponentially expanding components only, the resulting systems will represent strongly confining wave guides as the guided mode will disappear exponentially out of the wave guide. For $u$ containing the bounded oscillating components, weakly localizing wave guides will be obtained. We will consider both $\mathcal{P}_2\mathcal{T}$-symmetric models as well as non-$\mathcal{P}_2\mathcal{T}$-symmetric ones. In all the cases, we will set $L_1=1$ that, as we discussed below (\ref{LPWG}), complies with (\ref{wavepacketpreservation}).

\subsection{Examples: $\mathcal{P}_2\mathcal{T}$-symmetric deformations of the P\"oschl-Teller potential}
Our initial system will be the free particle again, so that we fix $S_0$ as in (\ref{FPEq}). 
We can identify the fundamental solutions of the stationary Schr\"odinger equation
with prescribed asymptotic behavior (\ref{set1}) and  (\ref{set2}) as
\begin{equation}
 \psi^+(\lambda,x)=\phi^-(\lambda,x)=e^{i\lambda x},\quad \phi^+(\lambda,x)=\psi^-(\lambda,x)=e^{-i\lambda x}.
\end{equation}
We shall construct $\mathcal{P}_2\mathcal{T}$-symmetric model. Taking into account (\ref{tildeFG}) and (\ref{PTFG}), the functions $F_j$ and $G_\ell$ with definite $\mathcal{P}_2\mathcal{T}$-parity have to be fixed in the following form
\begin{eqnarray}\label{FGepsilon}
 F_{j,\epsilon}&=&a^+_je^{k_jx}+\epsilon \overline{a_j^+}e^{-k_jx}=\begin{cases}|a_j^+|\cosh (k_jx+i\delta_j),\quad \epsilon=1,\\|a_j^+|\sinh (k_jx+i\delta_j),\quad \epsilon=-1, \end{cases}\quad a_{j}^+=|a_j^+|e^{i\delta_j},\quad \delta_j\in\mathbb{R},\\
G_{\ell,\epsilon}&=&c^+_\ell e^{ir_{\ell}x}+\epsilon \overline{c_\ell^+}e^{-ir_\ell x}=\begin{cases}|c_\ell^+|\cos (r_\ell x+i\mu_\ell),\quad \epsilon=1,\\|c_\ell^+|\sin (r_\ell x+i\mu_\ell),\quad \epsilon=-1, \end{cases}\quad c_{\ell}^+=|c_\ell^+|e^{i\mu_\ell},\quad \mu_j\in\mathbb{R},
\end{eqnarray}
that guarantees that 
 $u_{\epsilon}=\sum\limits_{j=0}^NF_{j,\epsilon}e^{ik_j^2z}+\sum\limits_{\ell=0}^MG_{\ell,\epsilon} e^{-ir_\ell^2z}$ will satisfy
\begin{equation}
\mathcal{P}_2\mathcal{T}u_{\epsilon}=\epsilon~u_{\epsilon}.
\end{equation}

In what follows, we will discuss the systems related to the following fixed form of the function $u$,
\begin{equation}\label{ue}
 u_\epsilon=F_{1,\epsilon}e^{ik_1^2z}+F_{2,\epsilon}e^{ik_2^2z}+G_{1,\epsilon}e^{-ir_1^2z}.
\end{equation}
As we argued below (\ref{G}), the character of the guided modes given by $v$ depends on the presence of $G_{1,\epsilon}$ in $u$. When it is absent in (\ref{ue}), i.e. $G_{1,\epsilon}\equiv 0$, then we can construct guided modes that are exponentially vanishing for large $|x|$. Otherwise, $\mathcal{L}v$ is asymptotically non-vanishing, bounded and oscillating. 

\subsubsection*{ $G_1= 0$: Strongly confining wave guides}
\begin{figure}[t!] 
	\centering
 \begin{subfigure}[b]{0.3\textwidth}
        \includegraphics[width=\textwidth]{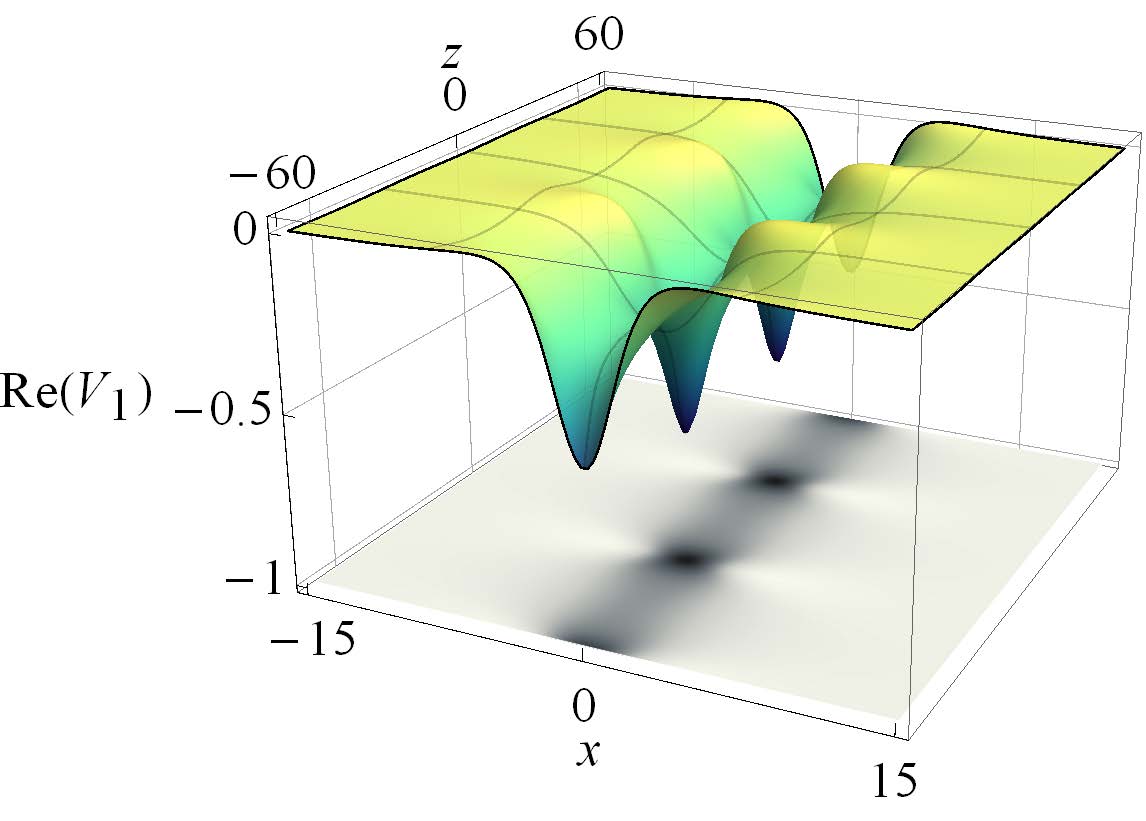}
        \caption{}
    \end{subfigure}
     \begin{subfigure}[b]{0.3\textwidth}
        \includegraphics[width=\textwidth]{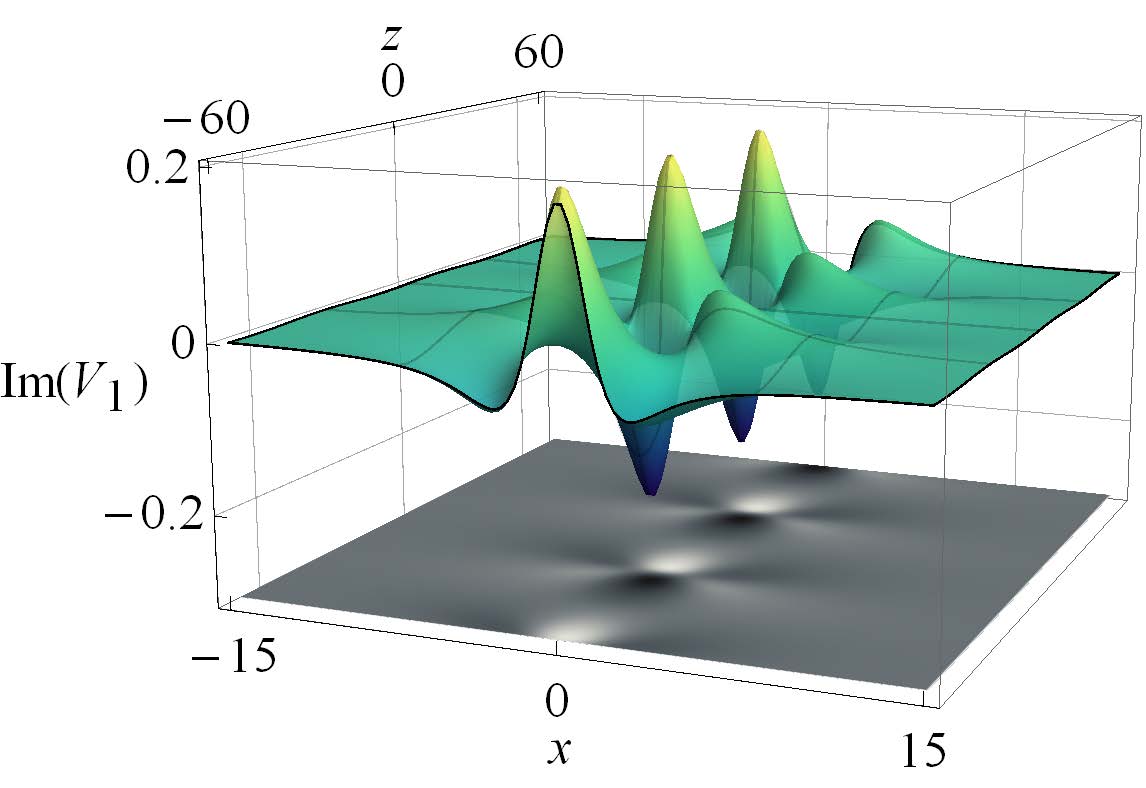}
        \caption{}
    \end{subfigure}
     \begin{subfigure}[b]{0.3\textwidth}
        \includegraphics[width=\textwidth]{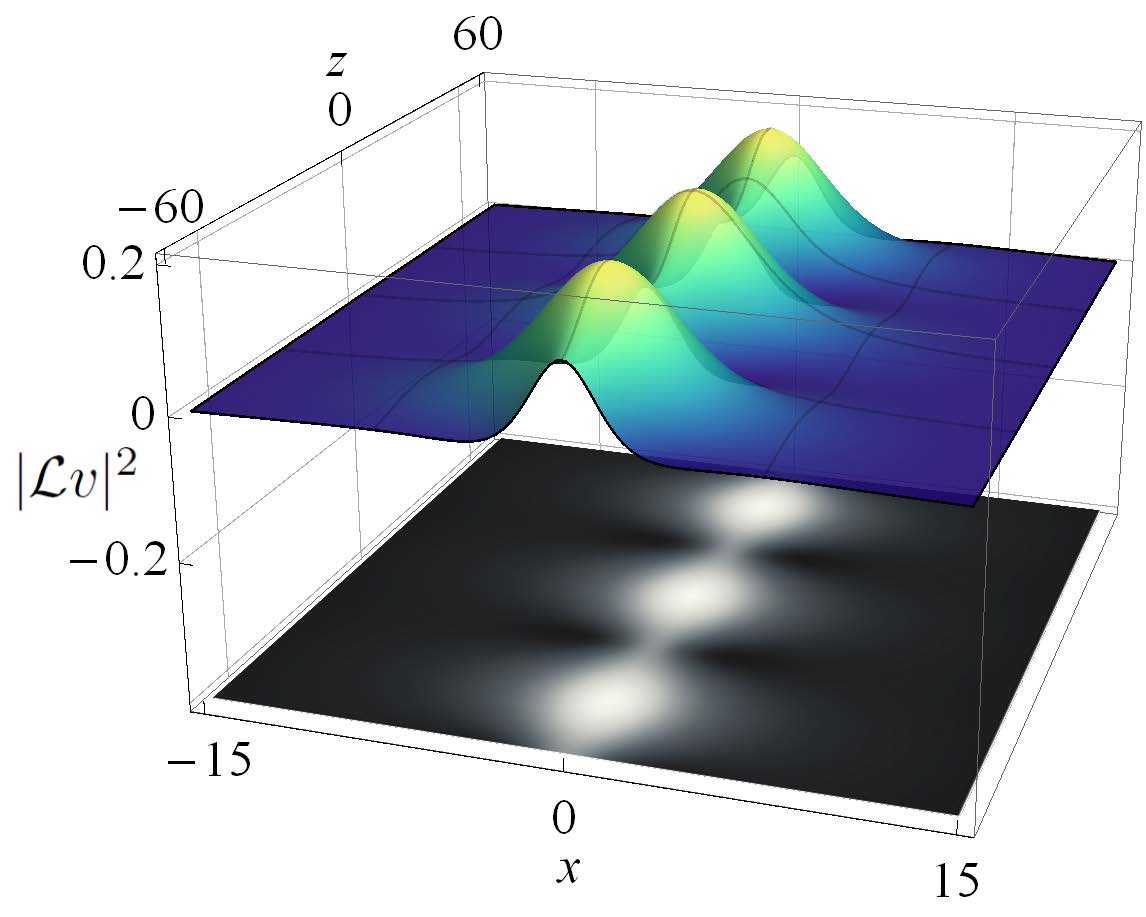}
        \caption{}
    \end{subfigure}\\
     \begin{subfigure}[b]{0.3\textwidth}
        \includegraphics[width=\textwidth]{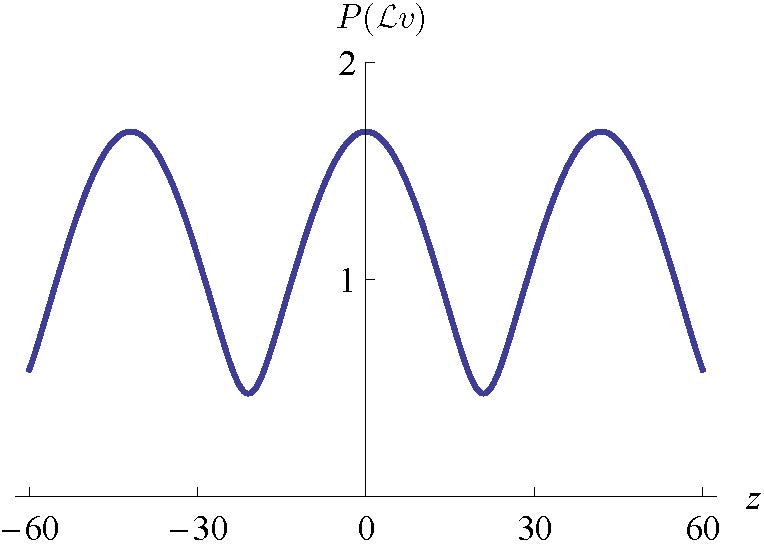}
        \caption{}
    \end{subfigure}
     \begin{subfigure}[b]{0.3\textwidth}
        \includegraphics[width=\textwidth]{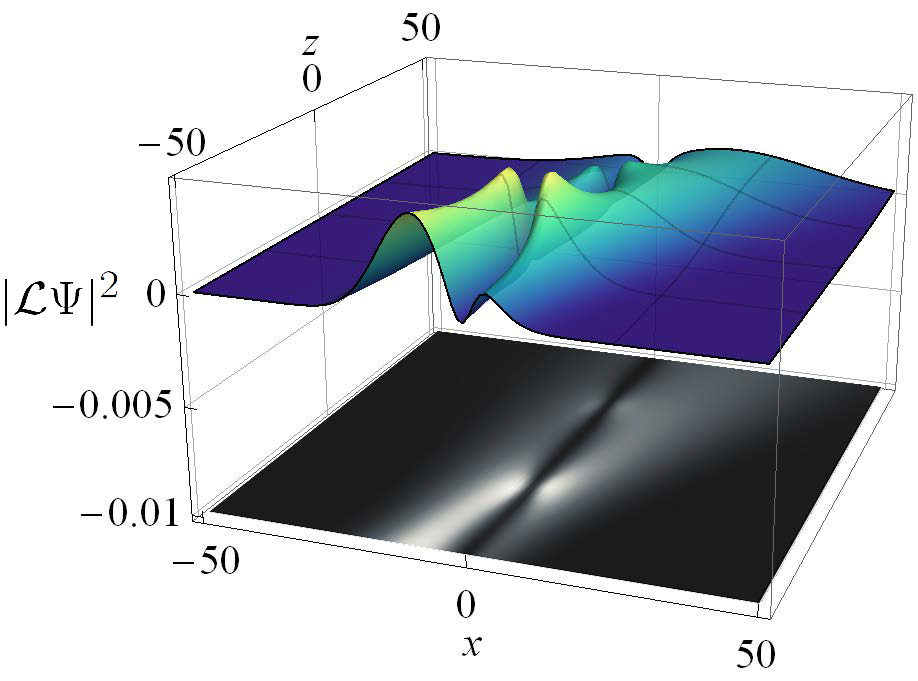}
        \caption{}
    \end{subfigure}
	\caption{A strongly confining waveguide. Plots of the real (a) and imaginary (b) parts of $V_1$, see \eqref{V1h1}, for the parameters  $k_1=0.4,~k_2=0.1,~\alpha=0.5$ are presented. Furthermore, the corresponding power density of the guided mode $ \mathcal{L}v$, see \eqref{guided V1h1}, is shown (c). The power $P(\mathcal{L}v)= \int_{-\infty}^{\infty}|\mathcal{L}v|^2 dx$  can be seen in (d), the power of the guided mode is oscillating. The power density $|\mathcal{L}\Psi_{x_0,z_0,v_0,\sigma}|^2$, for the parameters $x_0=-7$, $z_0=-50$, $v_0=-0.2$ and $\sigma=40$ is shown in (e).   } \label{FigV1h1}
\end{figure}

Let us discuss two choices of the functions $u$ and $v$.
First, we fix
\begin{eqnarray}
 &&u=\cosh k_1 xe^{i k_1^2z}+\alpha\cosh k_2 x e^{i k_2^2z},
\\
&&v=\sinh k_1 xe^{i k_1^2z}+\alpha\sinh k_2 x e^{i k_2^2z},\quad \mbox{where}\quad |\alpha|<1,\quad \alpha\in \mathbb{R}.
\end{eqnarray}
Comparing with (\ref{aa}), we can see that $c^+=1$ while $c^-=-1$. 
We can show that the function $u$ has no zeros. Indeed, the equation $u=0$ can be written as
$ \frac{\cosh k_1x}{\cosh k_2x}=-\alpha e^{i(k_2^2-k_1^2)z}.$
Its left side is greater than one (notice that $\cosh x$ is a monotonic function for $x>0$ or $x\leq0$), while the absolute value of the right-hand side can be only smaller than one. 
The potential $V_1$ and the guided mode $\mathcal{L}v$ acquire the following forms
\begin{equation}\label{V1h1}
 V_1=-2\frac{k_1^2+\alpha^2 k_2^2 e^{2i(k_2^2-k_1^2)z}}{\cosh^2 k_1x\left(1+ e^{i(k_2^2-k_1^2)z}\alpha\frac{\cosh k_2x}{\cosh k_1x}\right)^2}-2\frac{e^{i(k_2^2-k_1^2)z}\alpha\cosh k_2x\left((k_1^2+k_2^2)-2k_1k_2\tanh k_1x\tanh k_2x\right)}{\cosh k_1x\left(1+ e^{i(k_2^2-k_1^2)z}\alpha\frac{\cosh k_2x}{\cosh k_1x}\right)^2}
\end{equation}
and
\begin{equation} \label{guided V1h1}
 \mathcal{L}v=\frac{e^{2ik_1^2z}k_1+e^{2ik_2^2 z}k_2\alpha^2+e^{i(k_1^2+k_2^2)z}(k_1+k_2)\alpha\cosh (k_1-k_2)x}{e^{ik_1^2 z}\cosh k_1x\left(1+e^{i(k_2^2-k_1^2)z}\alpha\frac{\cosh k_2x}{\cosh k_1x}\right)}.
\end{equation}
The intertwining operator has the following action on the wave packets (\ref{wavepacket})
\begin{equation}
\mathcal{L}\Psi_{x_0,z_0,v_0,\sigma}=G(x,z)\Psi_{x_0,z_0,v_0,\sigma},
\end{equation}
where
\begin{equation}
G(x,z)=i\frac{x-x_0+iv_0\sigma}{2(z-z_0-i\sigma)}-\frac{e^{ik_2^2z}k_2\alpha\sinh k_2x+e^{ik_1^2z}k_1\sinh k_1x}{ie^{ik_2^2z}\alpha\cosh k_2x+e^{ik_1^2z}\cosh k_1x}.
\end{equation}
In Fig. \ref{FigV1h1} we present  the real (a) and imaginary (b)  parts of the potential as well as the power density $|\mathcal{L}v|^2$ (c) of the guided mode, for the parameters $k_1=0.4,~k_2=0.1,~\alpha=0.5$.  The power $P(\mathcal{L}v)= \int_{-\infty}^{\infty}|\mathcal{L}v|^2 dx$  can be seen in (d), the power of the guided mode is oscillating. The power density $|\mathcal{L}\Psi_{x_0,z_0,v_0,\sigma}|^2$, for the parameters $x_0=-7$, $z_0=-50$, $v_0=-0.2$ and $\sigma=40$, is shown in (e).
%%%%%%%%%%%%%%%%%%%%%%
%%%%%%%%%%%%%%%%%%%%%%
%%%%%%%%%%%%%%%%%%%%%%
%%%%%%%%%%%%%%%%%%%%%%

\begin{figure}[t!] 
	\centering
	 \begin{subfigure}[b]{0.3\textwidth}
        \includegraphics[width=\textwidth]{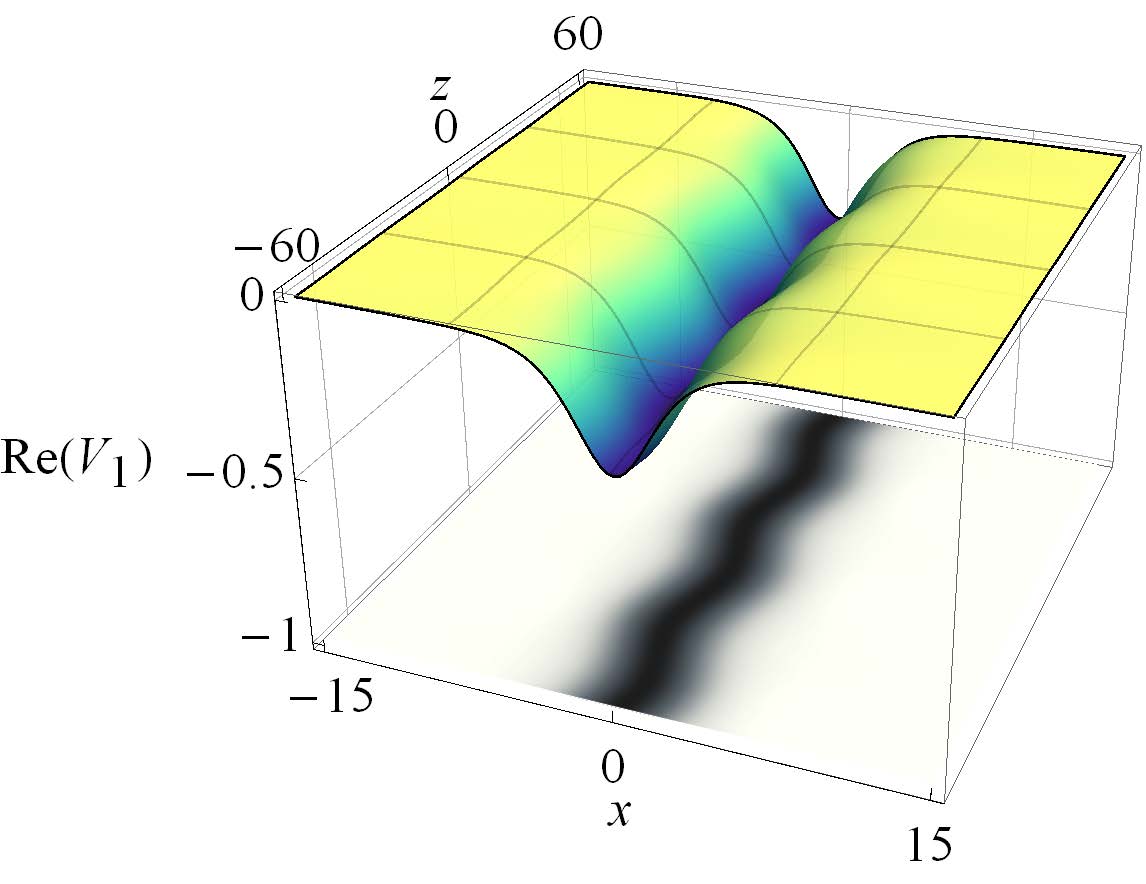}
        \caption{}
    \end{subfigure}
     \begin{subfigure}[b]{0.3\textwidth}
        \includegraphics[width=\textwidth]{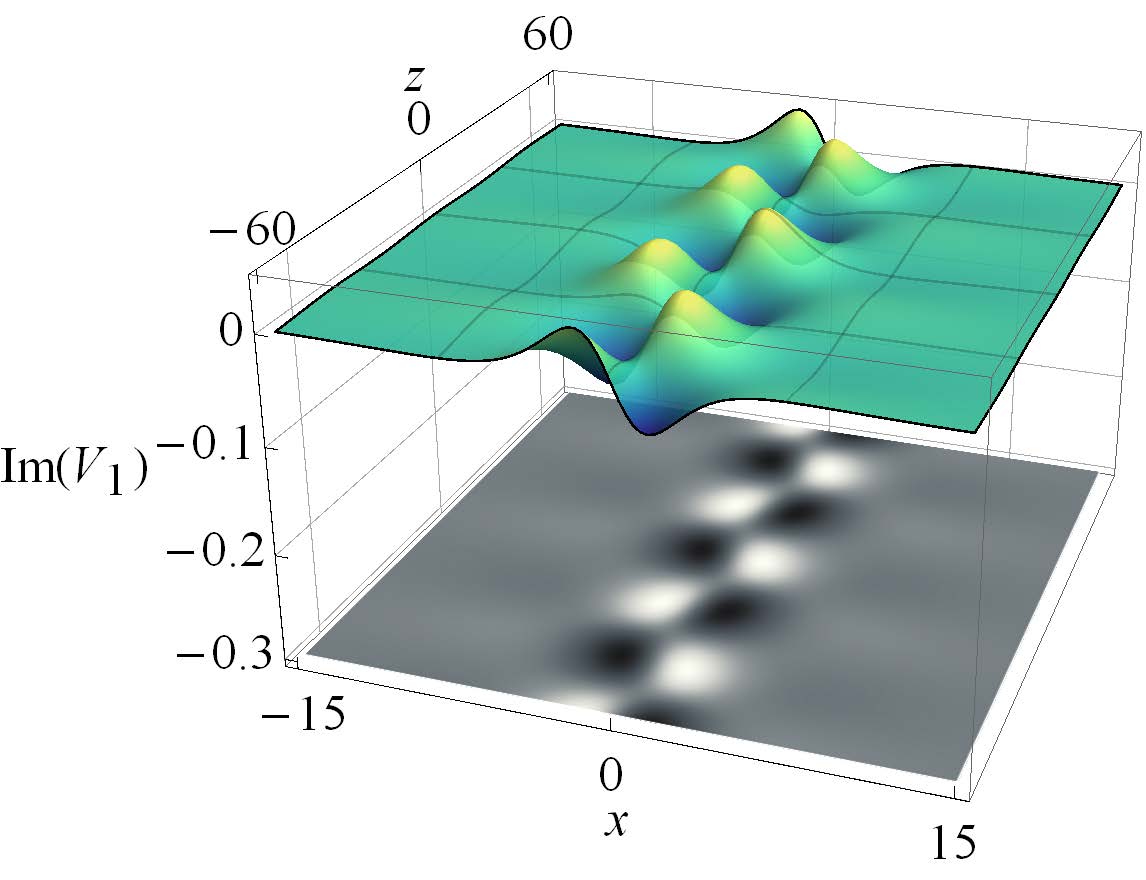}
        \caption{}
    \end{subfigure}
     \begin{subfigure}[b]{0.3\textwidth}
        \includegraphics[width=\textwidth]{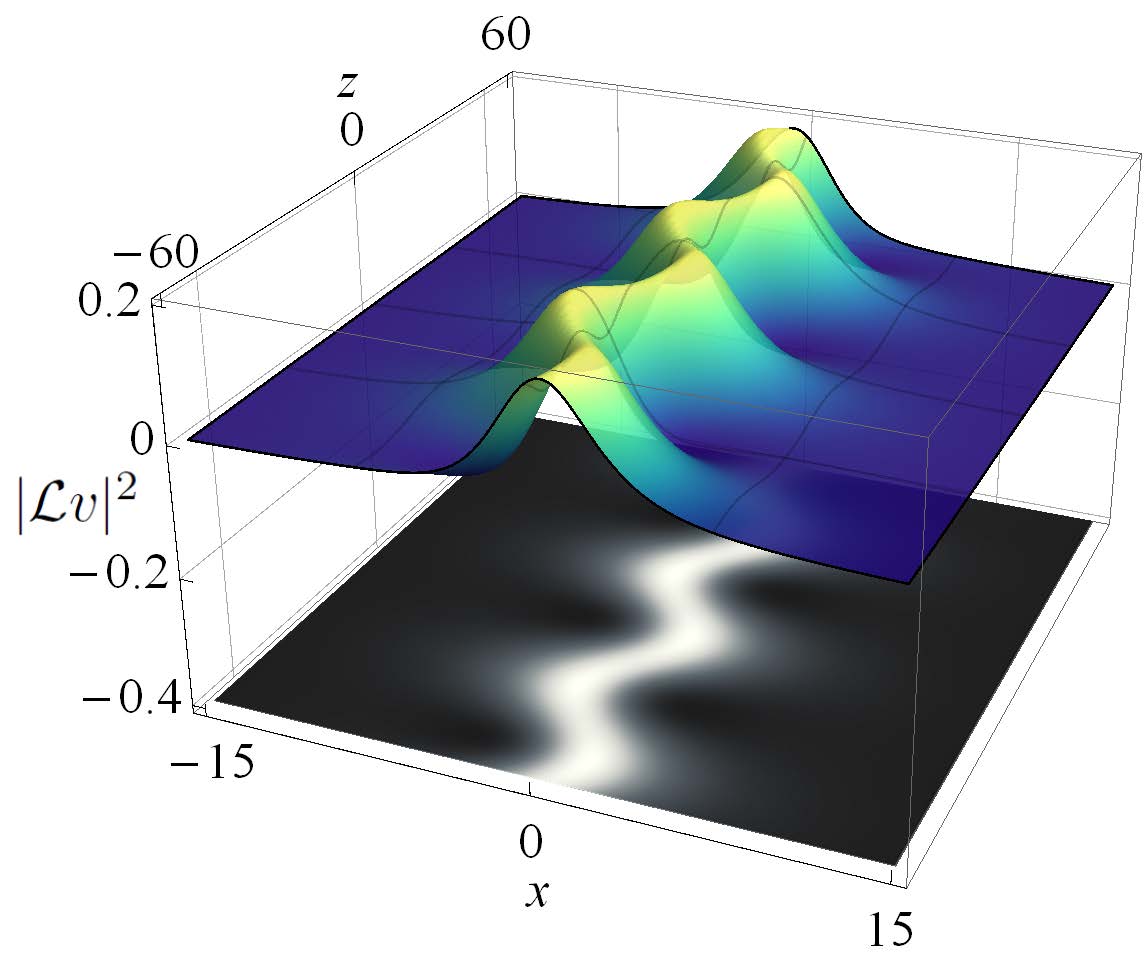}
        \caption{}
    \end{subfigure}\\
     \begin{subfigure}[b]{0.3\textwidth}
        \includegraphics[width=\textwidth]{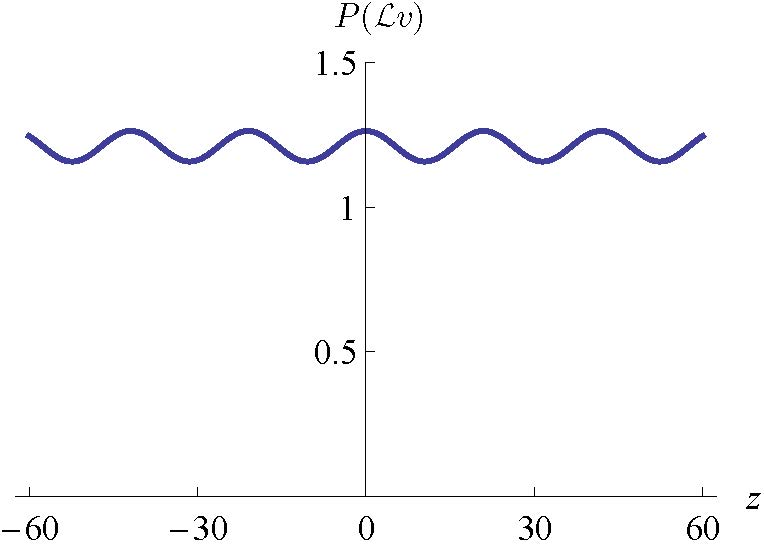}
        \caption{}
    \end{subfigure}
     \begin{subfigure}[b]{0.3\textwidth}
        \includegraphics[width=\textwidth]{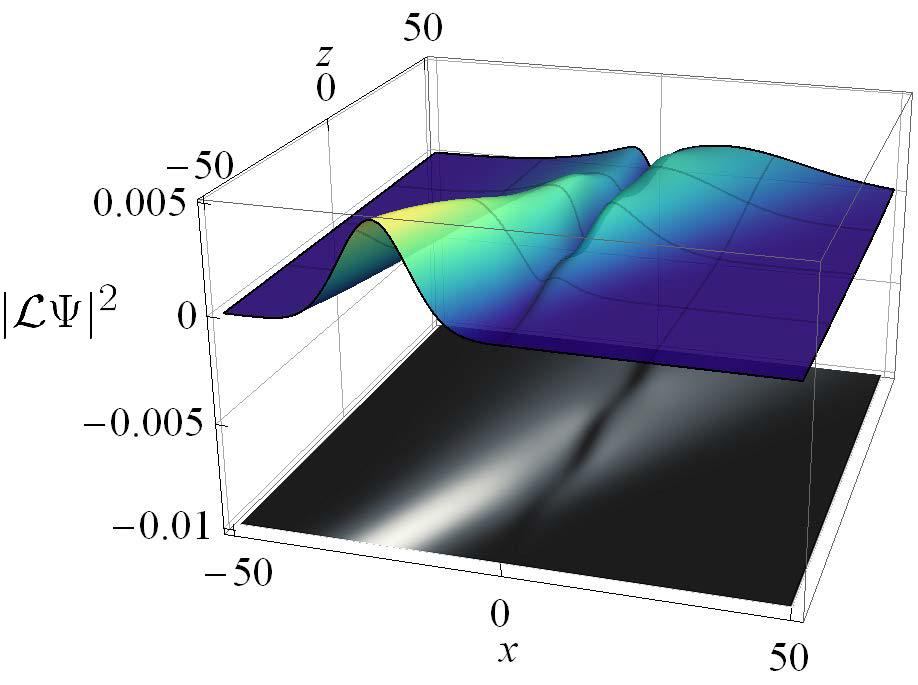}
        \caption{}
    \end{subfigure}
  
	\caption{A strongly confining waveguide. Plots of the real (a) and imaginary  (b) parts of $V_1$, see \eqref{V1h2}, for the parameters  $k_1=0.4,~k_2=0.1,~\alpha=0.5$ are presented. Furthermore, the corresponding power density of the guided mode $ \mathcal{L}v$, see \eqref{guided V1h2}, is shown (c). The power $P(\mathcal{L}v)= \int_{-\infty}^{\infty}|\mathcal{L}v|^2 dx$  can be seen in (d), the power of the guided mode is oscillating. The power density $|\mathcal{L}\Psi_{x_0,z_0,v_0,\sigma}|^2$, for the parameters $x_0=-20$, $z_0=-50$, $v_0=-0.3$ and $\sigma=40$ is shown in (e).    } \label{FigV1h2}
\end{figure}

Now, let us consider a different choice of the function $u$ and of the preimage $v$ of the guided mode, 
\begin{eqnarray}\label{uhii}
&& u=\cosh k_1 xe^{i k_1^2z}+i\alpha\sinh k_2 x e^{i k_2^2z},\\
&&v=\sinh k_1 xe^{i k_1^2z}+i\alpha\cosh k_2 x e^{i k_2^2z},\quad \mbox{where}\quad  |\alpha|<1.
\end{eqnarray}
We can prove that $u$ has no zeros. Indeed, the equation $u=0$ leads to $\frac{\sinh k_2x}{\cosh k_1x}=i\alpha^{-1}e^{i(k_1^2-k_2^2)z}$. We can show that the absolute value of the left-hand side is smaller then one: for $x>0$, we have $0<\frac{\sinh k_2x}{\cosh k_1x}<\frac{\sinh k_1x}{\cosh k_1x}=\tanh k_1x<1$. For $x\leq 0$, we have $0>\frac{\sinh k_2x}{\cosh k1 x}=\tanh k_1x+\frac{\sinh k_2x-\sinh k_1x}{\cosh k_1x}>\tanh k_1x>-1$. 
Hence, the potential $V_1$ is regular and it can be written as
\begin{equation}\label{V1h2}
 V_1=-2\frac{k_1^2+\alpha^2 k_2^2 e^{2i(k_2^2-k_1^2)z}}{\cosh^2 k_1x\left(1+i e^{i(k_2^2-k_1^2)z}\alpha\frac{\sinh k_2x}{\cosh k_1x}\right)^2}-2i\frac{e^{i(k_2^2-k_1^2)z}\alpha\cosh k_2x\left((k_1^2+k_2^2)\tanh k_2x-2k_1k_2\tanh k_1x\right)}{\cosh k_1x\left(1+i e^{i(k_2^2-k_1^2)z}\alpha\frac{\sinh k_2x}{\cosh k_1x}\right)^2}
\end{equation}
whereas the guided mode $\mathcal{L}v$ acquires the following form
\begin{equation}
 \mathcal{L}v=\frac{e^{2ik_1^2z}k_1+e^{2ik_2^2 z}\alpha^2k_2-ie^{i(k_1^2+k_2^2)z}(k_1+k_2)\alpha\sinh (k_1-k_2)x}{e^{ik_1^2 z}\cosh k_1x\left(1+ie^{i(k_2^2-k_1^2)z}\alpha\frac{\sinh k_2x}{\cosh k_1x}\right)}. \label{guided V1h2}
\end{equation}
Now, the wave packets get transformed in the following manner 
	\begin{equation}
	\mathcal{L}\Psi_{x_0,z_0,v_0,\sigma}=G(x,z)\Psi_{x_0,z_0,v_0,\sigma},
	\end{equation}
	where
	\begin{equation}
	G(x,z)=i\frac{x-x_0+iv_0\sigma}{2(z-z_0-i\sigma)}-\frac{ie^{ik_2^2z}k_2\alpha\cosh k_2x+e^{ik_1^2z}k_1\alpha\sinh k_1x}{ie^{ik_2^2z}\alpha\sinh k_2x+e^{ik_1^2z}\cosh k_1x}.
		\end{equation}
The potential $V_1$ as well as the guided mode together with $|\mathcal{L}\Psi_{x_0,z_0,v_0,\sigma}|^2$ are illustrated in Fig. \ref{FigV1h2}.  The power of the guided mode $P(\mathcal{L}v)= \int_{-\infty}^{\infty}|\mathcal{L}v|^2 dx$ (d) is also oscillating.
Let us remark that both the potentials (\ref{V1h1}) and (\ref{V1h2}) reduce to the $z$-independent P\"oschl-Teller potential for $\alpha=0$. 
%%%%%%%%%%%%%%%%%%%%%%
%%%%%%%%%%%%%%%%%%%%%%
%%%%%%%%%%%%%%%%%%%%%%
%%%%%%%%%%%%%%%%%%%%%%

\subsubsection*{$G_1\neq 0$: weakly confining wave guides}
\begin{figure}[t!] 
	\centering
	   \begin{subfigure}[b]{0.3\textwidth}
        \includegraphics[width=\textwidth]{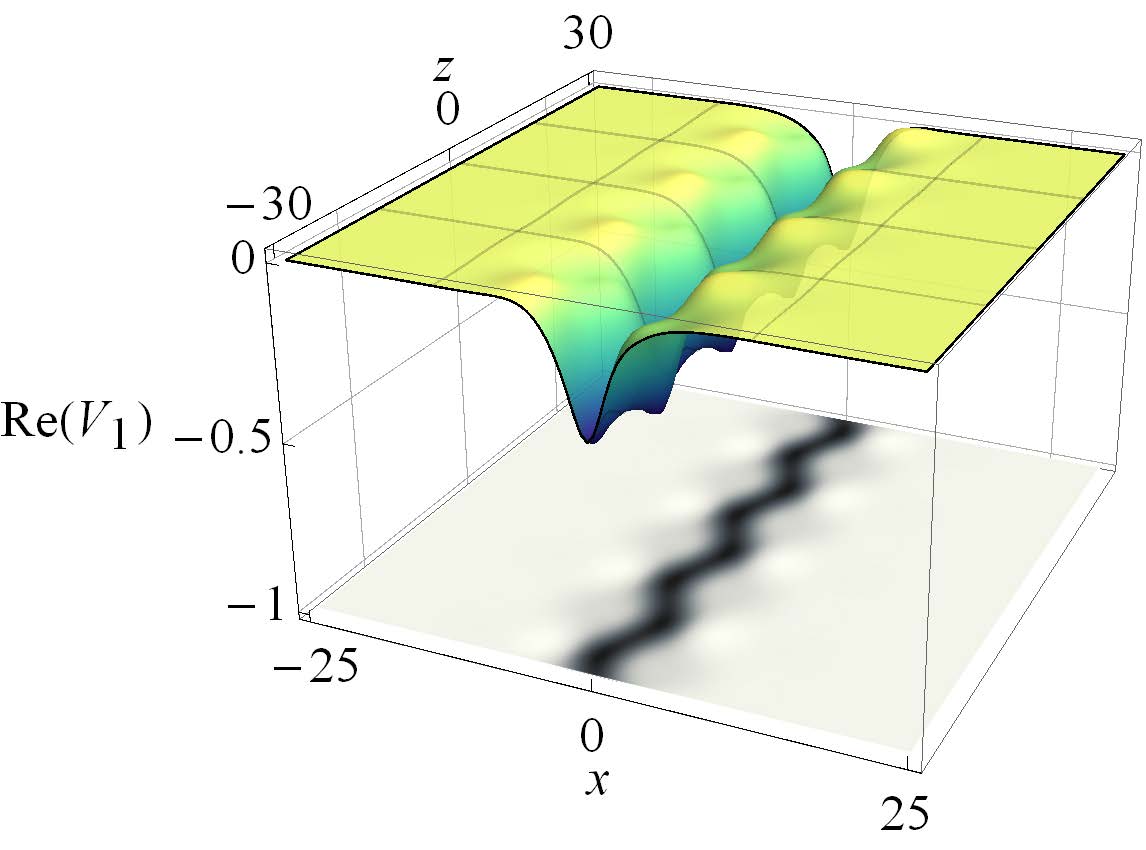}
        \caption{}
    \end{subfigure}
       \begin{subfigure}[b]{0.3\textwidth}
        \includegraphics[width=\textwidth]{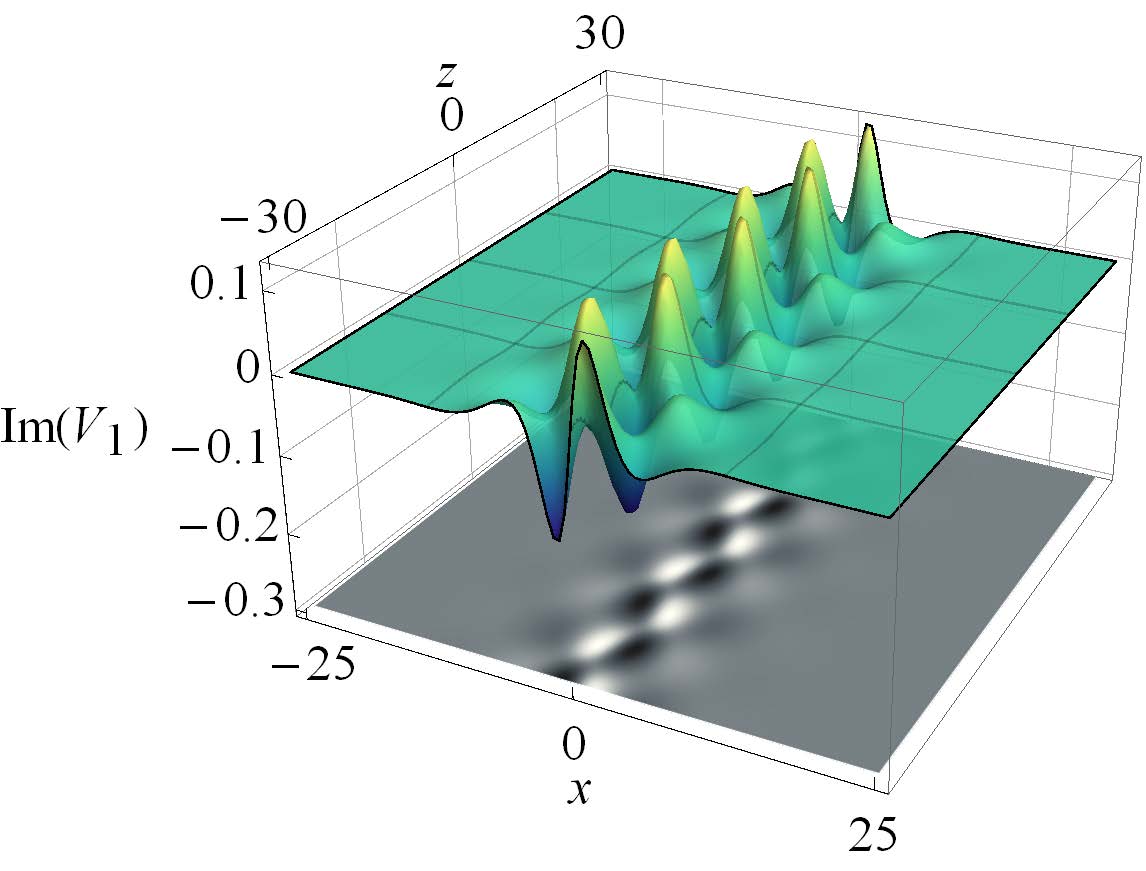}
        \caption{}
    \end{subfigure}
       \begin{subfigure}[b]{0.3\textwidth}
        \includegraphics[width=\textwidth]{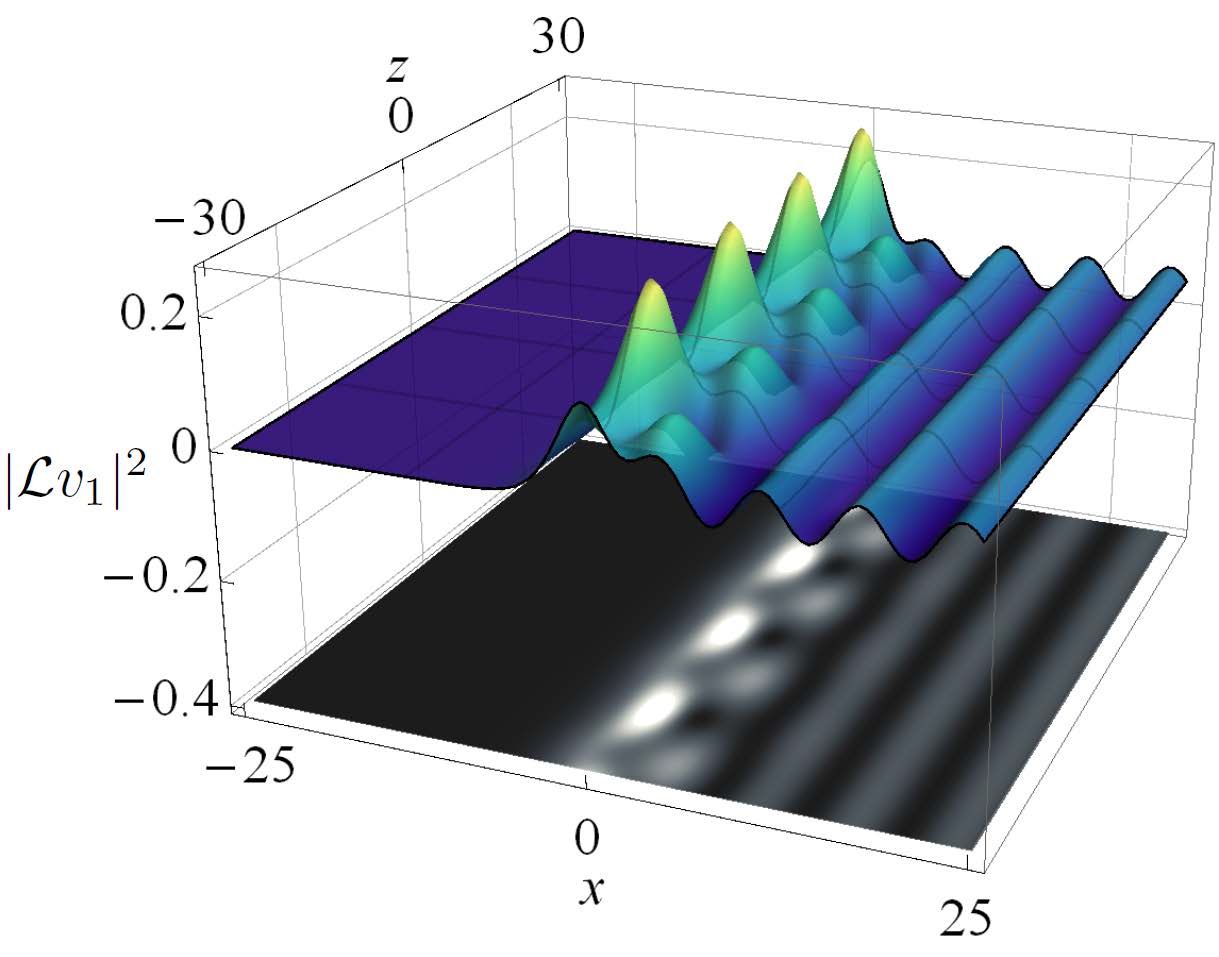}
        \caption{}
    \end{subfigure}\\
       \begin{subfigure}[b]{0.3\textwidth}
        \includegraphics[width=\textwidth]{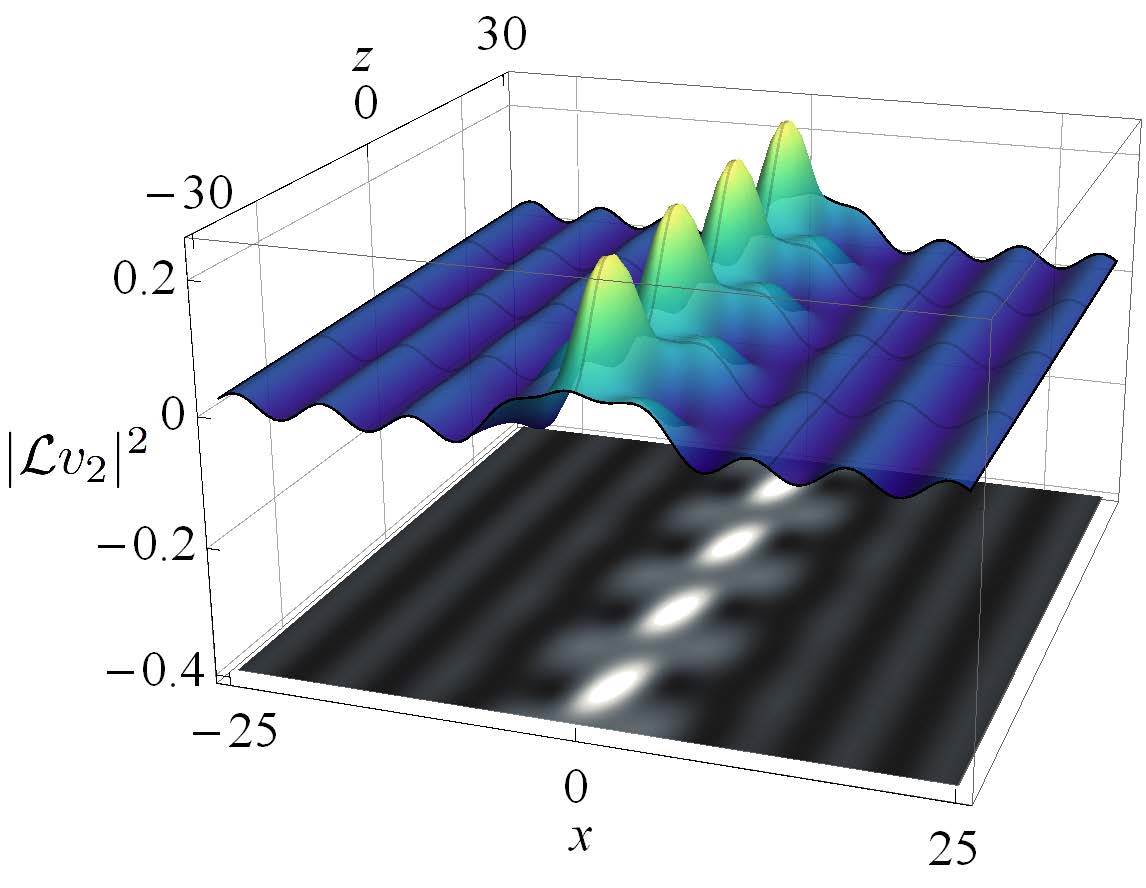}
        \caption{}
    \end{subfigure}
       \begin{subfigure}[b]{0.3\textwidth}
        \includegraphics[width=\textwidth]{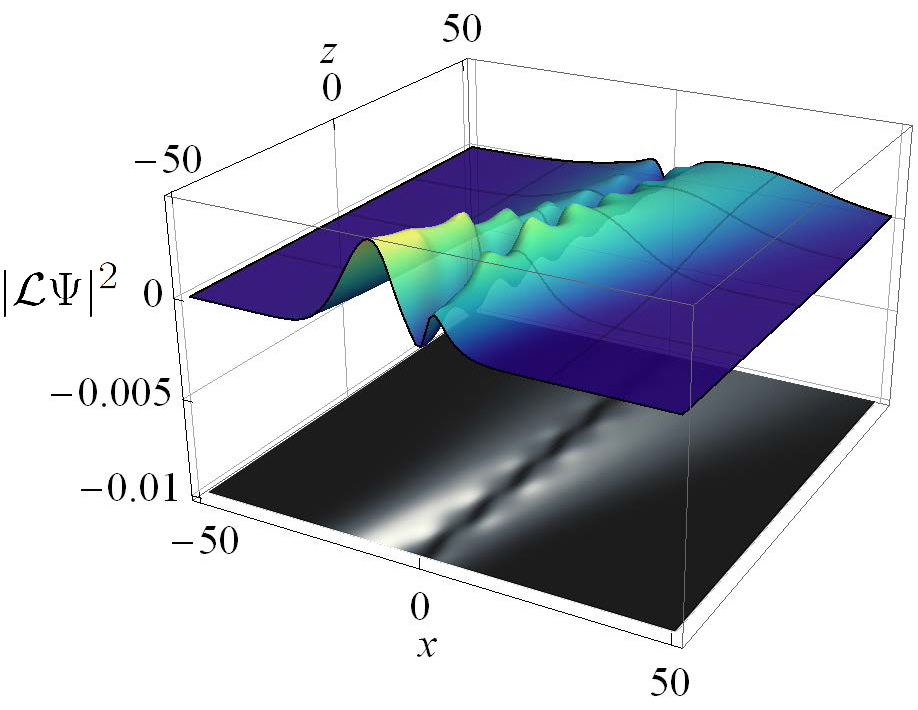}
        \caption{}
    \end{subfigure}
%    
%    
%	\includegraphics[width=.3\textwidth]{Periodic10b.jpg} 
%	\includegraphics[width=.3\textwidth]{Periodic11b.jpg} \\
%	\includegraphics[width=.3\textwidth]{Periodic12b.jpg}
%	\includegraphics[width=.3\textwidth]{Periodic13b.jpg}
%	%\includegraphics[width=.3\textwidth]{beamWG3.jpg} 
%	\includegraphics[width=.3\textwidth]{Fig81.jpg} 
	\caption{Weakly confining wave guides. Plots of the real (a) and imaginary (b) parts of $V_1$ when $k_1=0.4,~r_\ell=0.5,~\alpha=0.2$, see \eqref{V1t1}. The intensity densities of $\mathcal{L}v_1$ and $\mathcal{L}v_2$ are shown as well ((c) and (d), respectively).  The power density $|\mathcal{L}\Psi_{x_0,z_0,v_0,\sigma}|^2$, for the parameters $x_0=-7$, $z_0=-50$, $v_0=-0.2$ and $\sigma=40$ is shown in (e).   } \label{FigV1h3}
\end{figure}

We fix 
\begin{equation}\label{uti}
 u=\cosh k_1x e^{ik_1^2z}+i\alpha\sin r_1 x e^{-ir_1^2z},\quad \alpha\in(-1,1).
\end{equation}
The function has no zeros. Writing the equation $\cosh k_1x=-i\alpha\sin r_1 x e^{-i(r_\ell^2+k_1^2)z}$, we can see that the left hand side is always greater or equal to one, whereas the absolute value of the right-hand side is smaller or equal to $|\alpha|$. As the function (\ref{uti}) can be obtained from (\ref{uhii}) by the substitution $k_2\rightarrow i r_1$, the potential $V_1$ is related to (\ref{V1h2}) in the same manner, 
\begin{equation}\label{V1t1}
 V_1=-2\frac{k_1^2+\alpha^2 r_1^2 e^{-2i(r_1^2+k_1^2)z}}{\cosh^2 k_1x\left(1+i\alpha e^{-i(r_1^2+k_1^2)z}\frac{\sin r_1x}{\cosh k_1x}\right)^2}-2\frac{ie^{-i(r_1^2+k_1^2)z}\alpha\cosh k_1 x\left((k_1^2-r_1^2)\sin r_1x
-2k_1r_1\cos r_1x\tanh k_1x\right)}{\cosh^2 k_1x\left(1+ i e^{-i(r_1^2+k_1^2)z}\alpha\frac{\sin r_1x}{\cosh k_1x}\right)^2}.
\end{equation}
The action of the intertwining operator on the wave packets is $\mathcal{L}\Psi_{x_0,z_0,v_0,\sigma}=G(x,z)\Psi_{x_0,z_0,v_0,\sigma}$ where   
\begin{equation}
G(x,z)=\frac{i(x-x_0+iv_0\sigma)}{2(z-z_0-i\sigma)}-\frac{ir_1\alpha\cos r_1x+e^{i(k_1^2+r_1^2)z}k_1\sinh k_1x}{e^{i(k_1^2+k_2^2)z}\cosh k_1z+i\alpha\sin r_1x}.
\end{equation}
The transformed wave packet is illustrated in Fig.  \ref{FigV1h3}.

Let us consider the functions $v_1$ and $v_2$,
\begin{equation}
 v_1=\sinh k_1x e^{ik_1^2z}-i\alpha\sin r_1 x e^{-ir_1^2z},\quad v_2=\sinh k_1x e^{ik_1^2z}-\alpha\sin r_1 x e^{-ir_1^2z}
\end{equation}
The state $v_1$ fulfills the condition (\ref{G}) so that $\mathcal{L}v_1$ vanishes exponentially on one side of the potential barrier $V_1$. Yet it breaks manifestly $\mathcal{P}_2\mathcal{T}$ symmetry (it can be written as a linear combination of two $\mathcal{P}_2\mathcal{T}$-symmetric solutions). The solution $v_2$ has definite $\mathcal{P}_2\mathcal{T}$ parity but does not comply with (\ref{G}). Therefore, $\mathcal{L}v_2$ has non-vanishing oscillations on both sides of the barrier, see Fig. \ref{FigV1h3} for illustration.
%%%%%%%%%%%%%%%%%%%%%%
%%%%%%%%%%%%%%%%%%%%%%
%%%%%%%%%%%%%%%%%%%%%%
%%%%%%%%%%%%%%%%%%%%%%

\subsection{Coupled wave guides\label{coupled}}

To obtain coupled wave guides we can use a higher order Darboux transformation. In this example, we will use the second-order transformation defined in \eqref{s2} and \eqref{ir12} in order to produce a system with two coupled wave guides. We start out by fixing the transformation functions $u_1$ and $u_2$: 
\begin{equation}\label{N2sols}
u_1=\cosh k_1x e^{i k_1^2 z}+i \alpha \sinh k_3 x e^{i k_3^2 z},\quad u_2=\sinh k_2x e^{i k_2^2 z}
\end{equation}
where we suppose that the constants $k_1$, $k_2$, $k_3$ and $\alpha$ are all real. Moreover, we fix $L_1=L_2=1$  that respects (\ref{wavepacketpreservation}).
The explicit form of the new potential term $V_2=-2\partial_x^2\ln W(u_1,u_2)$ is not quite compact, so that we refer to (\ref{s2}) from which it can be obtained directly when substituting (\ref{N2sols}). For $\alpha=0$,  both $u_1$ and $u_2$ correspond to the stationary states of the free particle Hamiltonian and the Darboux transformation $\mathcal{L}_{12}$ renders $z$-independent potential 
\begin{equation}\label{V2indep}
V_2|_{\alpha=0}=\frac{(k_1^2-k_2^2)(k_2^2-k_1^2+k_1^2\cosh 2k_2x+k_2^2\cosh 2k_1x)}{(k_2\cosh k_1x\cosh k_2x-k_1\sinh k_1x\sinh k_2x)^2}.
\end{equation}
It corresponds to two parallel wave guides that were discussed in \cite{vega}.

\begin{figure}[t!] 
	\centering
\begin{subfigure}[b]{0.3\textwidth}
        \includegraphics[width=\textwidth]{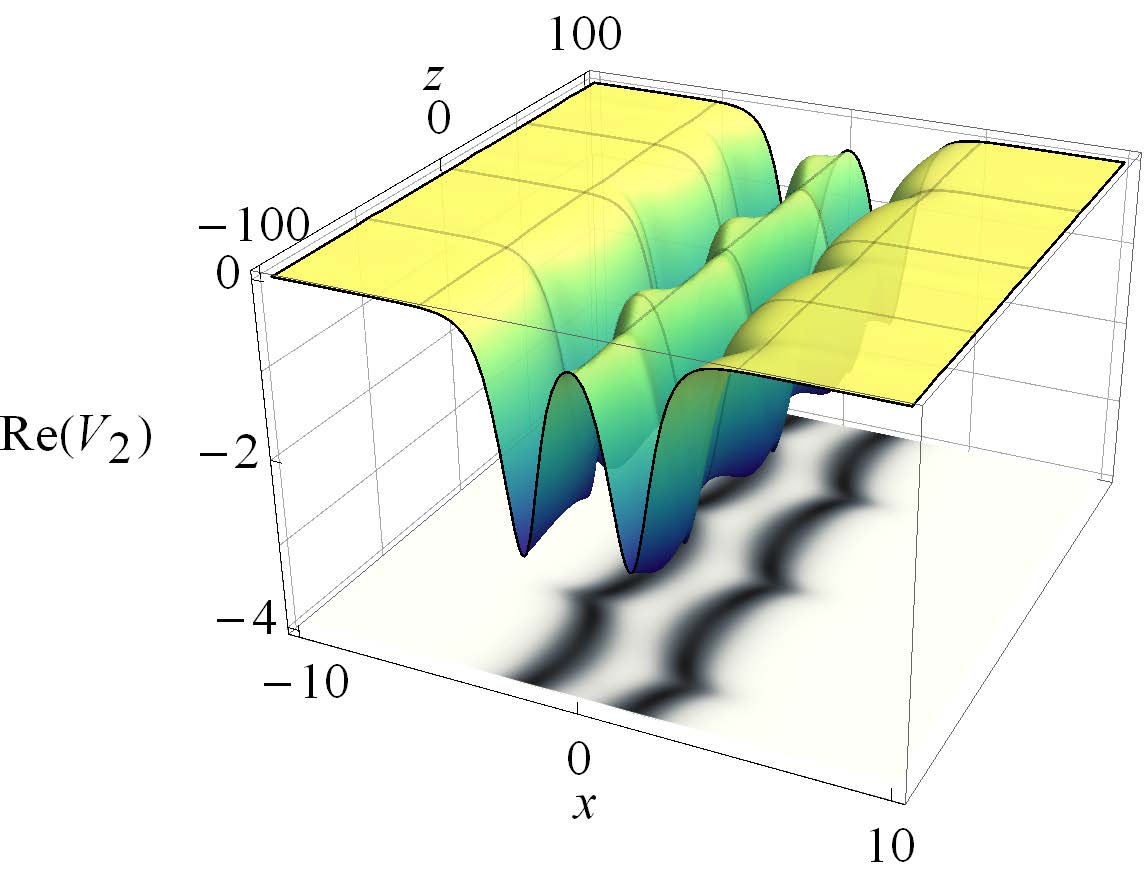}
        \caption{}
    \end{subfigure}
    \begin{subfigure}[b]{0.3\textwidth}
        \includegraphics[width=\textwidth]{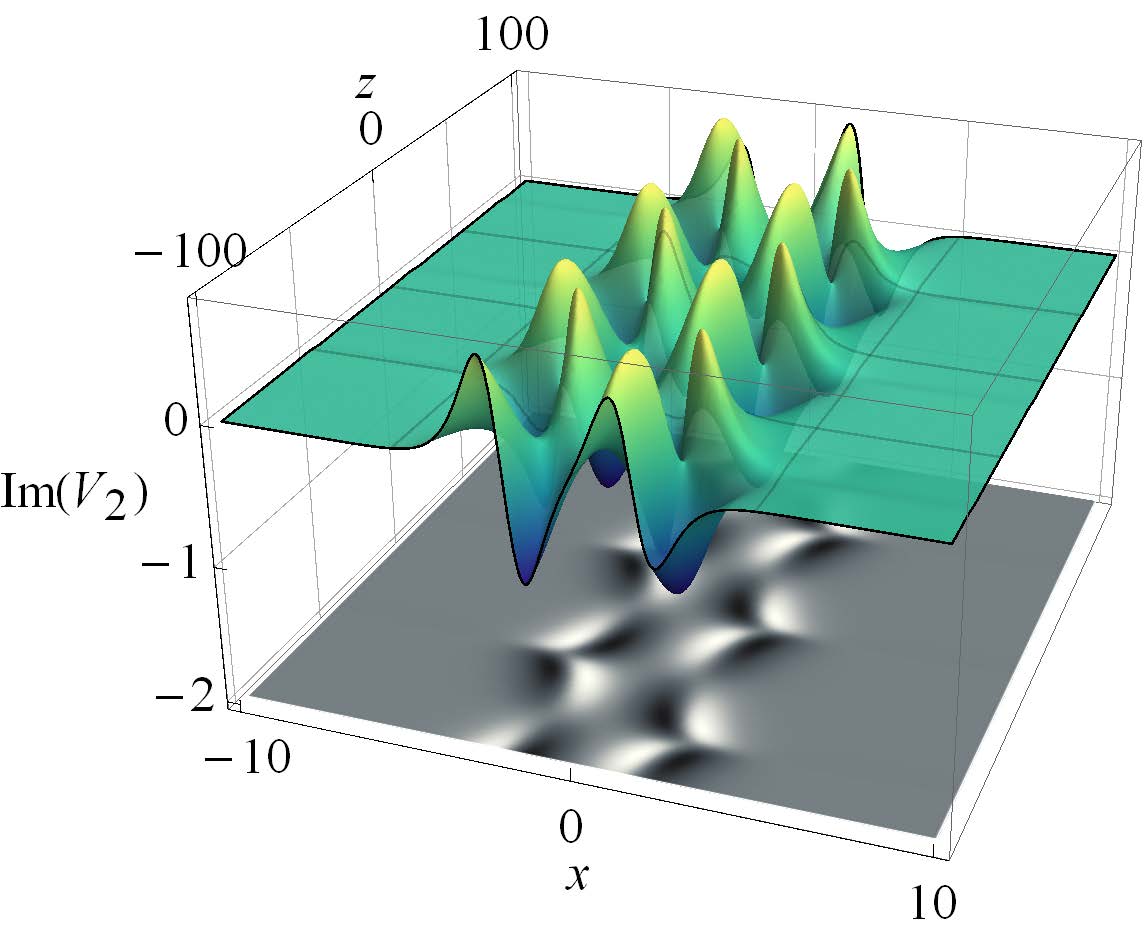}
        \caption{}
    \end{subfigure}\\
    \begin{subfigure}[b]{0.3\textwidth}
        \includegraphics[width=\textwidth]{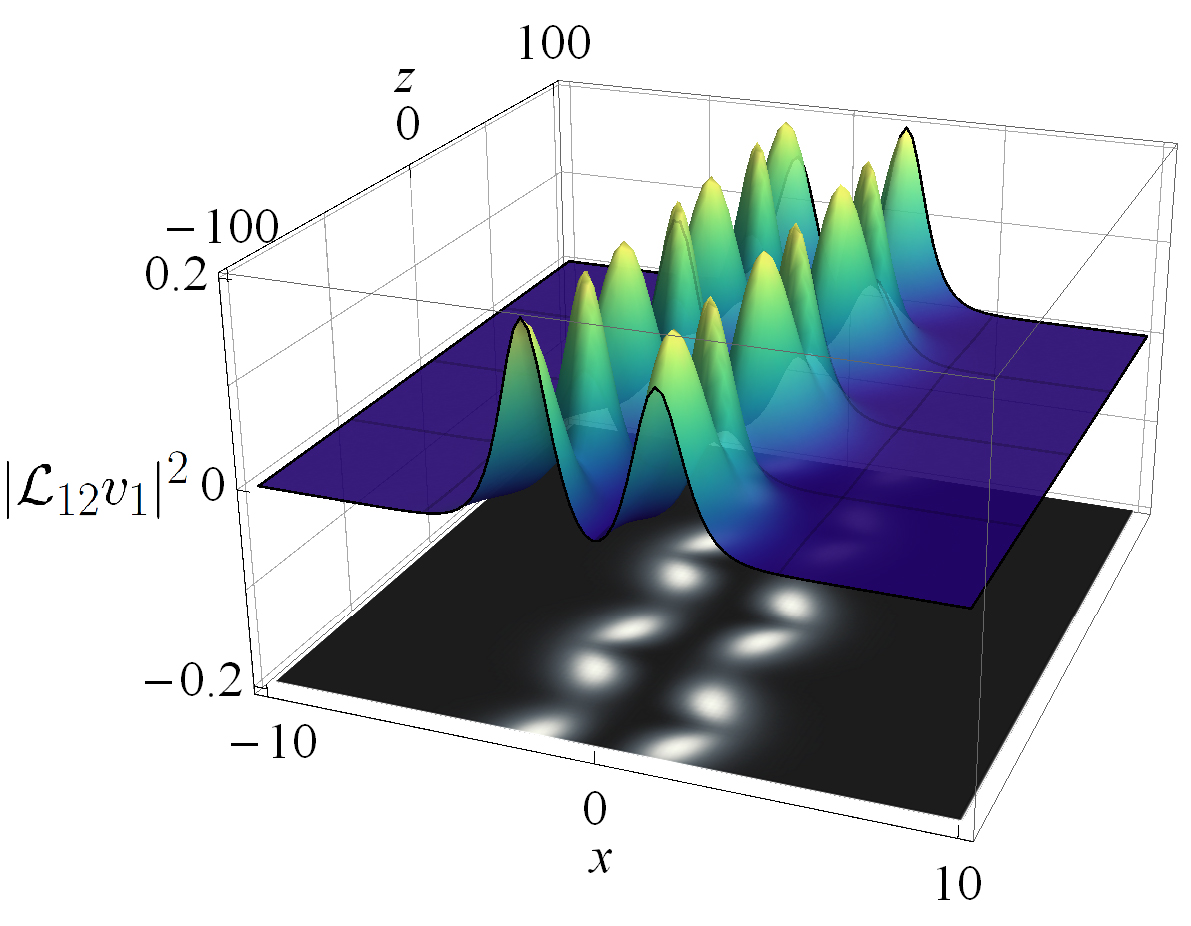}
        \caption{}
    \end{subfigure}
    \begin{subfigure}[b]{0.3\textwidth}
        \includegraphics[width=\textwidth]{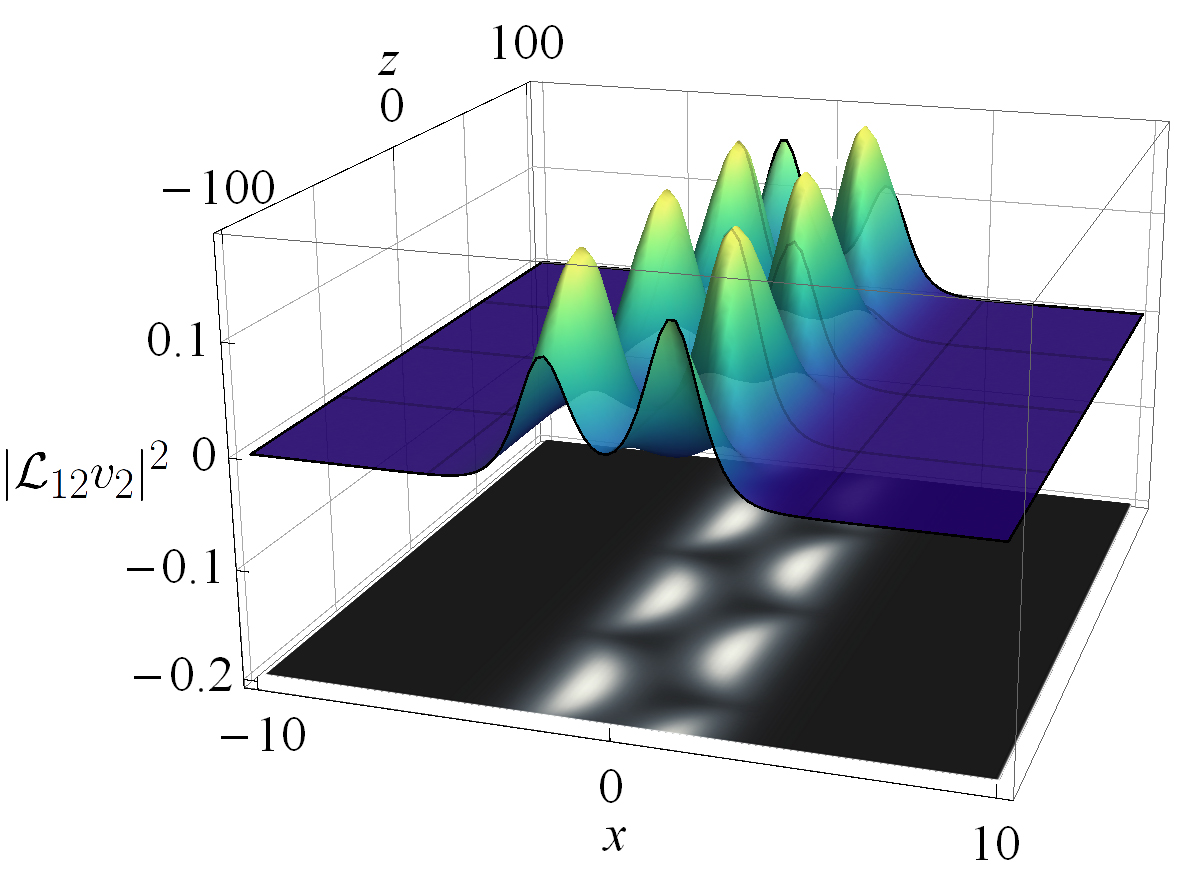}
        \caption{}
    \end{subfigure}	
	\caption{Coupled wave guides. The real (a) and imaginary (b) parts of $V_2(x,z)$, see \eqref{V2indep}, for the parameters $k_1 = 1,~ k_2 = 1.09,~ k_3 = 0.95,$ and $\alpha = 0.5$ are plotted. The intensity densities of the guided modes are displayed, $|\mathcal{L}_{12}v_1|^2$ (c) and $|\mathcal{L}_{12}v_2|^2$ (d), see \eqref{coupled v1} and \eqref{coupled v2}, respectively. } \label{coupled pot}
\end{figure}

We can find two guided modes of $S_2$. They can be obtained as $\mathcal{L}_{12}v_1$ and $\mathcal{L}_{12}v_2$ where the functions $v_1$ and $v_2$ are fixed as
\begin{equation}
v_1= \sinh k_1 x e^{i k_1^2 z}+i \alpha \cosh k_3 xe^{i k_3^2 z},\quad v_2=\cosh k_2xe^{ik_2z}.
\end{equation}
As one can see directly from their explicit form, 
\begin{eqnarray}
\mathcal{L}_{12}v_1&=&e^{ik_2^2z}\left\{\frac{e^{2ik_1^2z}k_1(k_1^2-k_2^2)+\alpha^2e^{2ik_3^2z}k_3(k_3^2-k_2^2)}{W(u_1,u_2)}\sinh k_2x\right. \nonumber \\&&\left. - \frac{i\alpha e^{i(k_1^2+k_3^2)}k_2(k_1^2-k_3^2)\left[\cosh k_2x\cosh (k_1-k_3)x +\frac{k_1k_3-k_2^2}{k_2(k_1-k_3)}\sinh k_2x\sinh(k_1-k_3)x\right]}{W(u_1,u_2)}\right\}, \label{coupled v1} \\
\mathcal{L}_{12}v_2&=&e^{2ik^2_2z}k_2\frac{e^{ik_1^2z}(k_2^2-k_1^2)\cosh k_1 x+i \alpha e^{ik_3^2z}(k_2^2-k_3^2)\sinh k_3x}{W(u_1,u_2)} ,\label{coupled v2}
\end{eqnarray}
they vanish rapidly outside the wave guide and represent the guided modes in the system. These states are illustrated in Fig. \ref{coupled pot}.
%%%%%%%%%%%%%%%%%%%%%%
%%%%%%%%%%%%%%%%%%%%%%
%%%%%%%%%%%%%%%%%%%%%%
%%%%%%%%%%%%%%%%%%%%%%

It is worth noticing that if we take $u_2$ as a first-step transformation function, the potential $V_1=-2\partial_x^2\ln u_2$ would be singular as $u_2$ vanishes at the origin. Hence, the current model is an example where the intermediate potential $V_1$ can be singular, however, the final one $V_2$ is regular.

Let us discuss regularity of the new system. 
The Wronskian $W(u_1,u_2)$ can be written as
\begin{align}
W(u_1,u_2)=e^{i(k_1^2+k_2^2)z}\left(k_1\sinh k_1x\sinh k_2x-k_2\cosh k_1x \cosh k_2x\right) \nonumber \\-i\alpha e^{i(k_2^2+k_3^2)z}\left(k_3\cosh k_3x\sinh k_2 x-k_2 \cosh k_2x \sinh k_3 x\right).
\end{align}
It should be free of zeros for all real $x$ and $z$ in order to have $V_2$ regular. It is convenient to consider its real and imaginary part separately: 
\begin{eqnarray}
\text{Re} \left( \frac{W(u_1,u_2)}{e^{i(k_1^2+k_2^2)z}} \right) &=&-k_2\cosh k_1x\cosh k_2x\left(1-\frac{k_1}{k_2}\tanh k_1x\tanh k_2x\right) \nonumber \\&&-\alpha k_2\sin((k_3^2-k_1^2)z)\cosh k_2x\cosh k_3x\left(\frac{k_3}{k_2}\tanh k_2x-\tanh k_3x\right),\\
\text{Im} \left( \frac{W(u_1,u_2)}{e^{i(k_1^2+k_2^2)z}} \right)&=&\alpha k_2\cos( (k_3^2-k_1^2)z)\cosh k_2x\cosh k_3x\left(\frac{k_3}{k_2}\tanh k_2x-\tanh k_3x\right).
\end{eqnarray}
First, let us focus on the imaginary part $\text{Im}( W(u_1,u_2)).$ One can show \footnote{We have $\partial_x\left(\frac{k_3}{k_2}\tanh k_2x-\tanh k_3x\right)=k_3(\mbox{sech}^2k_3x-\mbox{sech}^2k_2x)$. The monotonicity follows from  $\mbox{sech}^2x_1>\mbox{sech}^2x_2$ whenever $|x_1|<|x_2|$.} that the term in brackets is a monotonic function which is increasing for $|k_2|>|k_3|$, decreasing for $|k_3|>|k_2|$ and it has a single zero at $x=0$. Therefore, Im$( W(u_1,u_2))=0$ for $x=0$ and $z=\frac{(n+1/2)\pi}{k_3^2-k_1^2}$ where $n$ is an integer. 

Considering Re$(W(u_1,u_2))$, we can see that is is nonvanishing for $x=0$ for $k_2\neq 0$. For  $z=\frac{(n+1/2)\pi}{k_3^2-k_1^2}$, the zeros of Re$( W(u_1,u_2))$ coincide with the zeros of 
\begin{equation}\label{intermid}
\frac{\cosh k_1 x}{\cosh k_3 x}\left(1-\frac{k_1}{k_2}\tanh k_1x\tanh k_2x\right)+\alpha \left(\tanh k_3 x-\frac{k_3}{k_2}\tanh k_2x\right).
\end{equation}
Let us suppose that $|k_1|<|k_2|$. Then the first term is positive. We also take $|k_1|>|k_3|$. Then $\frac{\cosh k_1 x}{\cosh k_3 x}>1$ and we can see that the first term is bounded from below by $1-\frac{|k_1|}{|k_2|}$. The second term is bounded from below by $\alpha \left(-1-\frac{|k_3|}{|k_2|}\right)$ and from above by $\alpha \left(1+\frac{|k_3|}{|k_2|}\right)$. So that it is granted that the term (\ref{intermid}) is positive when $\left(1-\frac{|k_1|}{|k_2|}\right)>|\alpha|\left(1+\frac{|k_3|}{|k_2|}\right).$ However, this estimate is very rough and the term remains nodeless (and $V_2$ regular) for larger range of $\alpha$. In the Fig. \ref{coupled pot} we present plots of $V_2$, its real (a) and imaginary (b) parts and also the intensity densities of the guided modes: $|\mathcal{L}_{12}v_1|^2$ (c) and $|\mathcal{L}_{12}v_2|^2$ (d), for the parameters $k_1 = 1,~ k_2 = 1.09,~ k_3 = 0.95,$ and $\alpha = 0.5$.

\subsection{Non-$\mathcal{PT}$-symmetric systems}\label{NonPT}
\begin{figure}[t!] 
	\centering
\begin{subfigure}[b]{0.3\textwidth}
        \includegraphics[width=\textwidth]{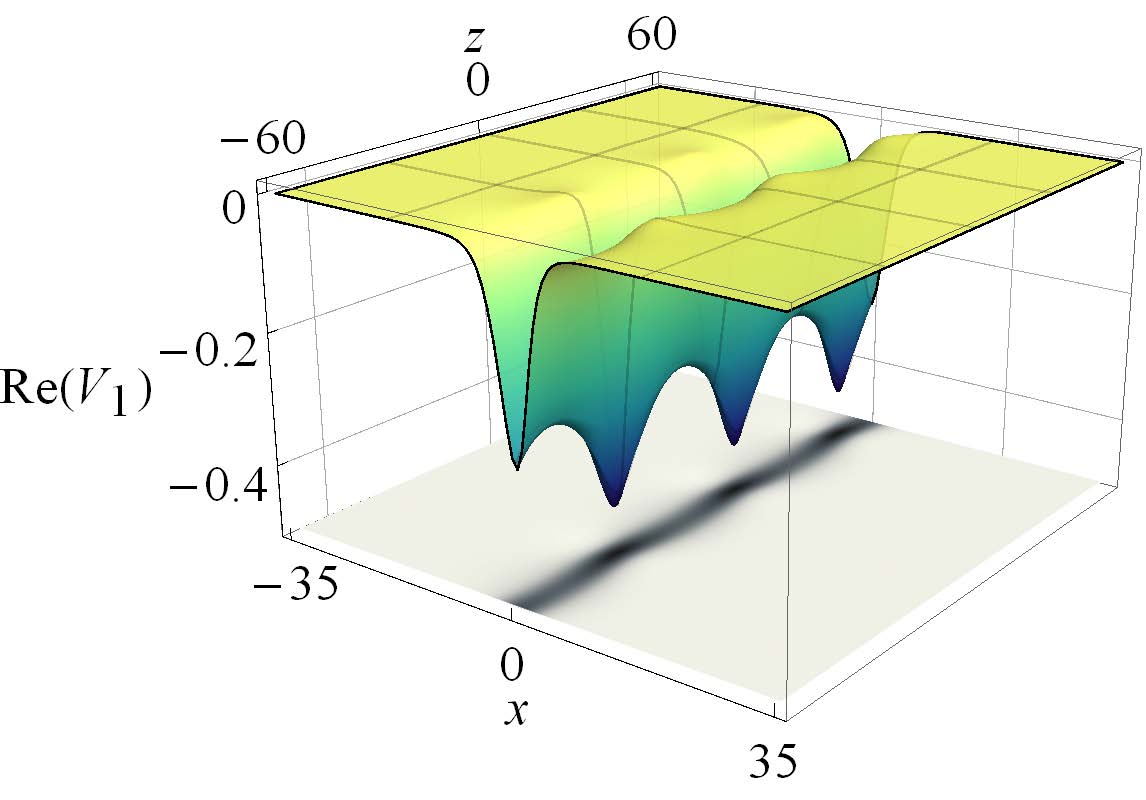}
        \caption{}
    \end{subfigure}
    \begin{subfigure}[b]{0.3\textwidth}
        \includegraphics[width=\textwidth]{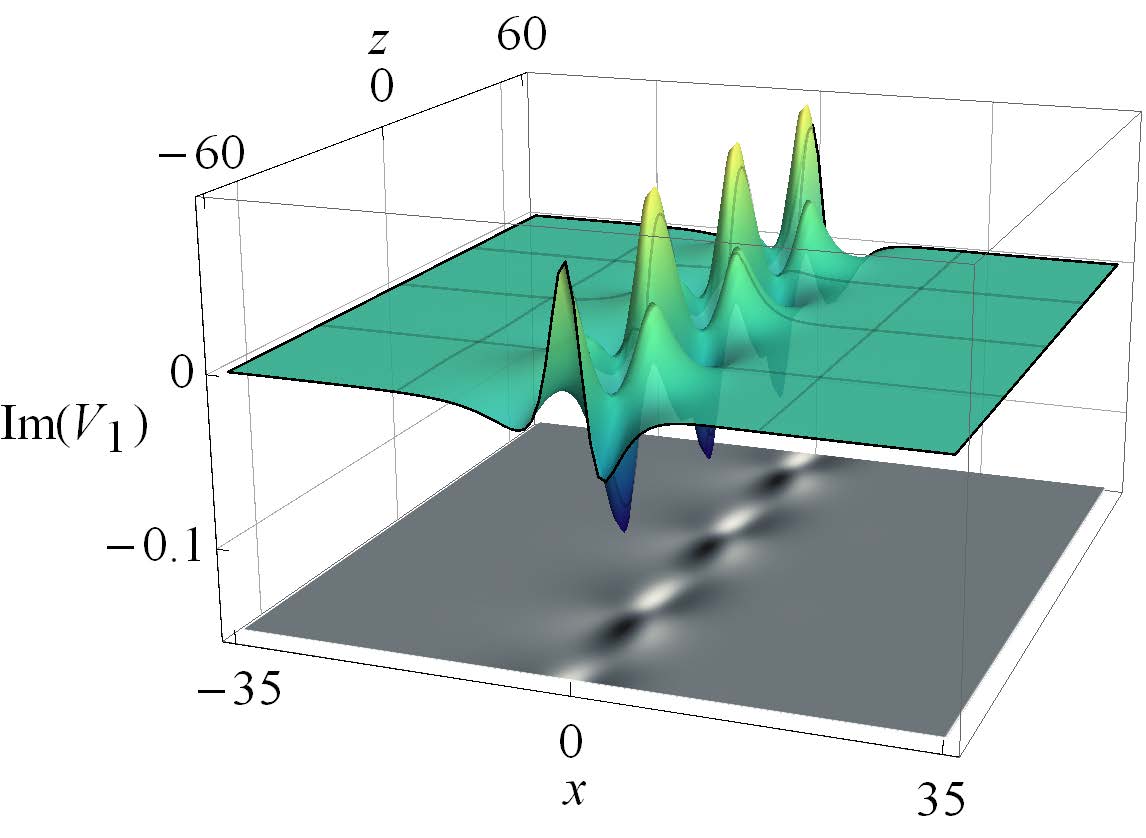}
        \caption{}
    \end{subfigure}\\
    \begin{subfigure}[b]{0.3\textwidth}
        \includegraphics[width=\textwidth]{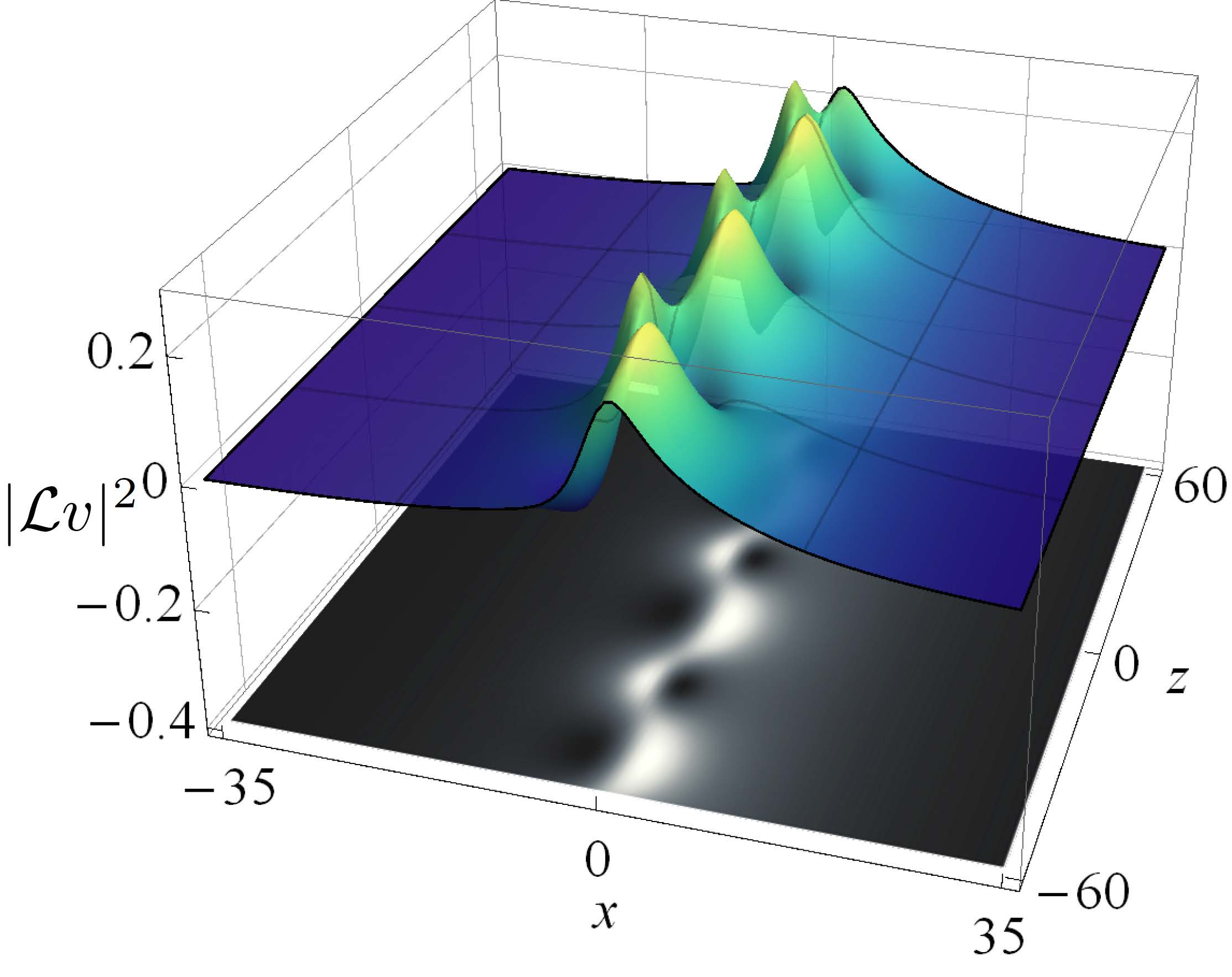}
        \caption{}
    \end{subfigure}
    \begin{subfigure}[b]{0.3\textwidth}
        \includegraphics[width=\textwidth]{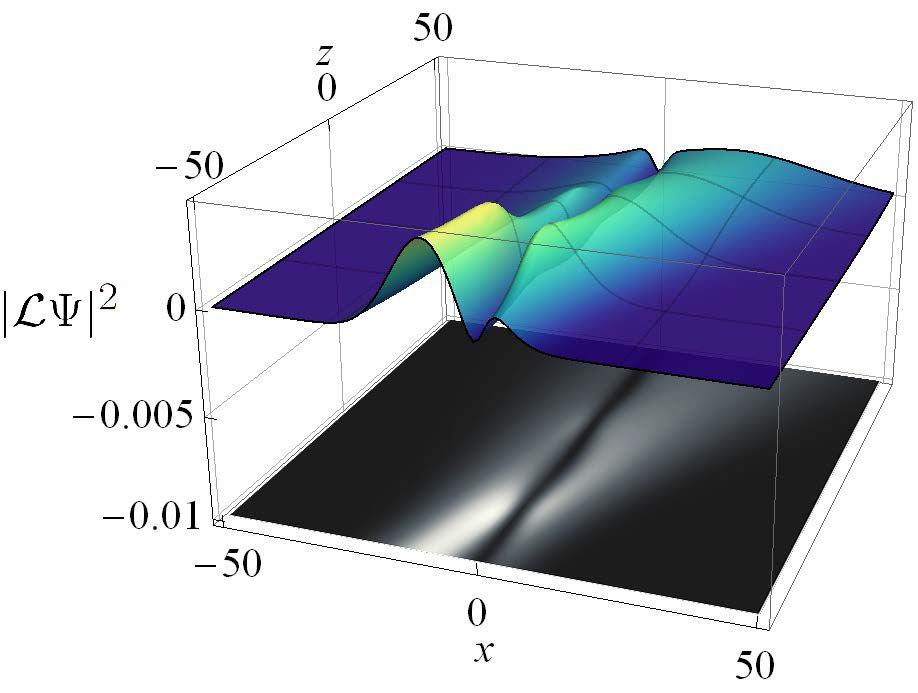}
        \caption{}
    \end{subfigure}	
%	
%	\includegraphics[width=.3\textwidth]{Periodic15b.jpg} 
%	\includegraphics[width=.3\textwidth]{Periodic16b.jpg}\\
%	\includegraphics[width=.3\textwidth]{Periodic17b.jpg}
%	%\includegraphics[width=.3\textwidth]{beamnonPT.jpg}  
%	\includegraphics[width=.3\textwidth]{Fig101.jpg}    
	\caption{A non-$\mathcal{PT}$-symmetric system. Plots of the real (a) and imaginary (b) parts of $V_1$, see \eqref{NonPT V1}, and the intensity density of the guided mode $\mathcal{L}v$ (c)  for the parameters $k_1=0.4, k_2=0.05~, \alpha= 0.1+0.8 i$, see \eqref{NonPT Lv}. The power density $|\mathcal{L}\Psi_{x_0,z_0,v_0,\sigma}|^2$, for the parameters $x_0=-7$, $z_0=-50$, $v_0=-0.2$ and $\sigma=40$ is shown in (d).  } \label{FigV1h4}
\end{figure}

The results of section \ref{4.1} are valid for large class of potentials, including those where the parity-time symmetry is manifestly broken. These systems, where guided modes still exist despite the lack of symmetry, can be constructed in the same vein as the $\mathcal{P}_2\mathcal{T}$-symmetric ones. Let us present briefly a simple example where the transformation function $u$ and the preimage $v$ of the guided mode are fixed as
\begin{equation}
 u=\cosh k_1x e^{ik_1^2z}+\alpha\sinh (k_2x+\delta)e^{ik_2^2z},\quad 
 v=\sinh k_1x e^{ik_1^2z}+\alpha\cosh (k_2x+\delta)e^{ik_2^2z},\quad \delta\in\mathbb{R}.
\end{equation}
We can see that when $\alpha\notin \mathbb{R}$ and $\delta\neq 0$, the function $u$ does not comply with (\ref{uPT2}) and, hence, the resulting potential $V_1$ ceases to be $\mathcal{P}_2\mathcal{T}$-symmetric. Analysis of the range of parameters where $u$ is nodeless can be performed similarly to preceding cases and we will not present it here explicitly. The new potential $V_1$ reads
\begin{eqnarray} \label{NonPT V1}
V_1&=&-2\frac{k_1^2-e^{2i(k_2^2-k_1^2)z}k_2^2\alpha^2}{\cosh^2k_1x\left(1+e^{i(k_2^2-k_1^2)z}\alpha\frac{\sinh (k_2x+\delta)}{\cosh k_1x}\right)^2}\nonumber\\&&-2\frac{\alpha e^{i(k_2^2-k_1^2)z}\cosh k_1x\cosh (k_2x+\delta)\left((k_1^2+k_2^2)\tanh(k_2x+\delta)-2k_1k_2\tanh k_1x\right)}{\cosh^2k_1x\left(1+e^{i(k_2^2-k_1^2)z}\alpha\frac{\sinh (k_2x+\delta)}{\cosh k_1x}\right)^2}.
\end{eqnarray}
The guided mode is obtained in the following form
\begin{eqnarray}
\mathcal{L}v=\frac{e^{2ik_1^2z}k_1-e^{2ik_2^2z}k_2\alpha^2-e^{i(k_1^2+k_2^2)z}(k_1+k_2)\alpha \sinh ((k_1-k_2)x-\delta)}{e^{ik_1^2z}\cosh k_1x\left(1+e^{i(k_2^2-k_1^2)z}\alpha\frac{\sinh (k_2x+\delta)}{\cosh k_1x}\right)}. \label{NonPT Lv}
\end{eqnarray}
Potential $V_1$,  the guided mode and the transformed wave packet $\mathcal{L}\Psi_{x_0,z_0,v_0,\sigma}$ are plotted in Fig.  \ref{FigV1h4}.
%%%%%%%%%%%%%%%%%%%%%%
%%%%%%%%%%%%%%%%%%%%%%
%%%%%%%%%%%%%%%%%%%%%%
%%%%%%%%%%%%%%%%%%%%%%

%\section{Point defects and coupled wave guides via second-order supersymmetry \label{cinco}}

\section{Summary}
The aim of the current article was to construct exactly solvable models of optical setting with complex refractive index, where propagation of light in paraxial approximation is governed by a two-dimensional, non-separable Schr\"odinger equation. We utilized the time-dependent Darboux transformation presented in \cite{Samsonov1}. In Sec. \ref{SUSY and PT}, we discussed its peculiar properties for general class of non-Hermitian systems. In particular, we focused on existence of the missing state and provided its generalized definition (\ref{um}). The framework allowed us to construct systems with localized defects of refractive index that can accommodate localized solutions, called by us light dots, or systems where periodically structured wave guides possess exponentially localized guided modes.

In order to get the models with parity-time symmetry, we considered two different definitions of parity operator; reflection with respect to the axis $x$ denoted as $\mathcal{P}_x$ and reflection with respect to the origin $\mathcal{P}_2$, see (\ref{Ps}). Actual choice of the parity operator determined the whole construction to a large extend. The transformation function $u$, the solution of the initial Schr\"odinger equation in terms of which both the Darboux transformation (\ref{S1L}) and the new potential (\ref{V_1}) were defined, had to comply with either (\ref{uPT1}) or (\ref{uPT2}), dependently on the definition of the parity operator. We showed that existence of the missing state (\ref{um}), which represents a localized state in the new systems, can be granted provided that the transformation function satisfies (\ref{Scond}) where the operator $\mathcal{S}$ is identified with $\mathcal{P}_x\mathcal{T}$.  

In section \ref{tres}, we focused on the construction of systems that can possess localized solutions, light dots. In order to get a transformation function $u$ of required properties, we utilized a mapping between Schr\"odinger equations of the harmonic oscillator and of the free particle. The wave packets  (\ref{u even}), (\ref{u odd}) obtained in this way served as the basis for construction of  solvable models. We presented solvable models of a wave guide with a localized defect (\ref{V1HO}), or with a localized defect of uniform refractive index (\ref{PTHO Example Potential}). We found the light dot solutions for these systems, see Fig.  \ref{Missing pot} and \ref{FigPTHOPotential} %\ref{Segunda estrella} 
for illustration. In this context, it is worth mentioning the Bohmian approach presented in \cite{bohmian}. The authors provided the framework that allows for construction of the potential term of the Helmholtz equation such it has desired solution, e.g. the light dot. When compared by the supersymmetric approach presented in this article, we greatly benefited from the solvability of the initial system. It allows us to obtain (possibly) infinitely many solutions of $S_1f=0$ by direct application of the intertwining operator $\mathcal{L}$.

In section \ref{cuatro}, we provided a general  construction of wave guides that are exponentially vanishing along $x$-axis and periodic along $z$-axis. Existence of the guided modes was analyzed. We showed that dependently on the choice of $u$, the wave guides differ by the strength of confinement. In the strong wave guides, the guided mode is vanishing exponentially in the perpendicular direction to the wave guide. In the weak wave guides, the guided modes leak from the wave guide and exhibit non-vanishing oscillations in transverse direction.  
We illustrated the general results on explicit examples of optical wave guides with two-dimensional fluctuations of refractive index, distinguished by different choices of the transformation function. Strong wave guides were generated in (\ref{V1h1}) and (\ref{V1h2}), a weak wave guide was presented in (\ref{V1t1}).  The presented solvable models were two-dimensional $\mathcal{P}_2\mathcal{T}$-symmetric generalizations of the reflectionless P\"oschl-Teller potential. Indeed, setting $\alpha=0$ in  (\ref{V1h1}), (\ref{V1h2}), (\ref{V1t1}), the expressions reduce to $V_1=-2k_1\mbox{sech}^2 k_1x$. 
We also constructed a system with two coupled wave guides (\ref{V2indep}) and calculated two associated guided modes.

In our constructions, the parity operator $\mathcal{P}_2$, which corresponds to the reflection with respect to origin, proved to be rather universal as all the presented parity-time symmetric systems possessed $\mathcal{P}_2\mathcal{T}$ symmetry. Only two of them, namely (\ref{V1HO}) and (\ref{PTHO Example Potential}) possessed both $\mathcal{P}_x\mathcal{T}$- and $\mathcal{P}_2\mathcal{T}$-symmetry.  
The presented construction of wave guides and guided modes is applicable to a large class of initial systems with an integrable potential.  When compared to the systems separable in Cartesian coordinates, see e.g. \cite{susyrandombands,longhicrossroad}, it allows for construction of the localized defects where the fluctuation of the refractive index is nonzero in a bounded region, or for construction of the straight wave guides where the fluctuation of the refractive index is exponentially vanishing in transverse direction to the wave guide and it is oscillating periodically along the wave guide. It is not restricted to parity-time-symmetric operator, so that it can be utilized for construction of systems where $\mathcal{PT}$ symmetry is manifestly broken. We exemplified construction of such setting in the end of Sec. \ref{NonPT}, see Fig. \ref{FigV1h4}. 

In the analysis of optical systems with separable evolution equations, supersymmetry techniques were used to study effectively one-dimensional settings, see e.g. \cite{MiriPRL}-\cite{Mathias2,LonghiBragg}. The intertwining operator provided a one-to-one mapping\footnote{Up to the state annihilated by the intertwining operator} between the stationary solutions of the original system and those of the new system, preserving the phase of the solutions. 
It was also used to extract a required mode (that corresponded to the kernel of the intertwining operator) such that its analog was missing in the superpartner system. Hence, the spectra of the two settings were either identical (in case of broken SUSY) or almost identical up to a single eigenvalue (the case of unbroken SUSY). The scattering properties of the superpartner systems were analyzed with the use of the intertwining operator whose superpotential was asymptotically constant. 
In our case, we dealt directly with the partial differential equation. The stationary states were not of primary importance in our work. The intertwining operator still provided the matching between the solutions of the original and of the new system. However, its structure was more complicated; the superpotential $\mathcal{W}(x,z)$ in (\ref{S1L}) is a two-dimensional function that is asymptotically non-constant in general, and, hence, it could have a profound impact on the properties of the transformed functions. This led us to implementation of the additional requirement  (\ref{wavepacketpreservation}) that granted boundedness of the transformed wave packets.  
In our work, the intertwining operator was not intended to annihilate a guided mode of the original system. Instead, it was defined such that the associated new system possessed an additional localized solution, the missing state. %The missing state gets annihilated by the inverse intertwining operator $\mathcal{L}^\sharp$, see (\ref{tildeL}). 
Nevertheless, the general framework of the time-dependent supersymmetry could be also utilized for extraction of guided modes from the system. However,  it goes beyond the scope of the current article.

\section*{Appendix: Light dots in  in curved wave guides}  
In the Appendix, we will illustrate the situation where we cannot construct the missing state, however, the system possesses other localized solutions. 
We use  \eqref{Superposition} for definition of $u$, fixing it as a linear combination of $\psi_n$ defined in (\ref{psi_n}). 
We shall consider some properties inherited from the eigenstates of the harmonic oscillator.  First, $\psi_0$ is the only solution without nodes. Second, for all odd $n$, $\psi_{n}(0,z)=0$, and third, two functions $\psi_{m}$ and $\psi_{n}$, $m\neq n$, can vanish simultaneously only at $x=0$. This information helps us to find a set of coefficients such that $u(x,z)\neq 0$. 

Let us take $u=\psi_j + i \alpha \psi_{j+1}$, where $\alpha$ is a real constant and $j$ is a positive integer number, (note that up to a global constant phase these are particular cases of \eqref{Superposition}) 
\begin{eqnarray}
u(x,z) &=  & \frac{1}{\sqrt{\sqrt{2 \pi } 2^j j!}} \frac{1}{(1+z^2)^{1/4}} \exp\left\{\frac{i}{4} \left[\frac{x^2}{z-i}  - 4 \left(j + \frac{1}{2} \right) \arctan(z)  
\right]  \right\} \nonumber \\
&\times & \left[H_j\left( \frac{x}{\sqrt{2 \left(z^2+1\right)}} \right)  + i\frac{\alpha}{\sqrt{2(j+1)}} \left(\frac{1 - iz}{\sqrt{1+z^2}} \right) H_{j+1}\left( \frac{x}{\sqrt{2 \left(z^2+1\right)}} \right)   \right]. \label{u estrellas}
\end{eqnarray}
This function is never zero. It follows from the fact that $H_{j}(x/\sqrt{2(z^2+1)})$ and $H_{j+1}(x/\sqrt{2(z^2+1)})$ are real functions and their zeros do not coincide. Indeed, the imaginary part of the linear combination of the Hermite polynomials in brackets is a real multiple of $H_{j+1}(\cdot)$ whereas the real part is a combination of $H_{j}(\cdot)$ and $H_{j+1}(\cdot)$, so that their zeros are mismatched.

The operator $\mathcal{L}$ in \eqref{S01L} reads 
\begin{eqnarray}
\mathcal{L}=L_1(z)\left( \partial_x +  \frac{x}{2(1+iz)} -  \partial_x \ln \left[H_j\left( \frac{x}{\sqrt{2 \left(z^2+1\right)}} \right)  + i\frac{\alpha}{\sqrt{2(j+1)}} \left(\frac{1 - iz}{\sqrt{1+z^2}} \right) H_{j+1}\left( \frac{x}{\sqrt{2 \left(z^2+1\right)}} \right)   \right]\right).
\end{eqnarray} 
It intertwines $S_0$ with the new Schr\"odinger operator whose potential $V_1$ defined in (\ref{V_1}) can then be written as 
\begin{eqnarray}
V_1=i\partial_z\ln L_1(z)+\frac{1}{1+iz} - 2 \partial_x^2 \ln \left[H_j\left( \frac{x}{\sqrt{2 \left(z^2+1\right)}} \right)  + i\frac{\alpha}{\sqrt{2(j+1)}} \left(\frac{1 - iz}{\sqrt{1+z^2}} \right) H_{j+1}\left( \frac{x}{\sqrt{2 \left(z^2+1\right)}} \right)   \right]. \label{V1 estrella}
\end{eqnarray}
The solutions of the corresponding Schr\"odinger equation can be written as $\phi_n = \mathcal{L} \psi_{n}$ . Using the property $H_n'(y)= 2 n H_{n-1}(y)$ of Hermite polynomials, we can write them as 
\begin{eqnarray}
\phi_n =L_1(z)\left( \frac{\sqrt{n}}{1+iz} \psi_{n-1} - \left( \partial_x \ln \left[H_j\left( \frac{x}{\sqrt{2 \left(z^2+1\right)}} \right)  + i\frac{\alpha}{\sqrt{2(j+1)}} \left(\frac{1 - iz}{\sqrt{1+z^2}} \right) H_{j+1}\left( \frac{x}{\sqrt{2 \left(z^2+1\right)}} \right)   \right] \right) \psi_n\right). \nonumber\label{Light dots sol}
\end{eqnarray}
It follows from the definition of $\mathcal{L}$ that $\phi_j= i \alpha \phi_{j+1}$ for $n\equiv j$. Indeed, on one side we have  $\phi_j= \mathcal{L} \psi_j  =  L_1(z)(\psi_j' - (\ln(\psi_j+i\alpha \psi_{j+1}))'\psi_j) = i L_1(z)\alpha ( \psi_j' \psi_{j+1}-\psi_j \psi_{j+1}')/(\psi_j+i\alpha \psi_{j+1}) $. On the other side, we get $\phi_{j+1}=  \mathcal{L} \psi_{j+1}=   L_1(z)(\psi_{j+1}' - (\ln(\psi_j+i\alpha \psi_{j+1}))'\psi_{j+1} )= L_1(z) ( \psi_j' \psi_{j+1}-\psi_j \psi_{j+1}')/(\psi_j+i\alpha \psi_{j+1})$.

The explicit form of $\mathcal{L}$ suggest we should fix $L_1(z)=O(z)$ in order to satisfy the condition (\ref{wavepacketpreservation}). We fix $L_1=1+iz$ in order to keep  $\mathcal{P}_2\mathcal{T}$-symmetry of the potential $V_1$. This choice also eliminates the first two terms in (\ref{V1 estrella}). The potential is 
a rational function in the $x$ variable for any (positive integer) $j$.  It is remarkable that $|V_1|$ follows a star-like pattern where the number of rays in the ``star-burst'' correlates with the value of $j$.
As we will show below, the asymptotic, star-like behavior of the potential can be understood explicitly with the use of the fact that $V_1$ is rational function in $x$ variable.

\begin{figure}[t]
	\begin{center}
 \begin{subfigure}[b]{0.3\textwidth}
        \includegraphics[width=\textwidth]{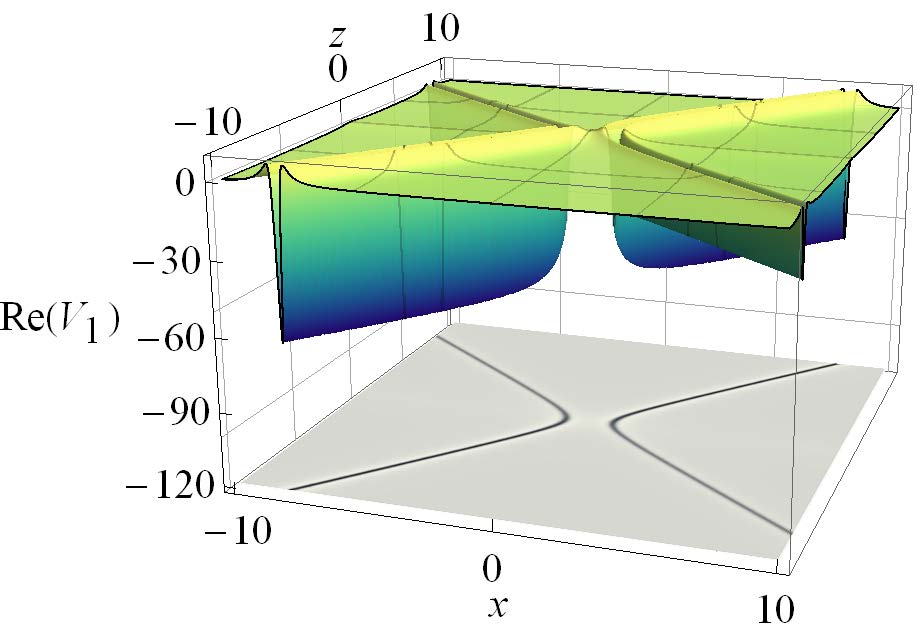}
        \caption{}
    \end{subfigure}	
     \begin{subfigure}[b]{0.3\textwidth}
        \includegraphics[width=\textwidth]{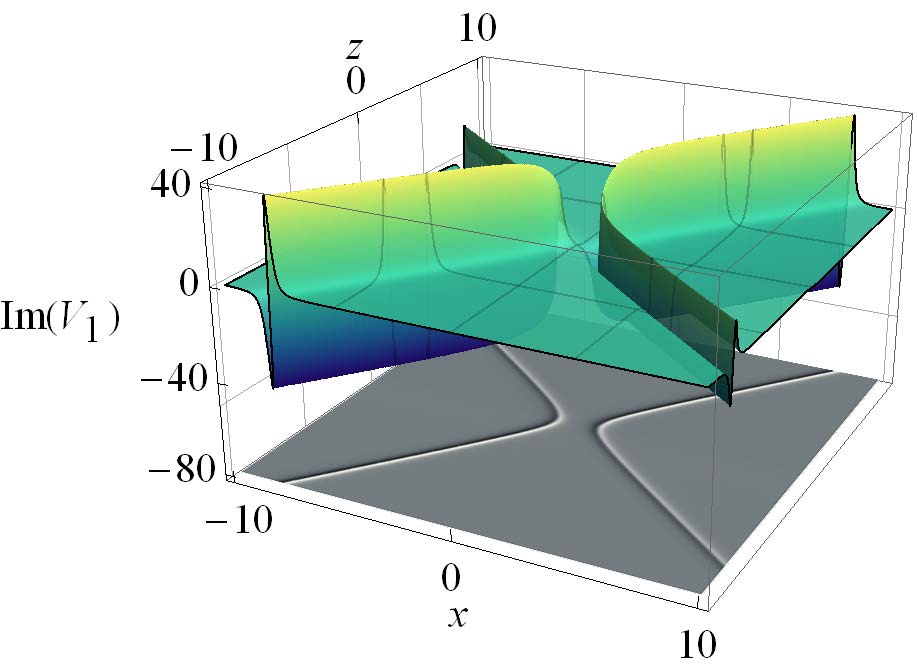}
        \caption{}
    \end{subfigure}	
     \begin{subfigure}[b]{0.3\textwidth}
        \includegraphics[width=\textwidth]{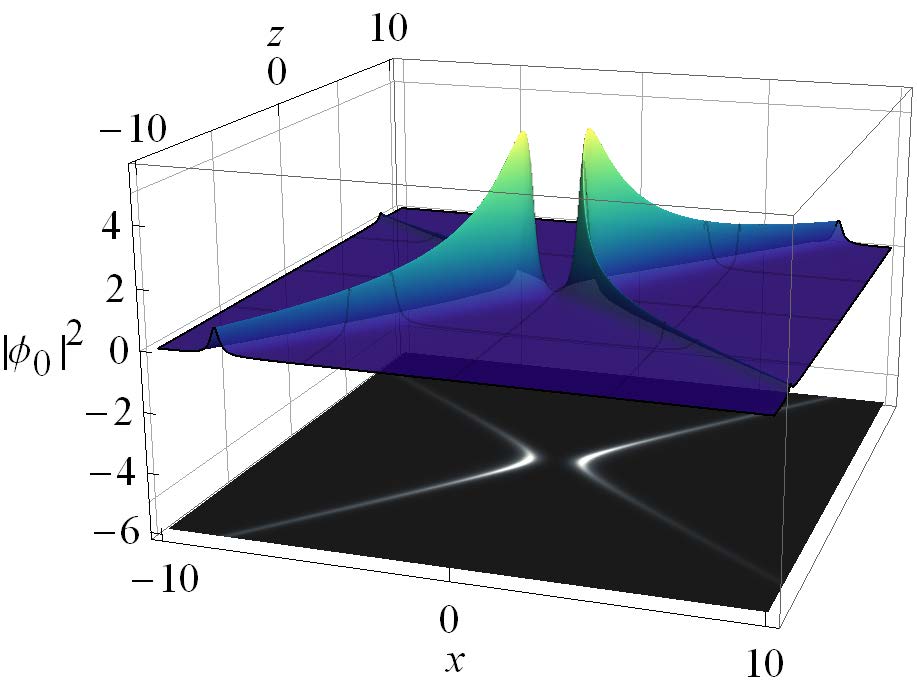}
        \caption{}
    \end{subfigure}	\\
     \begin{subfigure}[b]{0.3\textwidth}
        \includegraphics[width=\textwidth]{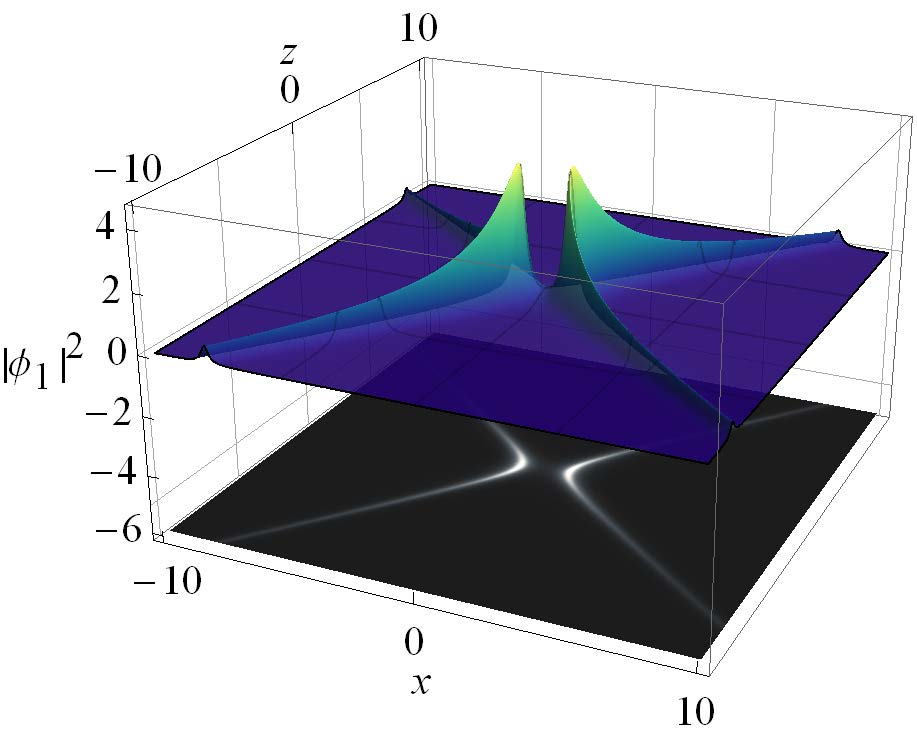}
        \caption{}
    \end{subfigure}	
     \begin{subfigure}[b]{0.3\textwidth}
        \includegraphics[width=\textwidth]{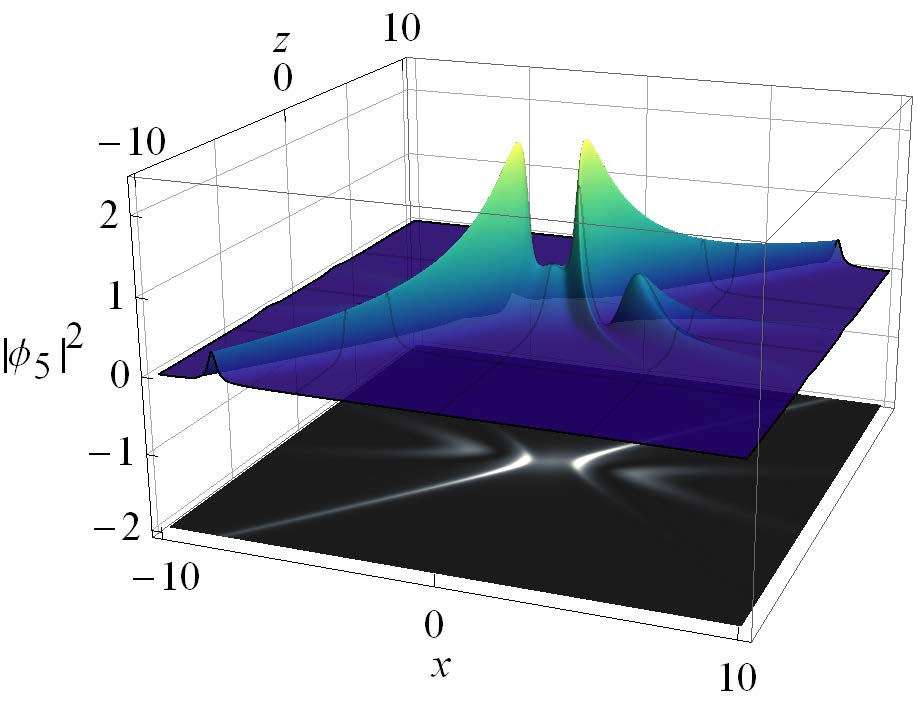}
        \caption{}
    \end{subfigure}		
		\caption{Plots of the real (a) and imaginary (b) parts of $V_1$ for $\alpha=3$, see \eqref{V1 estrellas 2}. Moreover, three different solutions, or light dots, $|\phi_0|^2$ (c), $|\phi_2|^2$ (d) and $|\phi_5|^2$ (e), see \eqref{sol light dots 2}, are shown.} \label{Segunda estrella}
	\end{center} 
\end{figure}

Let us set $u=\psi_1+ i \alpha  \psi_2$ and $L_1=1$. The corresponding potential term $V_1$ and the intertwining operator are:
\begin{eqnarray} \label{V1 estrellas 2}
V_1(x,z)&=&\frac{1}{1+iz}- \frac{4 \alpha}{\alpha x^2 + \sqrt{2}(z-i)x-(1+z^2)\alpha } + 2\left(\frac{2 \alpha x + \sqrt{2}(z-i)}{\alpha x^2 + \sqrt{2}(z-i)x-(1+z^2)\alpha } \right)^2, \nonumber \\
\mathcal{L}&= &\partial_x + \frac{x}{2(1+iz)}- \frac{2 \alpha x + \sqrt{2}(z-i)}{\alpha x^2 + \sqrt{2}(z-i)x-(1+z^2)\alpha}.
\end{eqnarray}
As the explicit formulas suggest,  $V_1$ is regular as the denominators cannot vanish. The potential represents two asymptotically straight wave guides that come together at a specific angle, but they avoid intersection, see the first two graphs ((a) and (b)) in Fig. \ref{Segunda estrella}. It is possible to understand the asymptotic behavior in the following manner: rewriting the potential as a single fraction, we substitute $x=a z+b$. This way, we get a polynomial of order four  in $z$ in the numerator, while there is a polynomial of order five in $z$ in the denominator. We require the polynomials to be of the same order, so that the coefficient of the leading term in the denominator has to vanish. This condition fixes the values of $a$ as $a_\epsilon=\frac{1}{\sqrt{2}\alpha}\left(-1+(-1)^\epsilon\sqrt{1+2\alpha^2}\right) $, $\epsilon=1,2$.
The asymptotic behavior of the potential can be then calculated as
\begin{eqnarray}
&&\lim\limits_{|z|\rightarrow\infty}V_1|_{x\rightarrow a_1 z+b}=\frac{4\alpha^2(1+2\alpha^2)}{(b\alpha\sqrt{2+4\alpha^2}-i(1+\sqrt{1+2\alpha^2}))^2},\\
&&\lim\limits_{|z|\rightarrow\infty}V_1|_{x\rightarrow a_2 z+b}=\frac{4\alpha^2(1+2\alpha^2)}{(b\alpha\sqrt{2+4\alpha^2}-i(-1+\sqrt{1+2\alpha^2}))^2}.
\end{eqnarray}
The asymptotic values are invariant with respect to conjugation joined by substitutions $z\rightarrow -z$ and $b\rightarrow-b$ which is just the manifestation of the $\mathcal{P}_2\mathcal{T}$-symmetry of the potential.

As the transformation function $u$ does not satisfy (\ref{uPT1}), we cannot construct the missing state via (\ref{um}). However, we can still find localized solutions
$\phi_n=\mathcal{L}\psi_n$ that are square integrable for fixed $z$  
\begin{eqnarray} \label{sol light dots 2}
\phi_n= \frac{\sqrt{n}}{1+iz} \psi_{n-1}- \frac{2 \alpha x + \sqrt{2}(z-i)}{\alpha x^2 + \sqrt{2}(z-i)x-(1+z^2)\alpha} \psi_n. 
\end{eqnarray}
They represent light dots that are concentrated at the bending of the wave guides, see Fig. \ref{Segunda estrella} (c)-(d) for illustration.
%%%%%%%%%%%%%%%%%%%%%%
%%%%%%%%%%%%%%%%%%%%%%
%%%%%%%%%%%%%%%%%%%%%%
%%%%%%%%%%%%%%%%%%%%%%

%%%%%%%%%%%%%%%%%%%%%%%%%%%%%%%%%%%%%%%%%%%%%%%%%%%%%%%%
%%%%%%%%%%%%%%%%%%%%%%%%%%%%%%%%%%%%%%%%%%%%%%%%%%%%%%%%%
%%%%%%%%%%%%%%%%%%%%%%%%%%%%%%%%%%%%%%%%%%%%%%%%%%%%%%%%
%%%%%%%%%%%%%%%%%%%%%%%%%%%%%%%%%%%%%%%%%%%%%%%%%%%%%%%%%%5

%\section*{Acknowledgements}

%

\end{document}